\documentclass{aa}

\usepackage{graphicx}
\usepackage{amsmath,epsfig,rotating,amssymb}

\begin{document}

\def \rot{{\rm {\bf rot} }}
\def \grad{{\rm {\bf grad} }}
\def \div{{\rm div}}
\def \cha{\widehat}
\def \pr{{\it permanent}  regime }


\author{ Hennebelle, P.\inst{1}, Commer{\c c}on, B.\inst{2}, Joos, M.\inst{1}, Klessen, R., S.\inst{3}, 
Krumholz, M.\inst{4}, Tan, J., C. \inst{5}, Teyssier, R.\inst{6,7} }

\authorrunning{Hennebelle et al.}

\institute{ Laboratoire de radioastronomie, UMR 8112 du CNRS, 
\newline {\'E}cole normale sup{\'e}rieure et Observatoire de Paris, 24 rue Lhomond,
\newline 75231 Paris cedex 05,
France 
\and 
Max Planck Institute for Astronomy, K{\" o}nigsthul 17, 69117 Heidelberg, Germany
\and
Zentrum f\"ur Astronomie der Universit\"at Heidelberg, Institut f\"ur Theoretische Astrophysik, Albert-Ueberle-Str. 2, 69120
Heidelberg, Germany
\and
Department of Astronomy and Astrophysics, University of California, Santa Cruz, CA 95064, USA
\and
Department of Astronomy, University of Florida, Gainesville, FL 32611, USA
\and
Laboratoire AIM, Paris-Saclay, CEA/IRFU/SAp - CNRS - Universit\'e Paris Diderot, 91191
Gif-sur-Yvette Cedex, France
\and
Institute for theoretical Physics, University of Z\"urich, CH-8057 Z\"urich, Switzerland
}

\offprints{ P. Hennebelle  \\
{\it e-mail:} patrick.hennebelle@ens.fr}   

\title{Collapse, outflows and fragmentation of massive, turbulent and magnetized prestellar barotropic cores}

\titlerunning{Fragmentation of massive prestellar cores}

\abstract{Stars, and more particularly massive stars, have a drastic impact on 
galaxy evolution. Yet the conditions in which they form and collapse are still not 
fully understood.}
{In particular, the influence of the magnetic field on the collapse of massive clumps is relatively 
unexplored, it is therefore of great relevance in the context of the formation of 
massive stars to investigate its impact. }
{We perform high resolution, MHD simulations of the collapse of one hundred solar masses, turbulent and 
magnetized clouds, with the adaptive mesh refinement code RAMSES. We compute various quantities
such as mass distribution, magnetic field, and angular momentum within the 
collapsing core and study the episodic outflows and the fragmentation that 
occurs during the collapse.}
{The magnetic field has a drastic impact on the cloud evolution.
We find that  magnetic braking  is able to substantially reduce the angular 
momentum in the inner part of the collapsing cloud. Fast and episodic 
outflows are being 
launched with typical velocities of the order of 1-3 km s$^{-1}$, although 
the highest velocities can be as high as 20-40 km s$^{-1}$. The fragmentation in several
objects 
is reduced in substantially magnetized clouds with respect to hydrodynamical
ones by a factor of the order of 1.5-2.}
{We conclude that magnetic fields have a significant impact on the evolution 
of massive clumps. In combination with radiation, magnetic fields 
largely determine  the outcome of massive core collapse. We stress that numerical 
convergence of MHD collapse is a challenging issue. In particular, numerical diffusion
appears to be important at high density and therefore could possibly lead to an 
overestimation of the number of fragments.}

\maketitle

\setlength{\unitlength}{1cm}
\begin{figure*} 
\begin{picture} (0,11)
\put(0,0){\includegraphics[width=7.5cm]{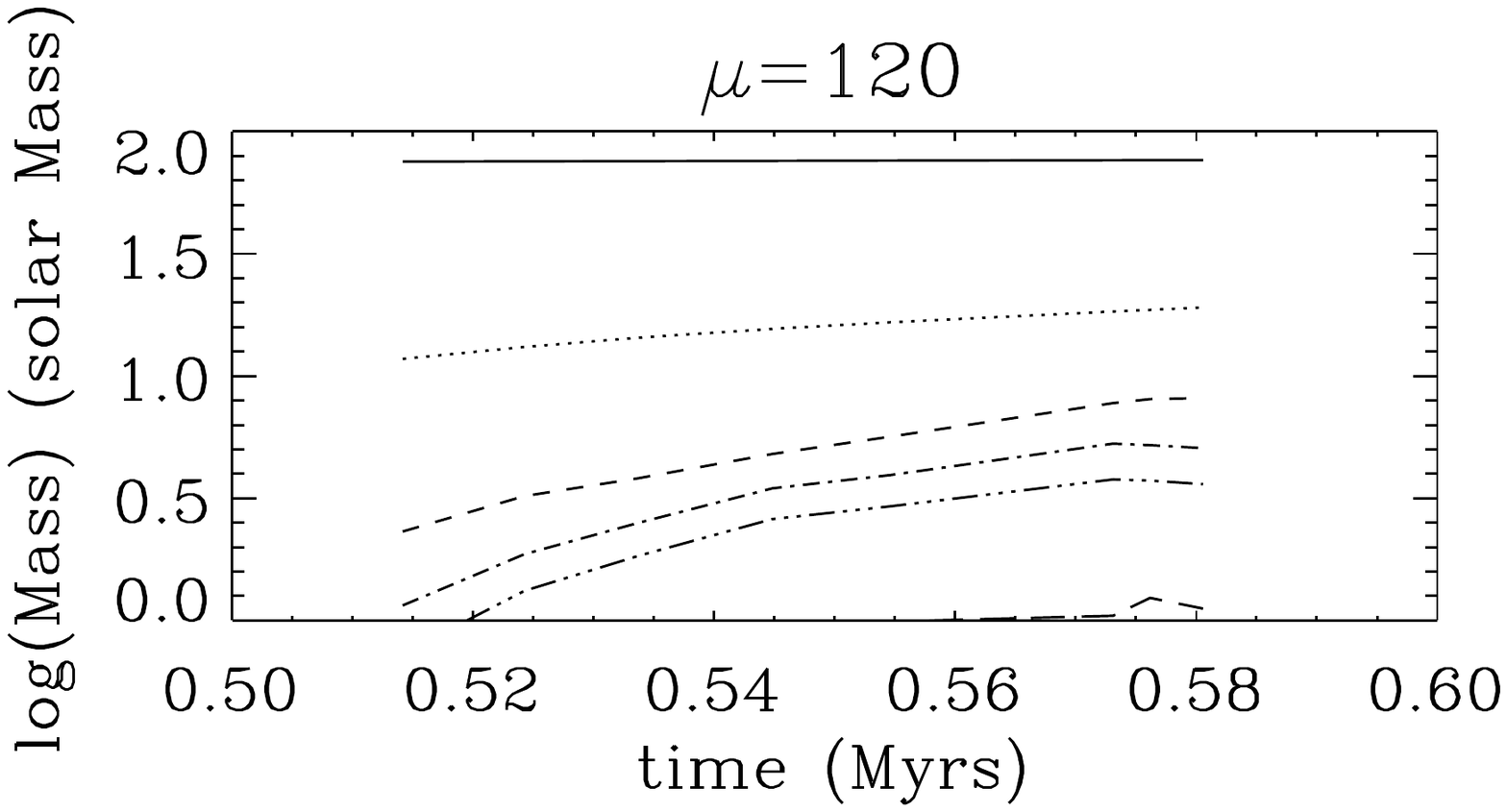}}
\put(8,0){\includegraphics[width=7.5cm]{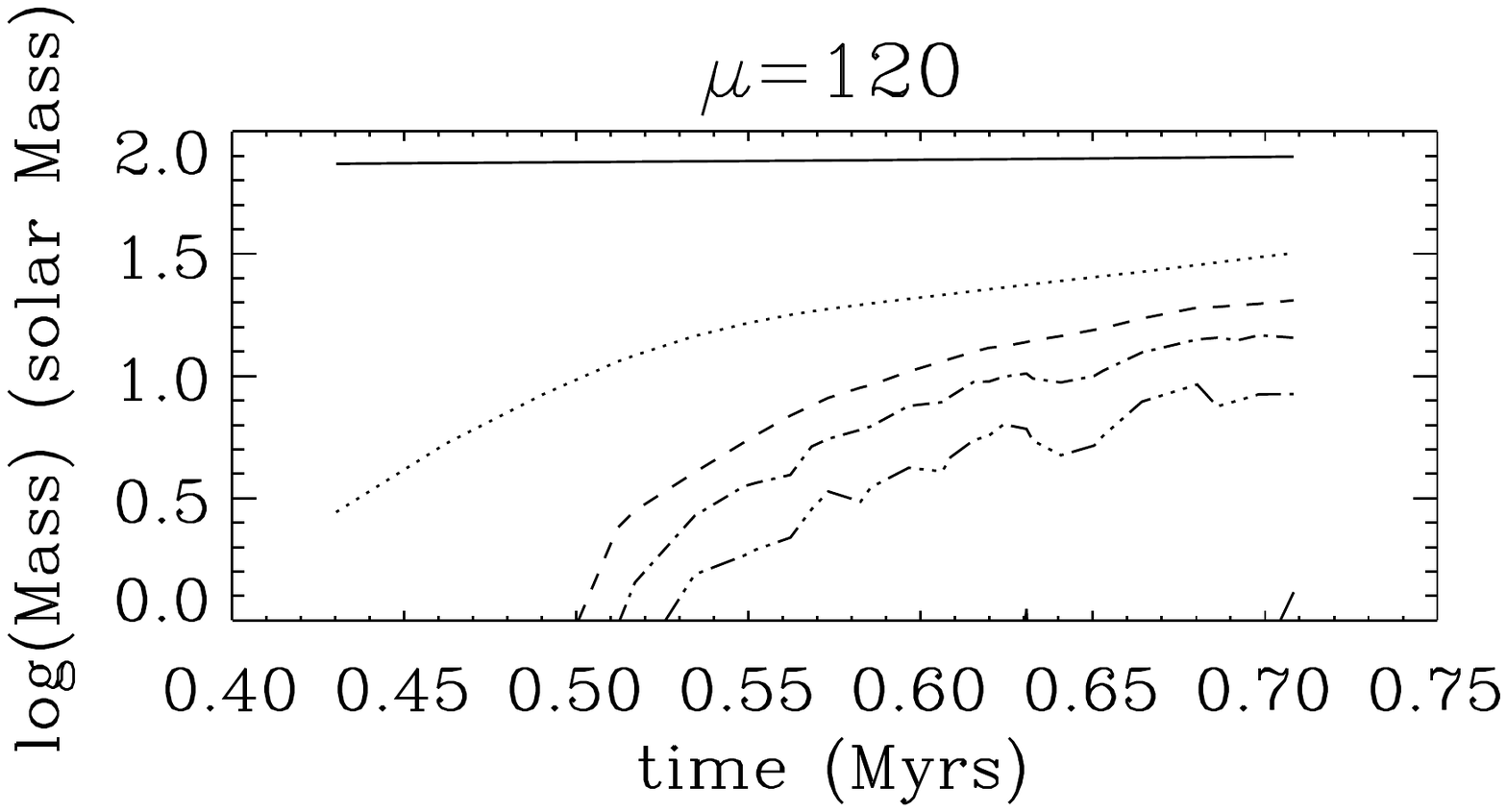}}
\put(0,3.5){\includegraphics[width=7.5cm]{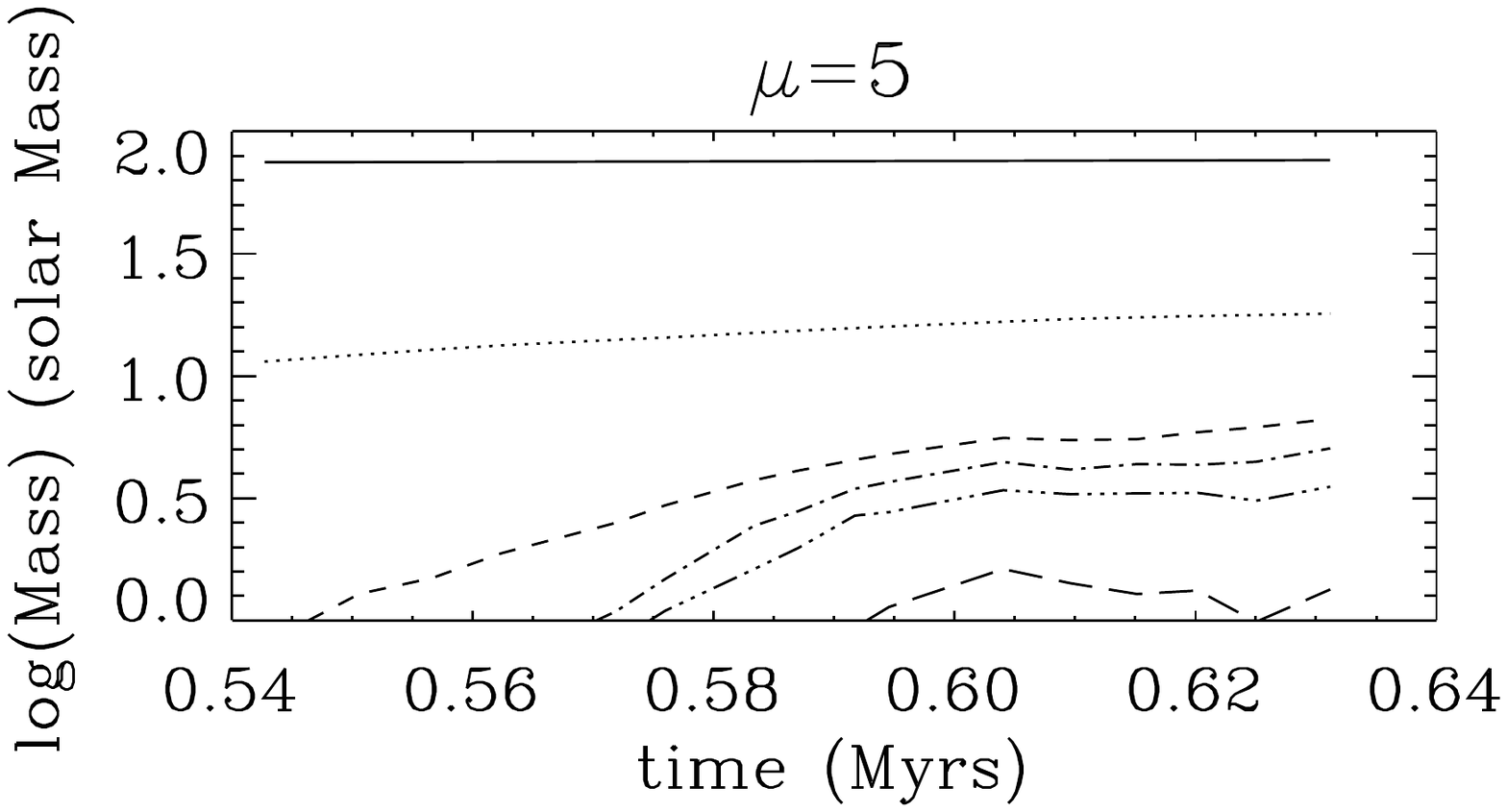}}
\put(8,3.5){\includegraphics[width=7.5cm]{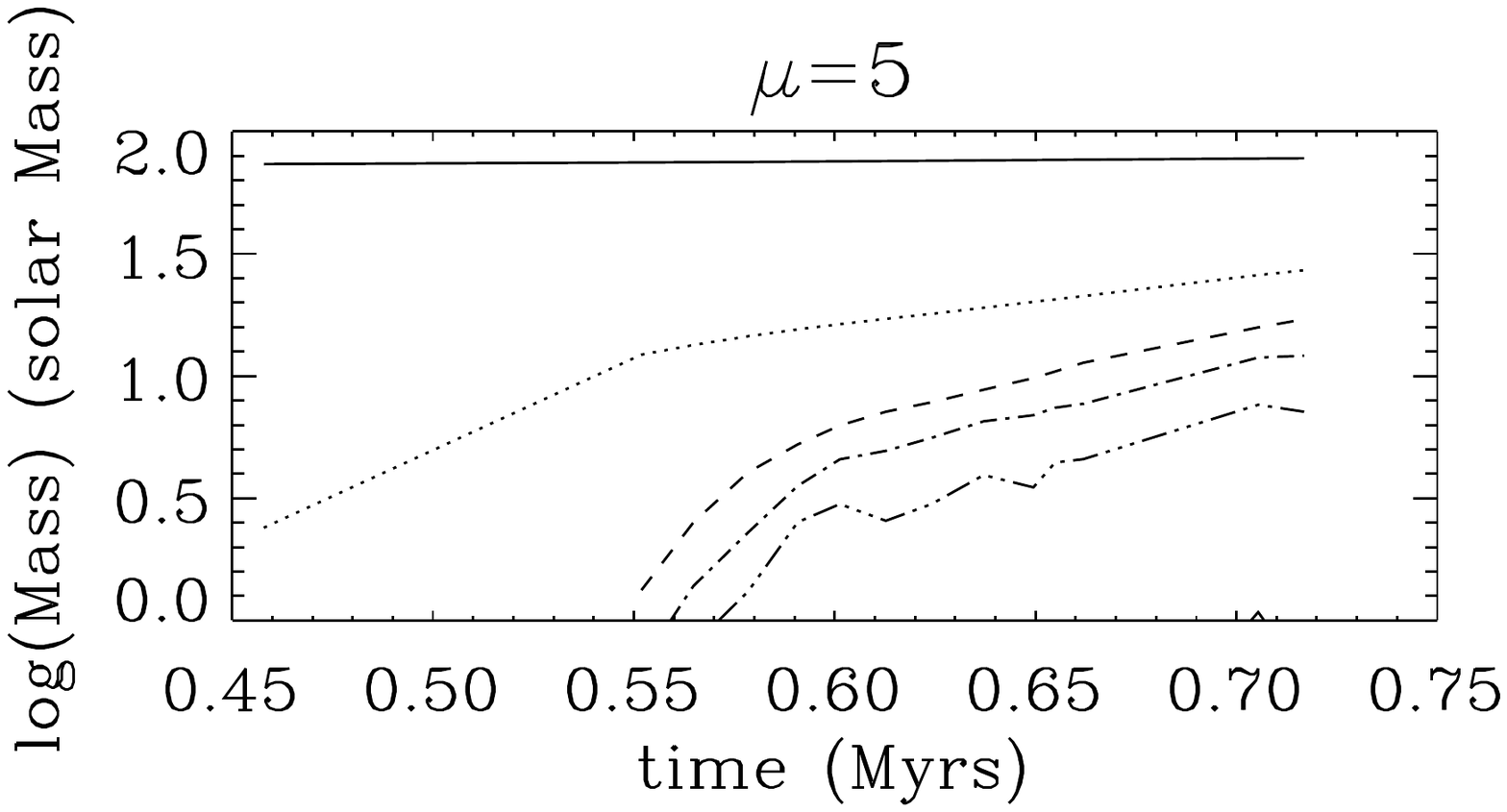}}
\put(0,7){\includegraphics[width=7.5cm]{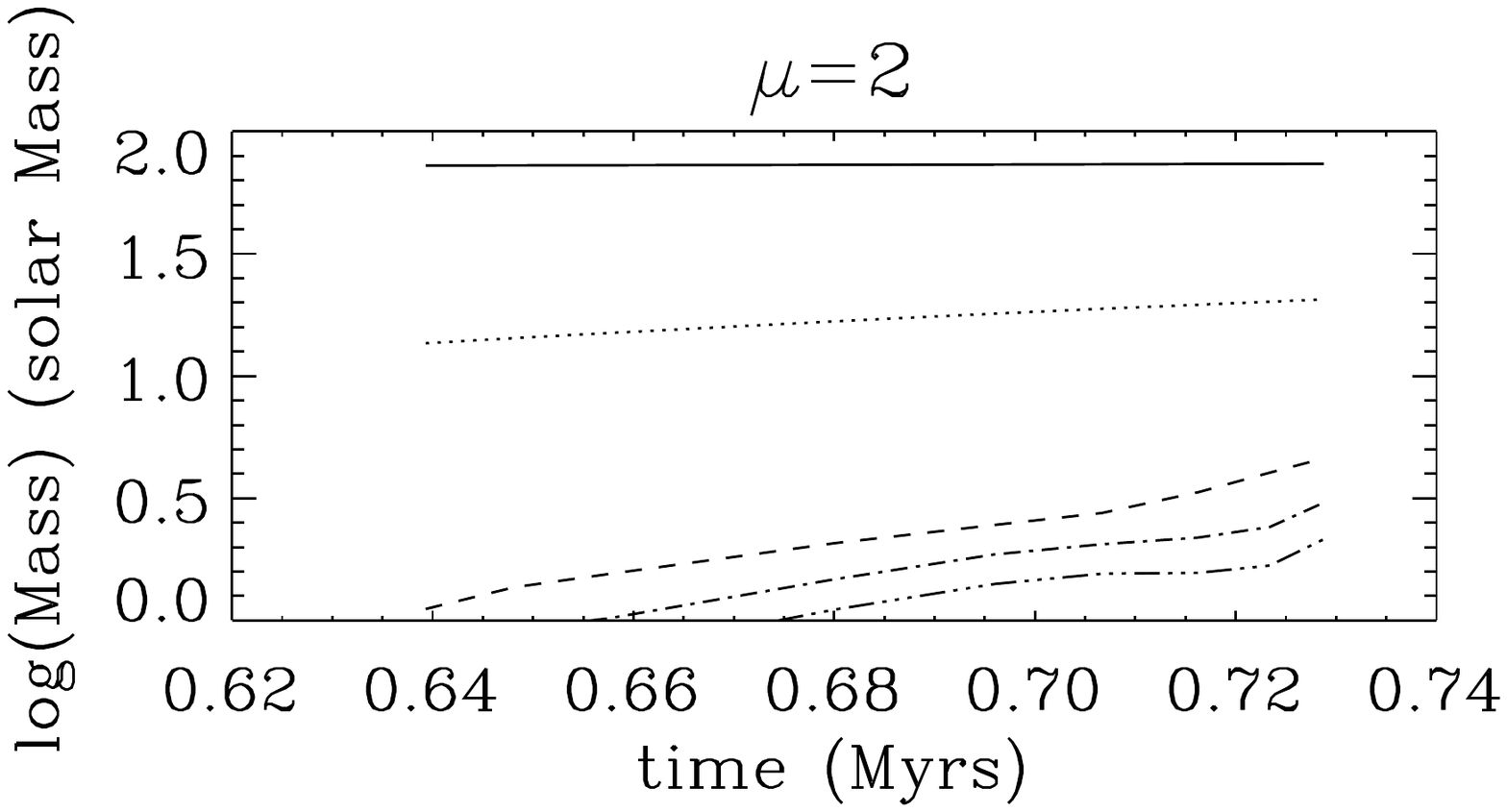}}
\put(8,7){\includegraphics[width=7.5cm]{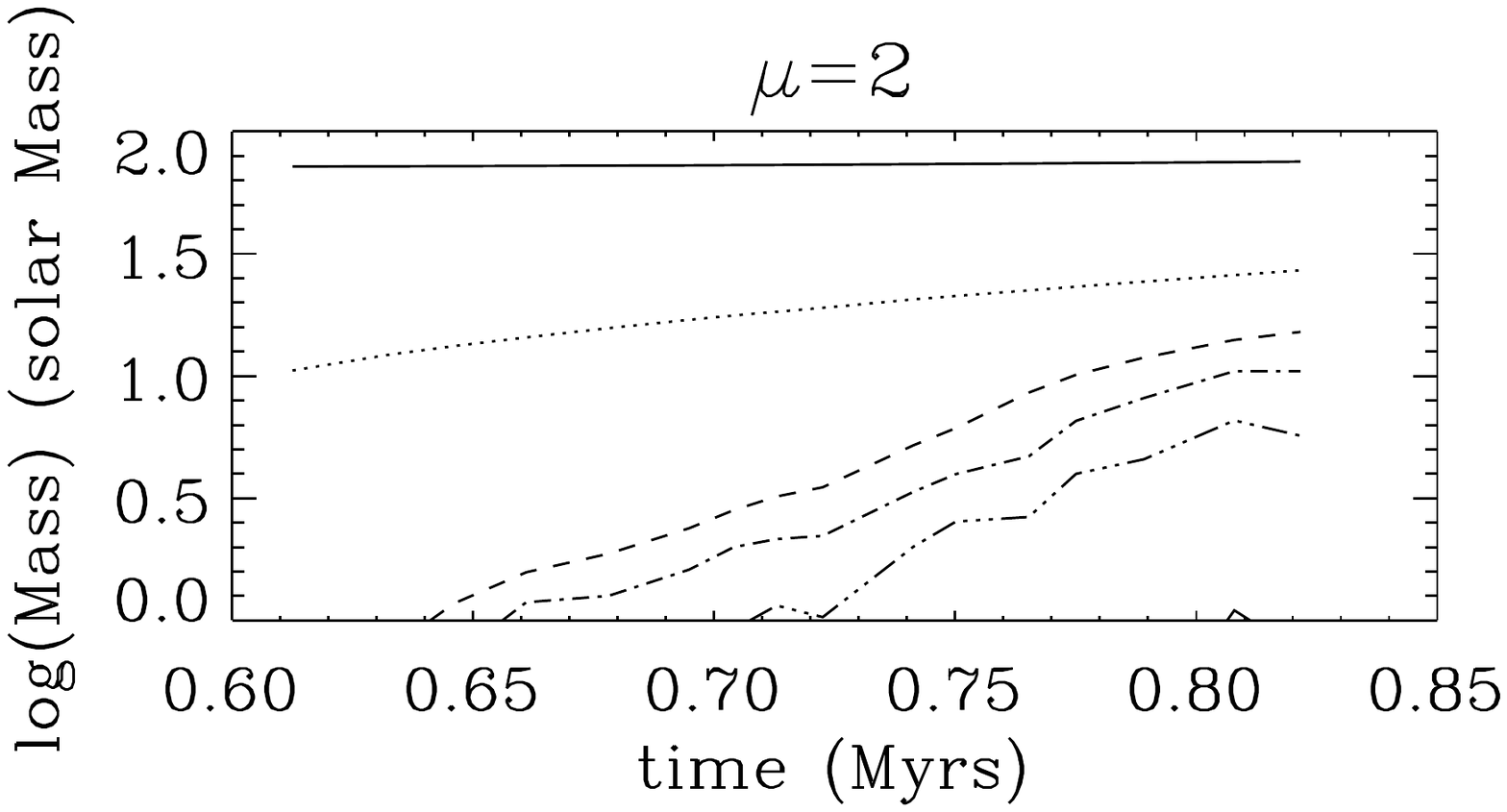}}
\end{picture}
\caption{Total mass above various density thresholds in the 
simulations as a function of time. Solid lines correspond to a density threshold of
$10^3$ cm$^{-3}$, dotted lines to $10^5$ cm$^{-3}$, dashed to $10^{7}$ cm$^{-3}$, 
dot-dashed to 
$10^{9}$ cm$^{-3}$, triple dot-dashed to $10^{11}$ cm$^{-3}$ and
long-dashed to $10^{13}$ cm$^{-3}$.
 The left column shows the high resolution simulations, while 
the right column shows the lower resolution.
Top panels display the $\mu=2$ case, middle panels the $\mu=5$ ones, while
bottom panels display the $\mu=120$ case.}
\label{mass}
\end{figure*}

\section{Introduction}

It is believed that stars form during the collapse of prestellar cores
inside molecular clouds. Understanding this  process is of great 
relevance as it determines the initial conditions of the protostars 
as well as the properties of accretion disks which form in their 
vicinity.
 It is also during this process that the fragmentation, that is
the formation of  binaries and clusters rather than a single object, 
may occur. During the last decades, many studies have been 
investigating the fragmentation of dense cores using either 
smooth particles hydrodynamic (SPH) based codes  or grid codes (see e.g. Matsumoto \& Hanawa, 2003, 
Commer{\c c}on et al. 2008, or Goodwin et al. 2007 for a review). 
Until recently, most works have been neglecting the magnetic field 
and assume an isothermal  equation of state until the gas becomes 
optically thick. Under these conditions, various studies infer
that the massive cores fragment into several objects. Simulations like the
ones performed by  Bonnell et al. (2004), Klessen \& Burkert (2000, 2001) and
Dobbs et al. (2005) generally  find that the number of fragments 
is comparable with or even exceeds  the number of initial 
Jeans masses within the clouds, which implies that a massive core
may result in a cluster that contains tens objects or even more.

Observationally the question as to whether massive cores are fragmented 
is difficult to investigate because of the large distances at which 
these objects are located. Preliminary investigations  
do not appear to show such high  
levels of fragmentation. For example, Bontemps et al. (2010) report  
1700 AU-resolution observations using PdBI of IR-quiet massive cores  
in Cygnus X, and find that although one of them does break up to some  
degree when observed at high resolution, most of them do not have most  
of their  collapsed mass in low mass objects. 
Some of them do not break up at  
all, and remain single compact objects even at 1700 AU resolution.  
Recent submillimeter array  observations by Longmore et al. (2010) reach similar  
conclusions: there is some fragmentation in massive cores,   but 
the number of objects remains limited.
Although higher resolution observations need to be performed before
definite conclusions can be reached, 
it is important to investigate which physical processes could  
reduce fragmentation substantially.

Although it has early been recognized that the magnetic field 
and the stellar feedback, e.g. the heating or even ionization of the gas caused by  the 
radiation emanating from the protostars should both play an active role in 
the cloud evolution in particular regarding the fragmentation, 
it is only recently that the progresses
of numerical algorithms and the increase of the computing power
have permitted  this problem to be addressed numerically. 
The impact of radiative feedback on fragmentation has been investigated analytically
by Krumholz (2006) and Krumholz \& McKee (2008), and numerically by Krumholz et al.
(2007, 2010), Bate (2009), Offner et al. (2009), Urban et al. (2010), Kuiper et al. (2010)
and Peters et al. (2010abcd). All authors agree that the radiative heating increases the Jeans mass
and changes the effective equation of state, reducing the degree of fragmentation
and leads to the formation of higher-mass stars. The quantitative effect on the
stellar cluster formation, however, differs substantially among the simulations. It
is unclear to what extent these differences result from differences in the numerical
schemes used to treat the radiation and to what extent it reflects differing initial
conditions (Girichidis et al. 2010).
 For low mass stars, while Offner et al. (2009)
conclude that the protostellar feedback is still 
sufficient to heat the gas substantially and 
therefore stabilizing the disk efficiently, Stamatellos  et al. (2009)
conclude that the disks are fragmenting. The differences of 
these studies are not clear yet and could be owing to the 
initial conditions or the absence of feedback in Stamatellos et al. (2009),
as recently suggested by Offner et al. (2010).

The effect of the magnetic field on the low mass core fragmentation,
assuming ideal MHD, has 
been considered by Hosking \& Whitworth (2004),
 Machida et al. (2005), Price \& Bate (2007),
Hennebelle \& Teyssier (2008) and Duffin \& Pudritz (2009).  
They generally conclude that even modest values of the magnetic field 
corresponding to high values of the mass-to-flux over critical 
mass-to-flux ratio, $\mu$, can deeply impact the fragmentation 
and even suppress it when density perturbation of modest amplitude
are initially seeded in the core. This is because in typical 
hydrodynamical simulations of low mass cores, the dominant modes
of fragmentation are rotationally driven, i.e. 
induced by the formation of massive strongly gravitationally unstable disks.
The magnetic field can efficiently suppress this mode of fragmentation 
because i) magnetic braking extracts the angular momentum possibly 
suppressing the disk formation  ii) when 
the field is too weak to prevent  disk formation, the azimuthal 
component of the magnetic field is quickly  amplified by the 
differential rotation, which stabilizes the disk. 
Few studies have explored the influence of non ideal MHD 
effects. Machida et al. (2008) include ohmic dissipation and find that 
binaries may form during the second collapse, while Duffin \& Pudritz 
(2009) consider ambipolar diffusion and find that in a highly 
rotating case, two fragments instead of one 
form when  ambipolar diffusion is included.

In the context of massive cores (Beuther et al. 2002a, Motte et al. 2007,
Wu et al. 2010, Csengeri et al. 2010), the influence of the magnetic field 
is not extensively explored yet because only few studies have been performed
(e.g. Banerjee \& Pudritz 2007) in spite of the measurement which suggest 
that it reaches substantial values (Crutcher 1999, Falgarone et al. 2008, 
Girart et al. 2009). 
Studying the impact that in particular magnetic fields 
 may  have in this context  is important as well because 
the massive cores present differences with regard to the low mass ones. 
First, massive cores are expected to contain initially more 
Jeans masses because the thermal over gravitational energy ratio 
is lower in these objects. Second, massive cores are expected 
to be much more turbulent (e.g. McKee \& Tan 2003, Wu et al. 2010) 
than low mass cores 
in which usually sonic or subsonic motions are observed. 
We also stress that no one has yet been investigating in detail 
the influence that magnetic braking may have in a turbulent core.
 Note nevertheless that Matsumoto \& Hanawa (2010)  recently 
investigated the collapse of low mass, magnetized, and turbulent cores.
The purpose of this paper is to address these issues for  massive 
cores assuming a simple barotropic equation of state. 
While there is little doubt that radiative transfer 
is playing a major role in the evolution of massive cores (although 
the exact influence of an outflow cavity along which the radiation
may escape remains to be understood e.g. Krumholz et al. 2005), 
it seems necessary to consider the various effects separately before
treating them altogether. We note that recently Commer{\c c}on 
et al. (2010), Tomida et al. (2010) and Peters et al. (2010d)
 have performed the first 
simulations of collapsing low mass cores at small scales
with grid techniques,
 which simultaneously consider the 
magnetic field and the radiative transfer, while Price \& Bate (2009)
have been performing these simulations on larger scales with SPH.

The plan of the paper is as follows. In the second section we
describe the initial conditions and the numerical method
we use. In the third section we discuss the evolution 
of the various core properties, such as density, angular momentum,
 and magnetic field during collapse. The fourth section is devoted
to the study of the outflows, which are spontaneously launched 
in our calculations, while in the fifth section we investigate
 the fragmentation, which occurs in the cores. In the sixth section, 
we discuss the various restrictions of this work that we will
need to improve. The seventh section concludes the paper.

\setlength{\unitlength}{1cm}
\begin{figure} 
\begin{picture} (0,21)
\put(0,14){\includegraphics[width=7.5cm]{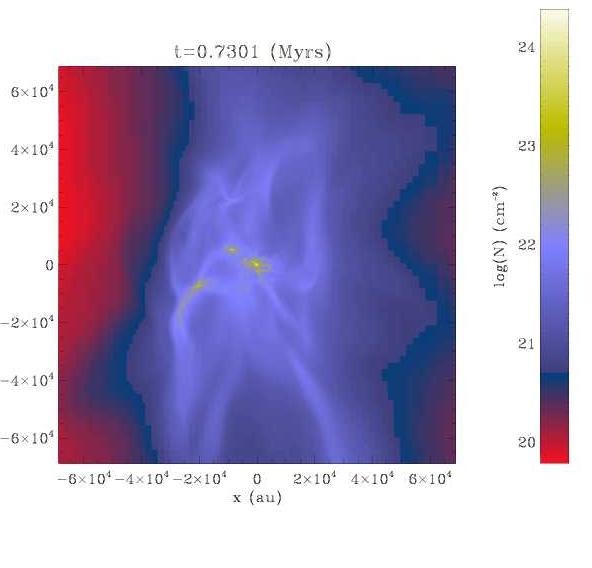}}
\put(0,7){\includegraphics[width=7.5cm]{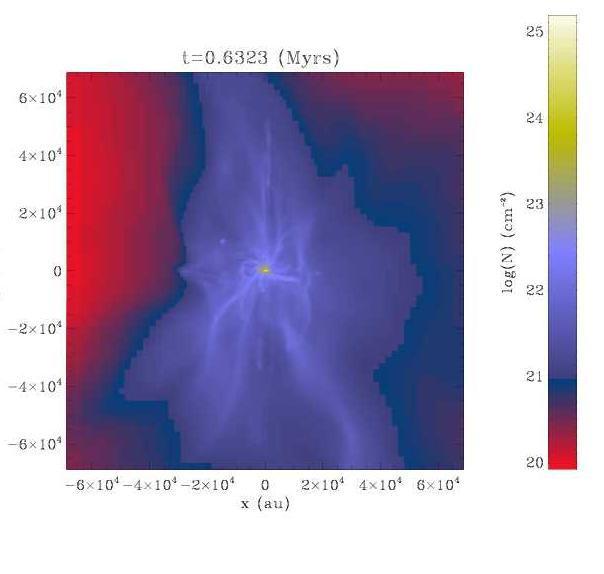}}
\put(0,0){\includegraphics[width=7.5cm]{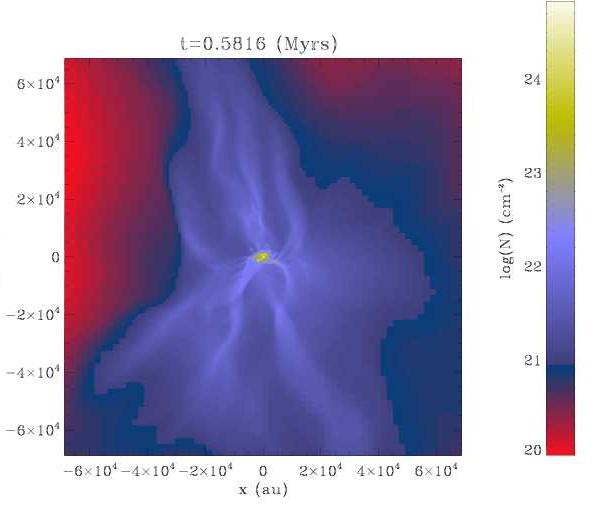}}
\end{picture}
\caption{Core column density for $\mu=120$  (bottom panel),
$\mu=5$  (middle panel) and $\mu=2$  (top panel)  along the z-axis.}
\label{col_dens}
\end{figure}

\setlength{\unitlength}{1cm}
\begin{figure} 
\begin{picture} (0,11)
\put(0,0){\includegraphics[width=7.5cm]{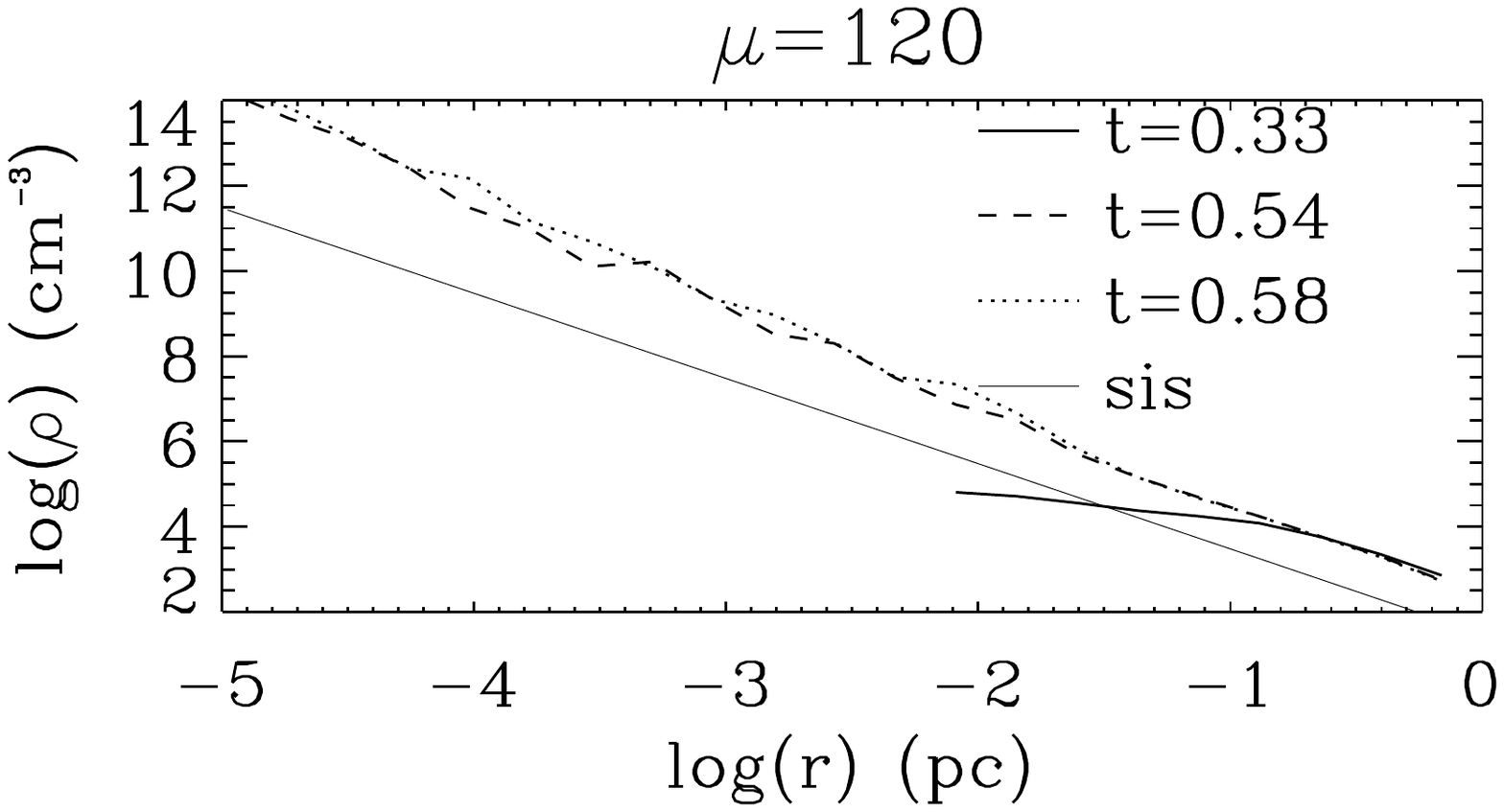}}
\put(0,3.5){\includegraphics[width=7.5cm]{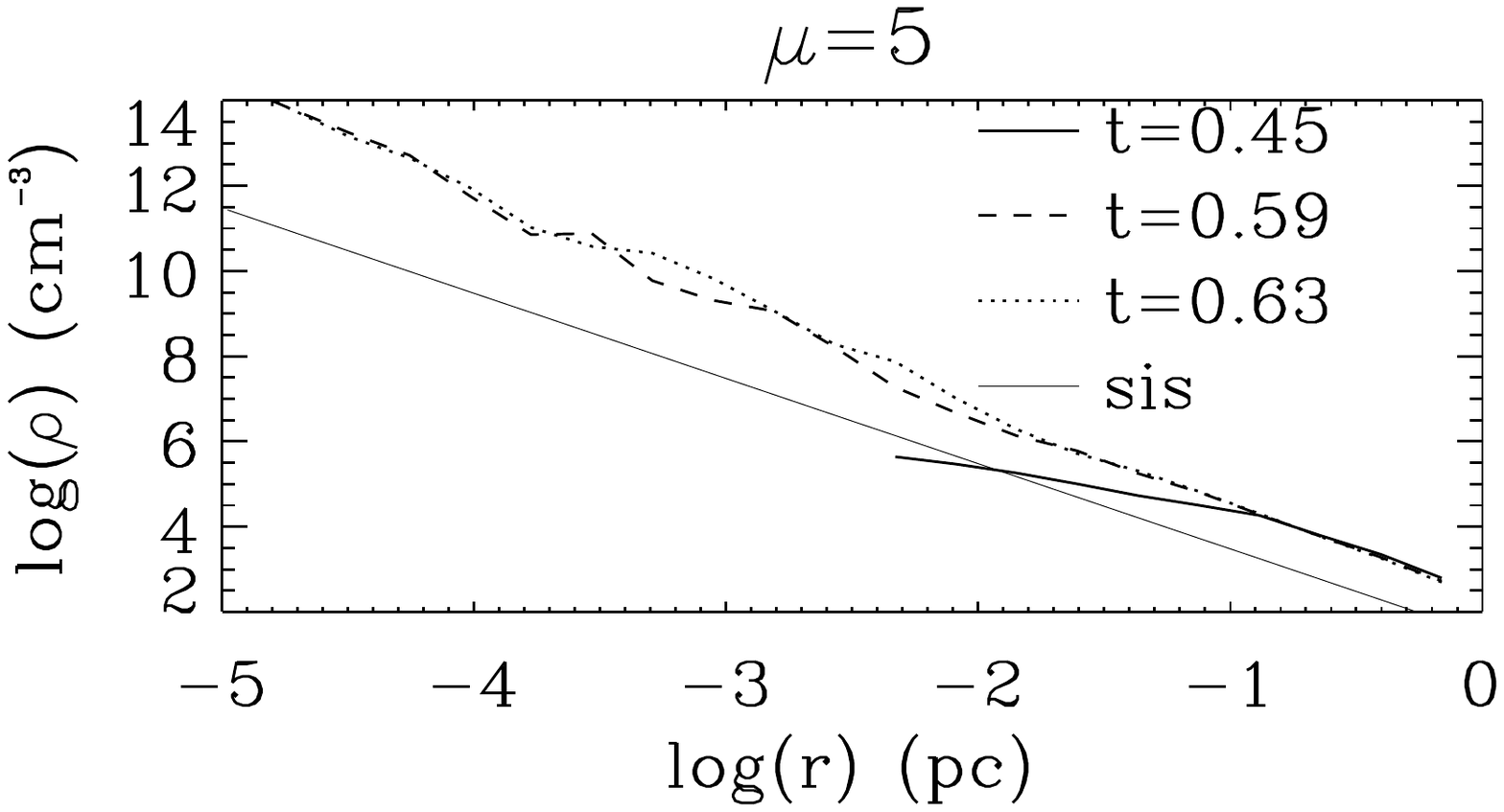}}
\put(0,7){\includegraphics[width=7.5cm]{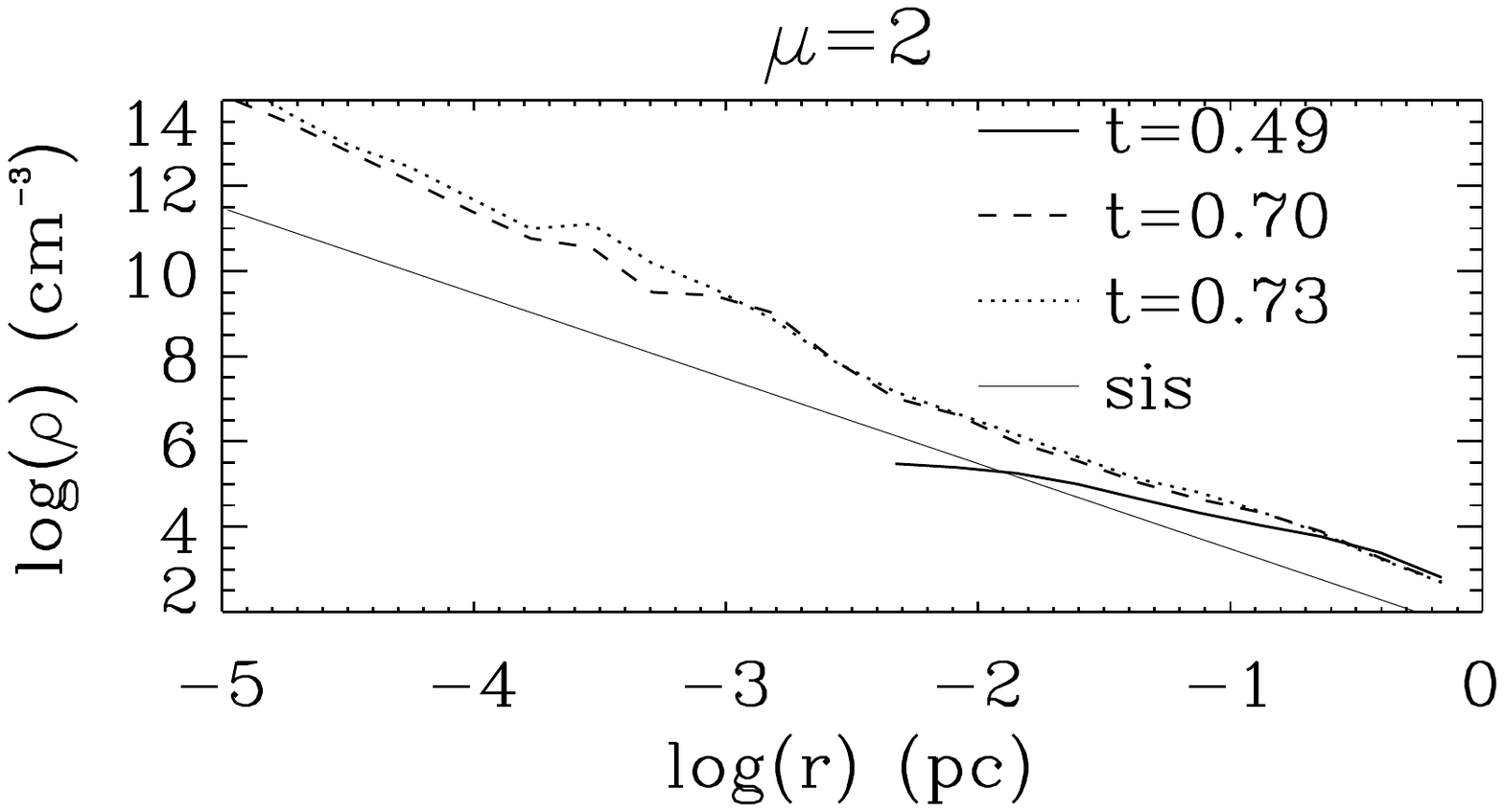}}
\end{picture}
\caption{Mean gas density within a sphere of radius $r$ as a function 
of $r$ for three different timesteps  of the high resolution runs.
The solid line is before the 
protostar formation while dotted and dashed lines correspond to 
later times. The straight line corresponds to the 
density of the singular isothermal sphere. The times are in Myr.}
\label{density}
\end{figure}

\setlength{\unitlength}{1cm}
\begin{figure} [t]
\begin{picture} (0,11)
\put(0,0){\includegraphics[width=7.5cm]{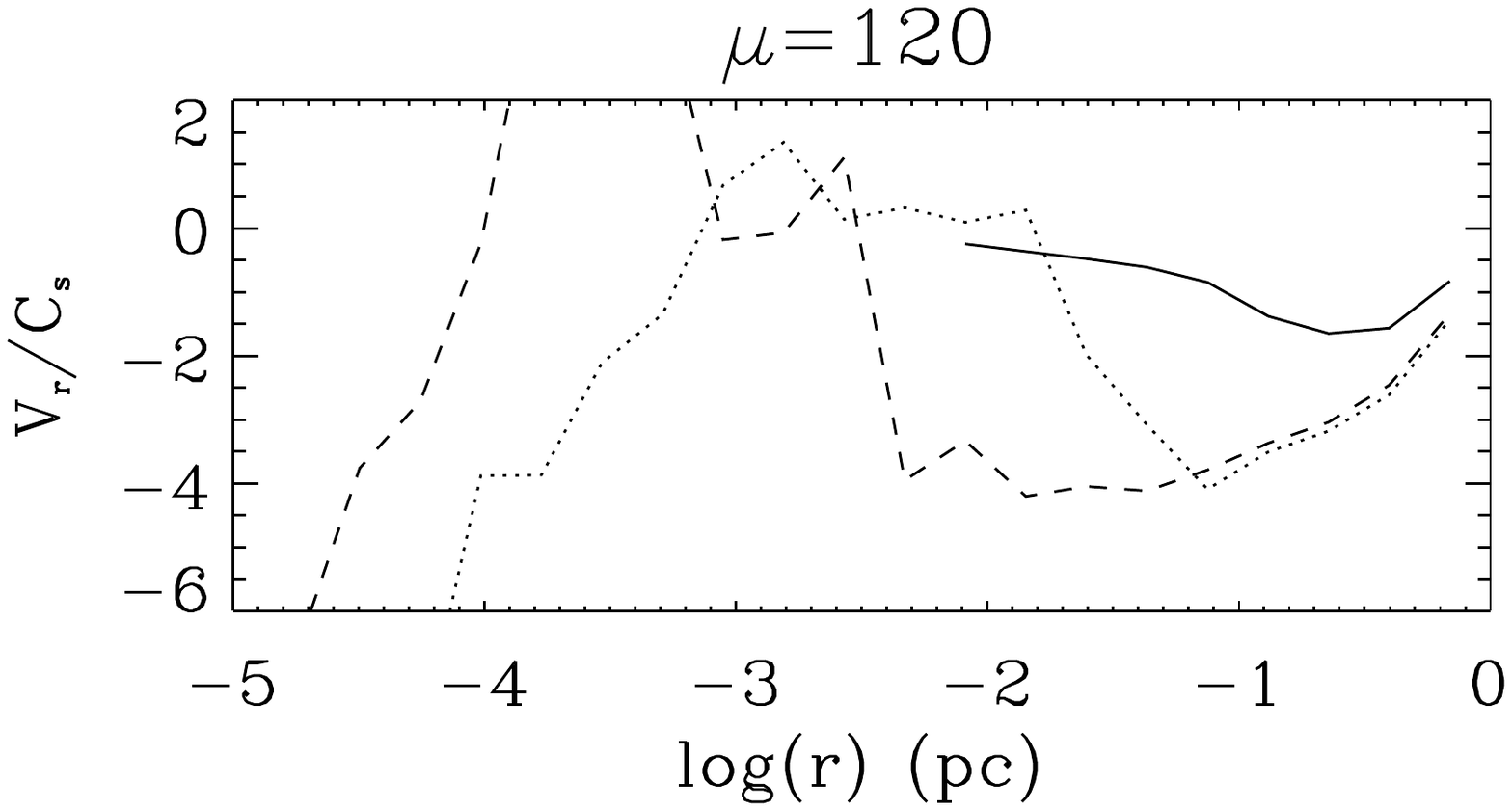}}
\put(0,3.5){\includegraphics[width=7.5cm]{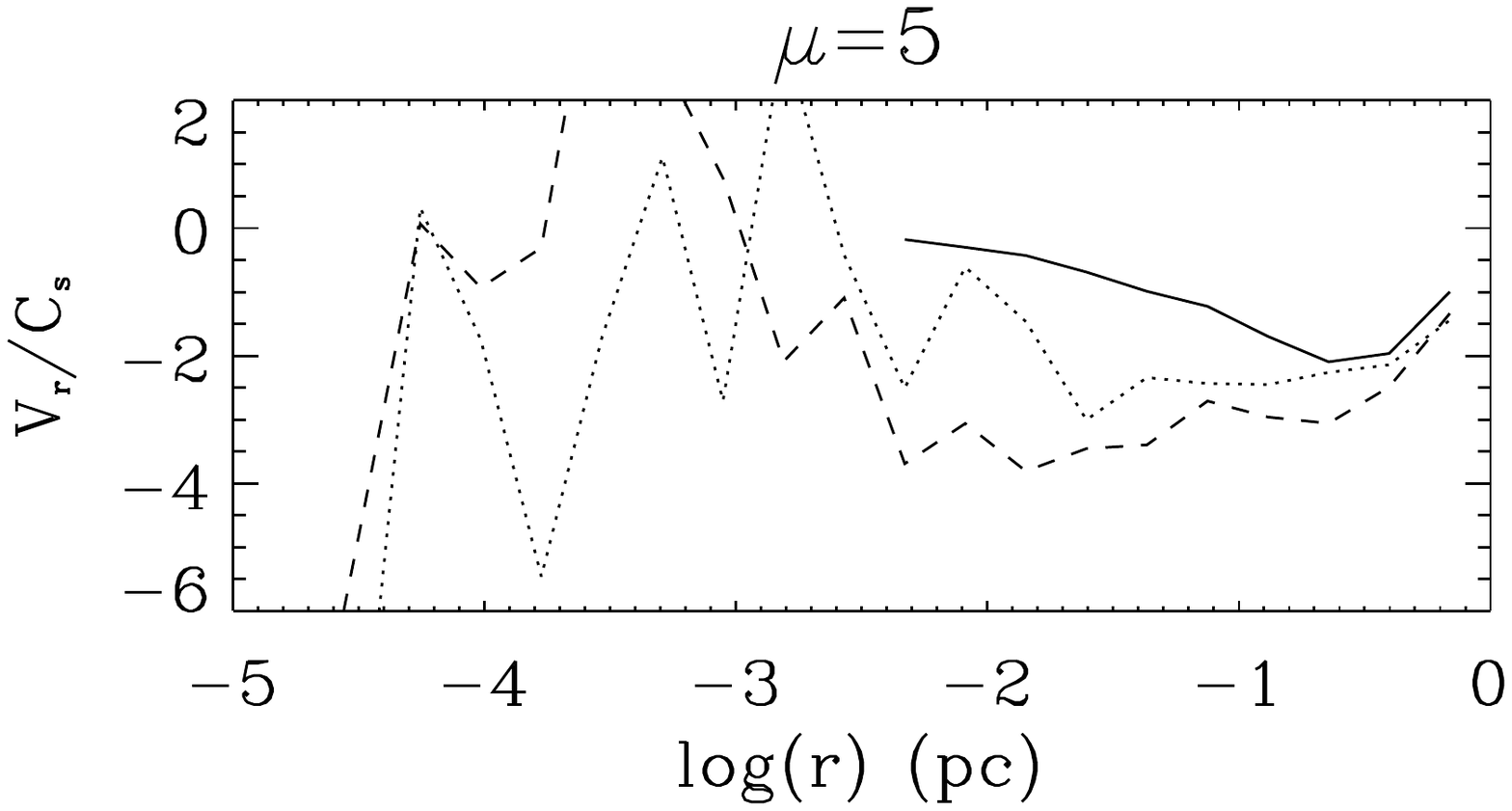}}
\put(0,7){\includegraphics[width=7.5cm]{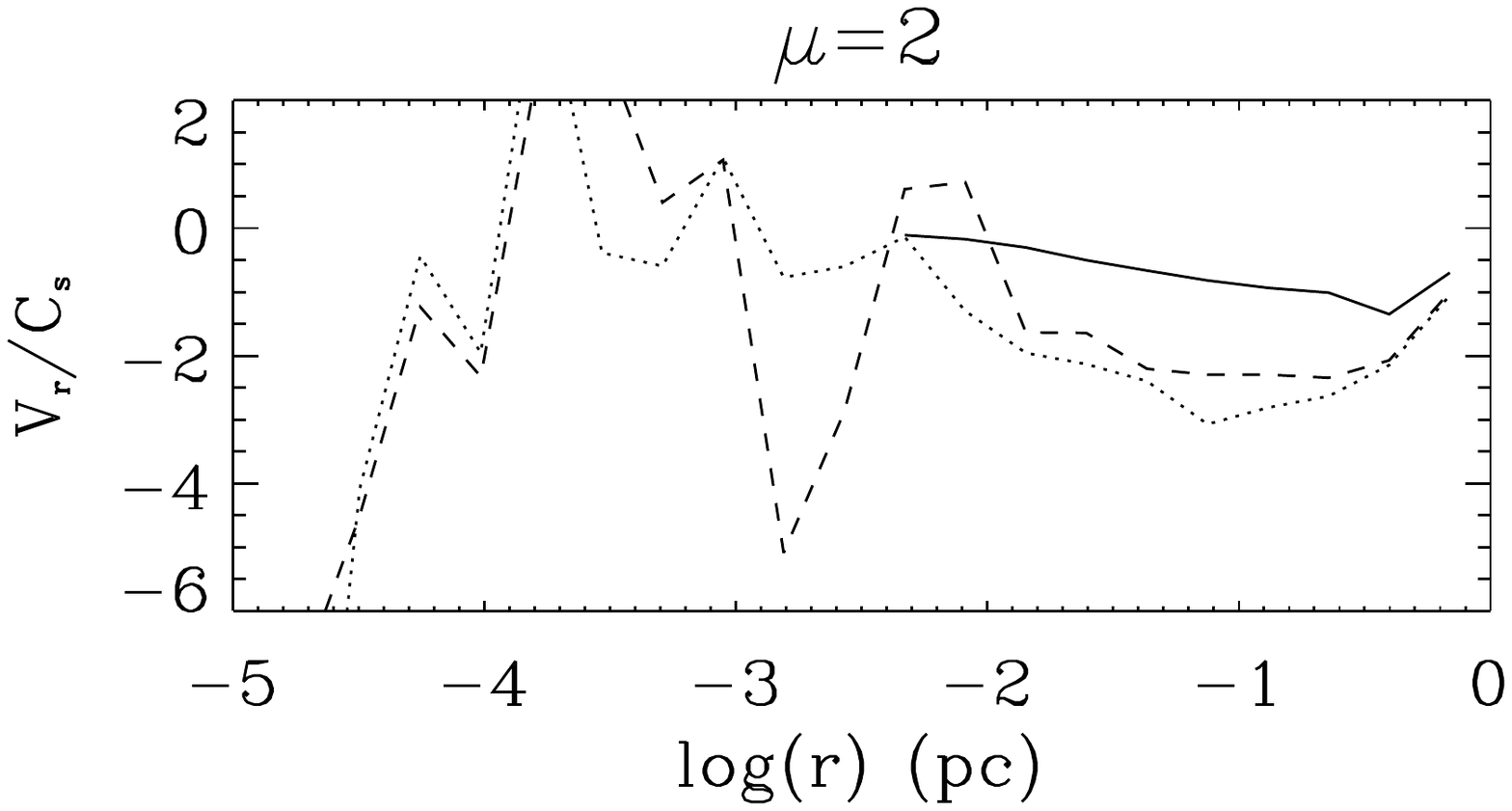}}
\end{picture}
\caption{Same as Fig.~\ref{density}, except that the radial velocity 
is displayed.}
\label{radial_velocity}
\end{figure}

\setlength{\unitlength}{1cm}
\begin{figure*} [t]
\begin{picture} (0,11)
\put(0,0){\includegraphics[width=7.5cm]{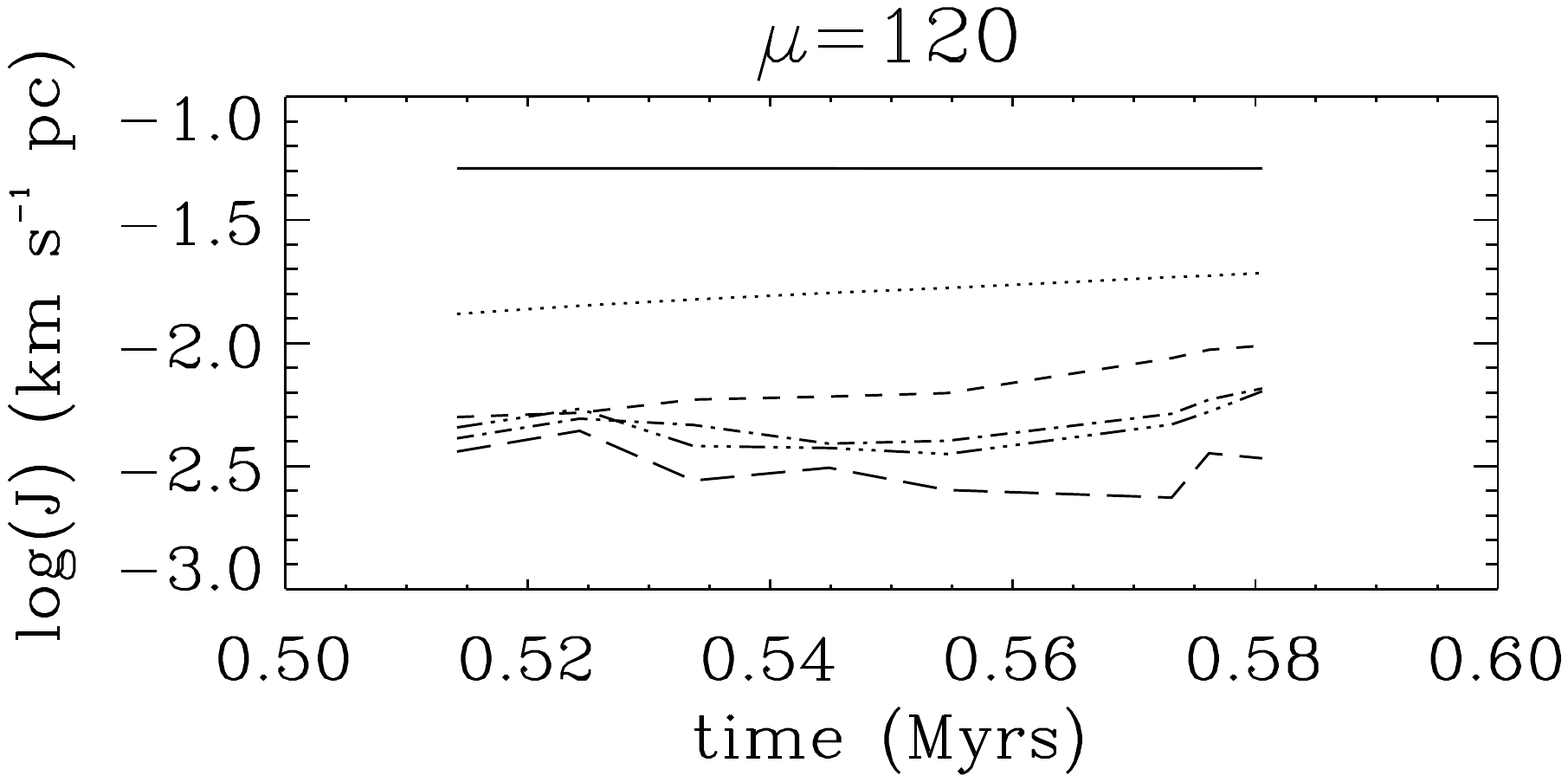}}
\put(8,0){\includegraphics[width=7.5cm]{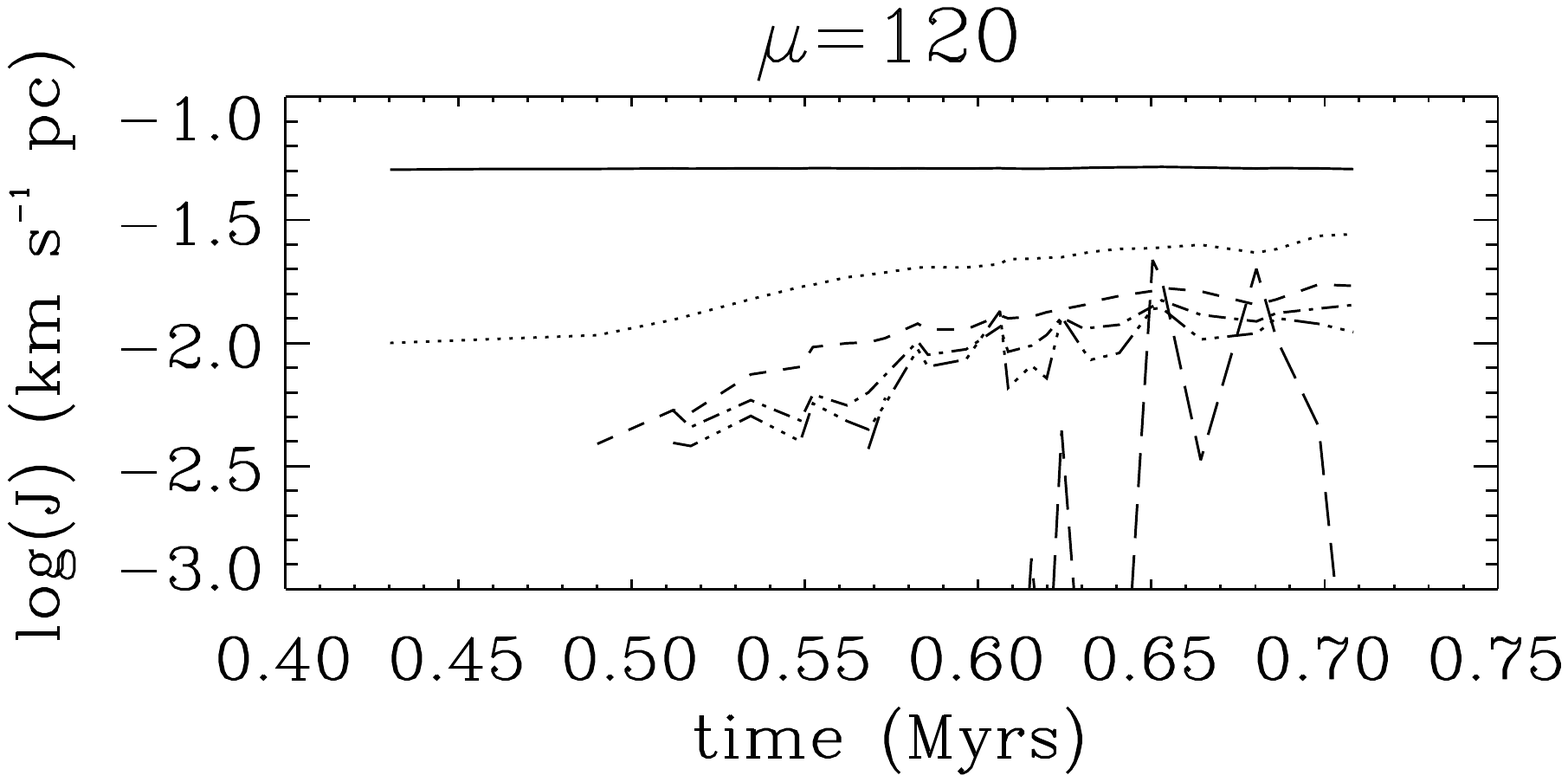}}
\put(0,3.5){\includegraphics[width=7.5cm]{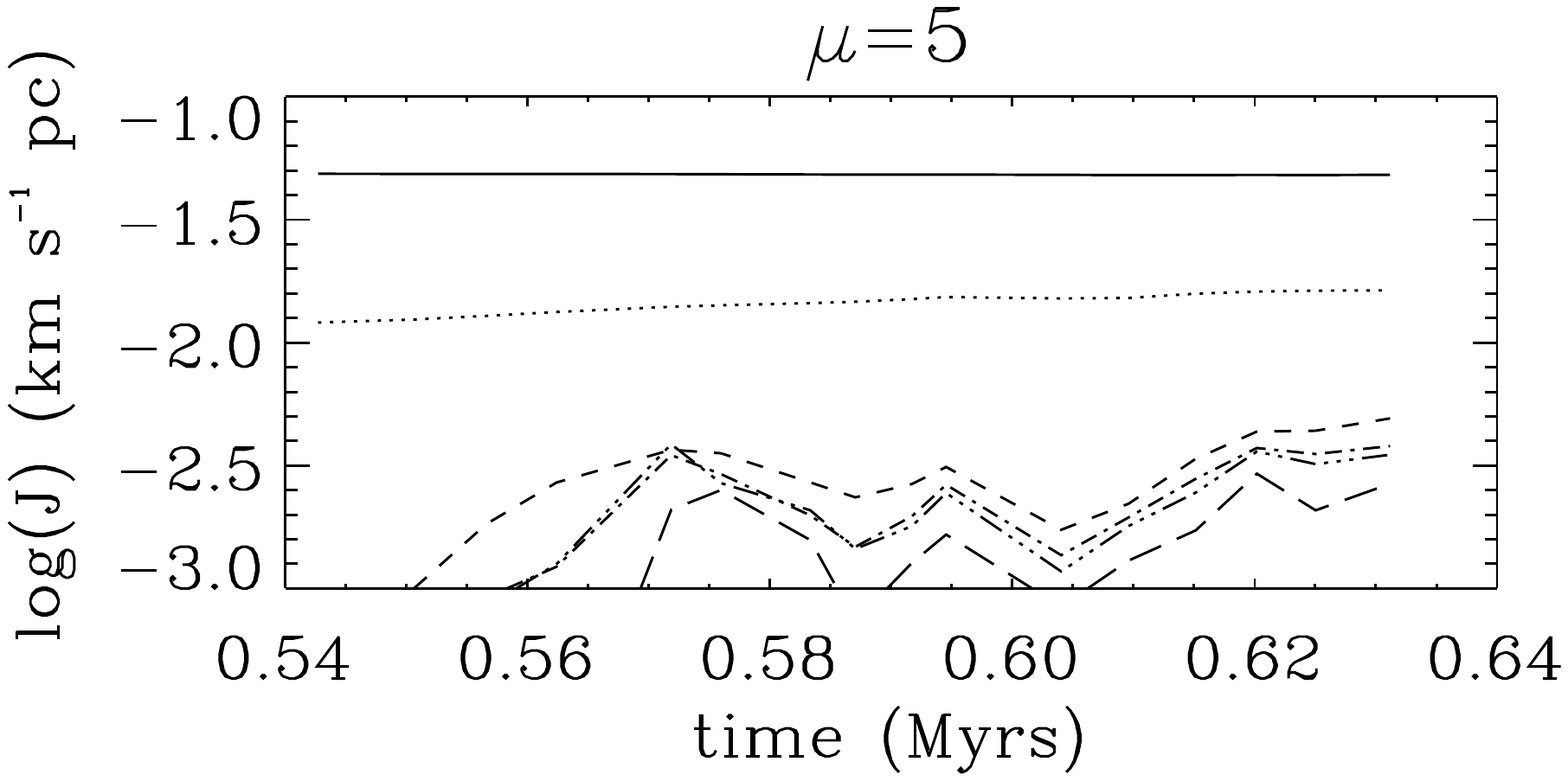}}
\put(8,3.5){\includegraphics[width=7.5cm]{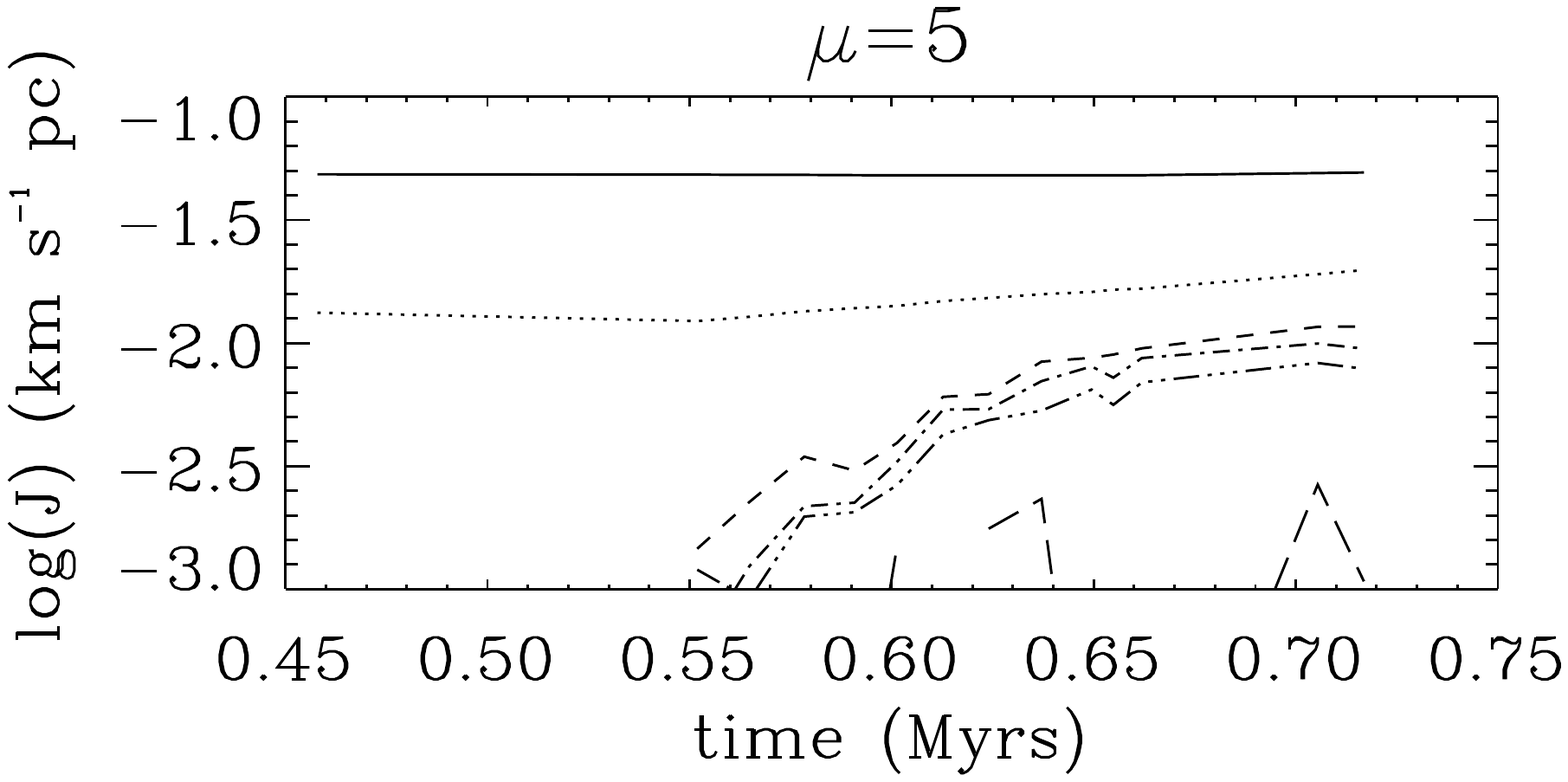}}
\put(0,7){\includegraphics[width=7.5cm]{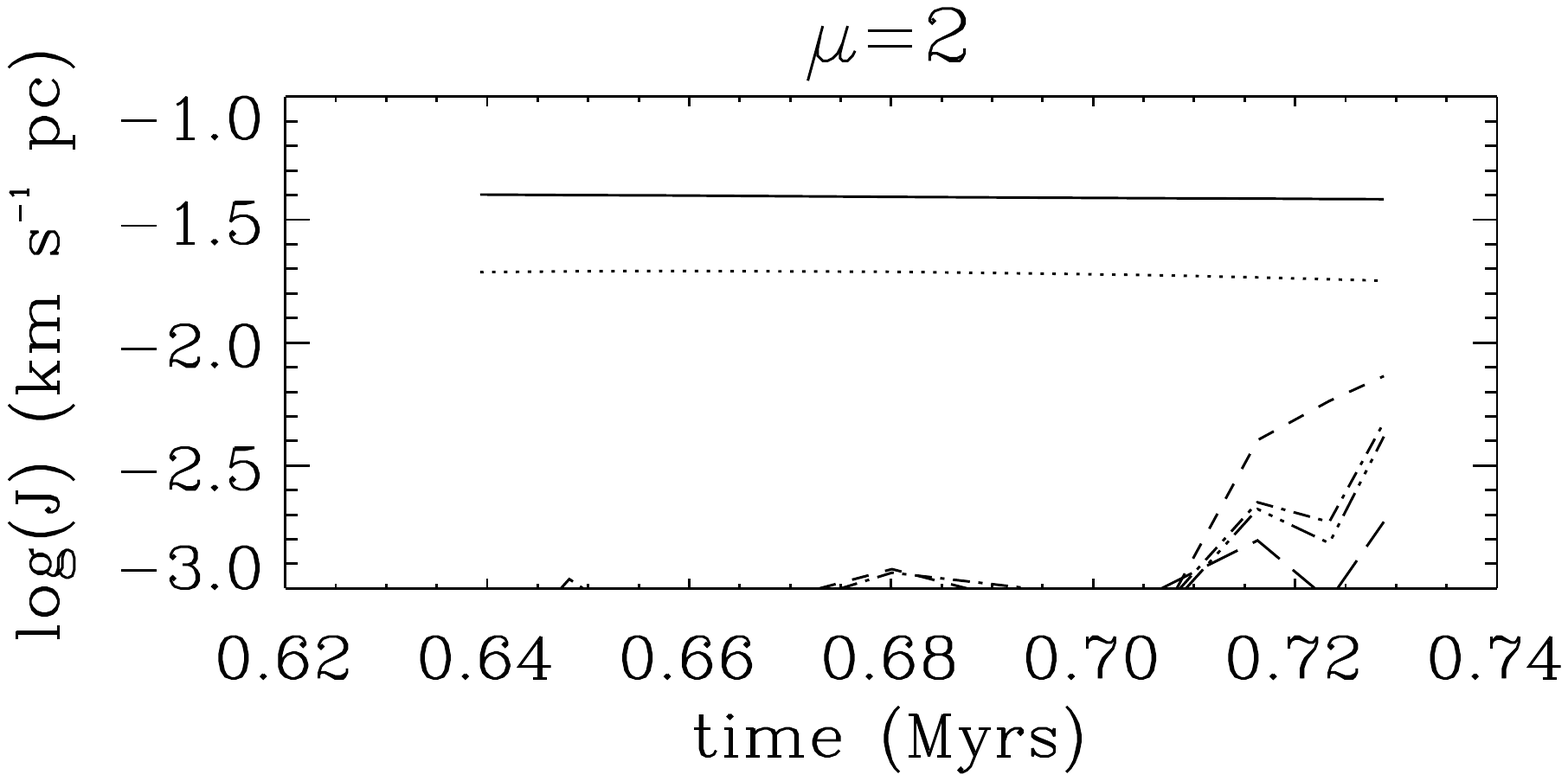}}
\put(8,7){\includegraphics[width=7.5cm]{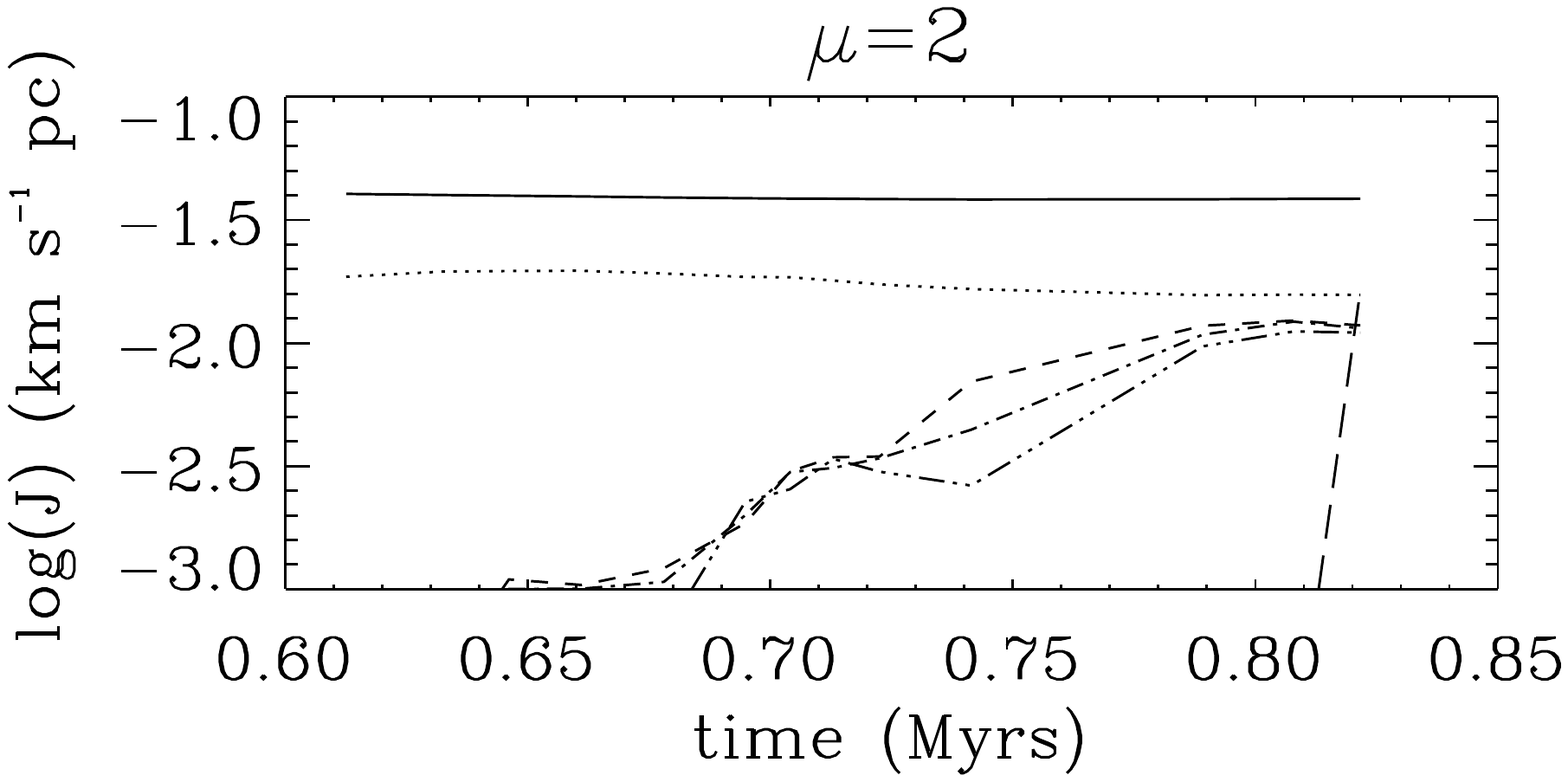}}
\end{picture}
\caption{Same as Fig.~\ref{mass}, except that the specific angular momentum 
is displayed.}
\label{mom}
\end{figure*}

\setlength{\unitlength}{1cm}
\begin{figure} [t]
\begin{picture} (0,11)
\put(0,0){\includegraphics[width=7.5cm]{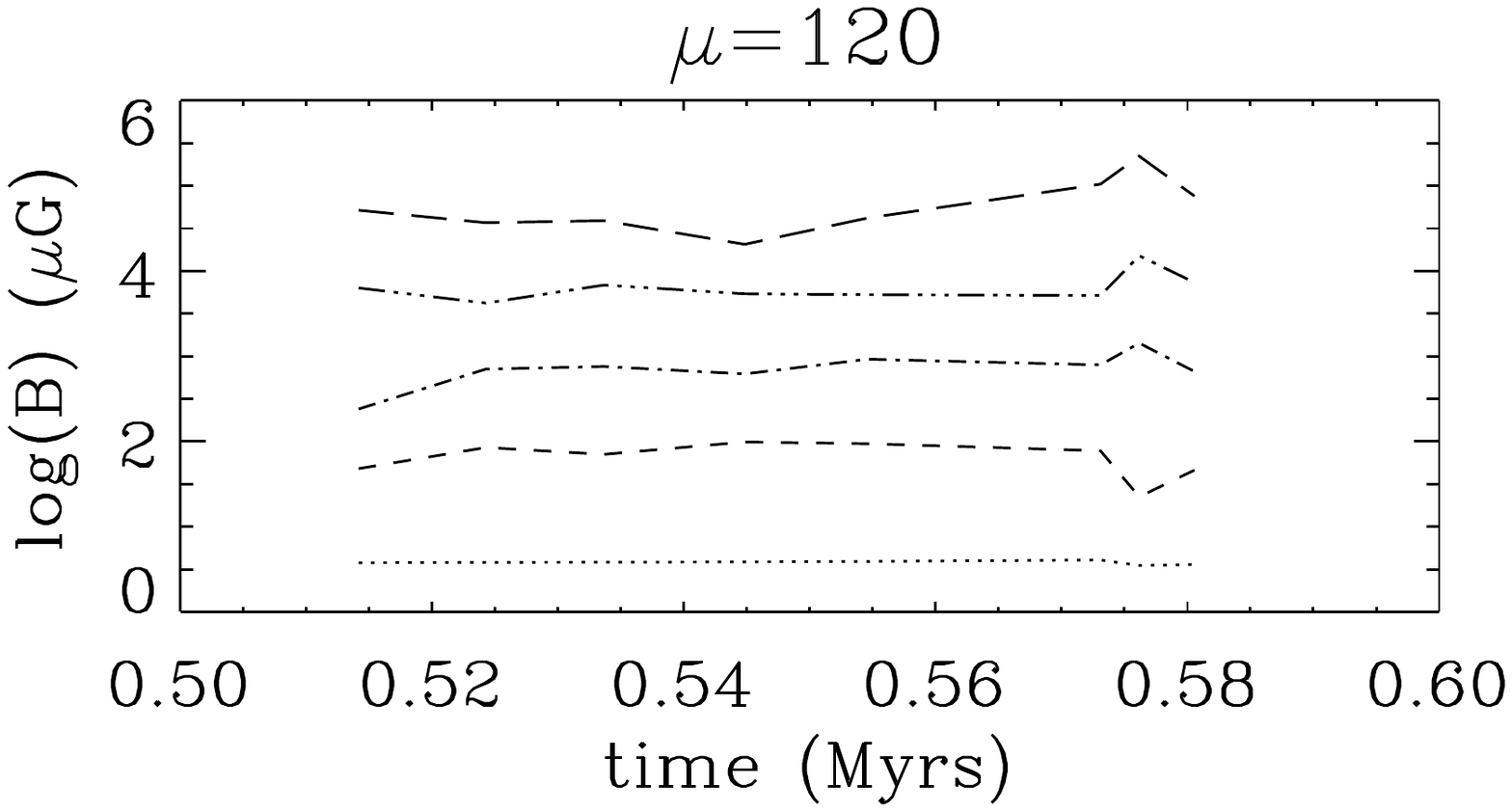}}
\put(0,3.5){\includegraphics[width=7.5cm]{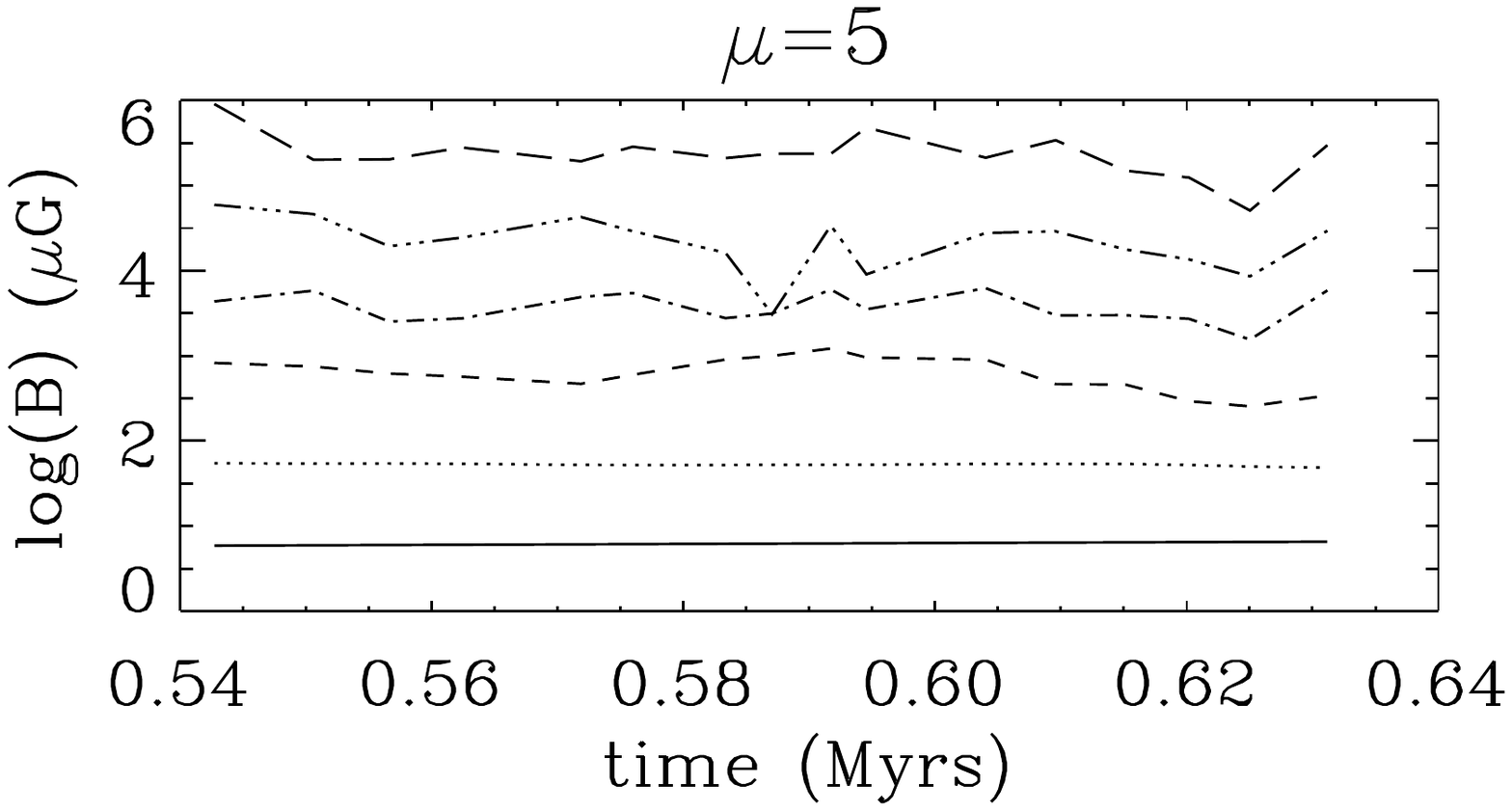}}
\put(0,7){\includegraphics[width=7.5cm]{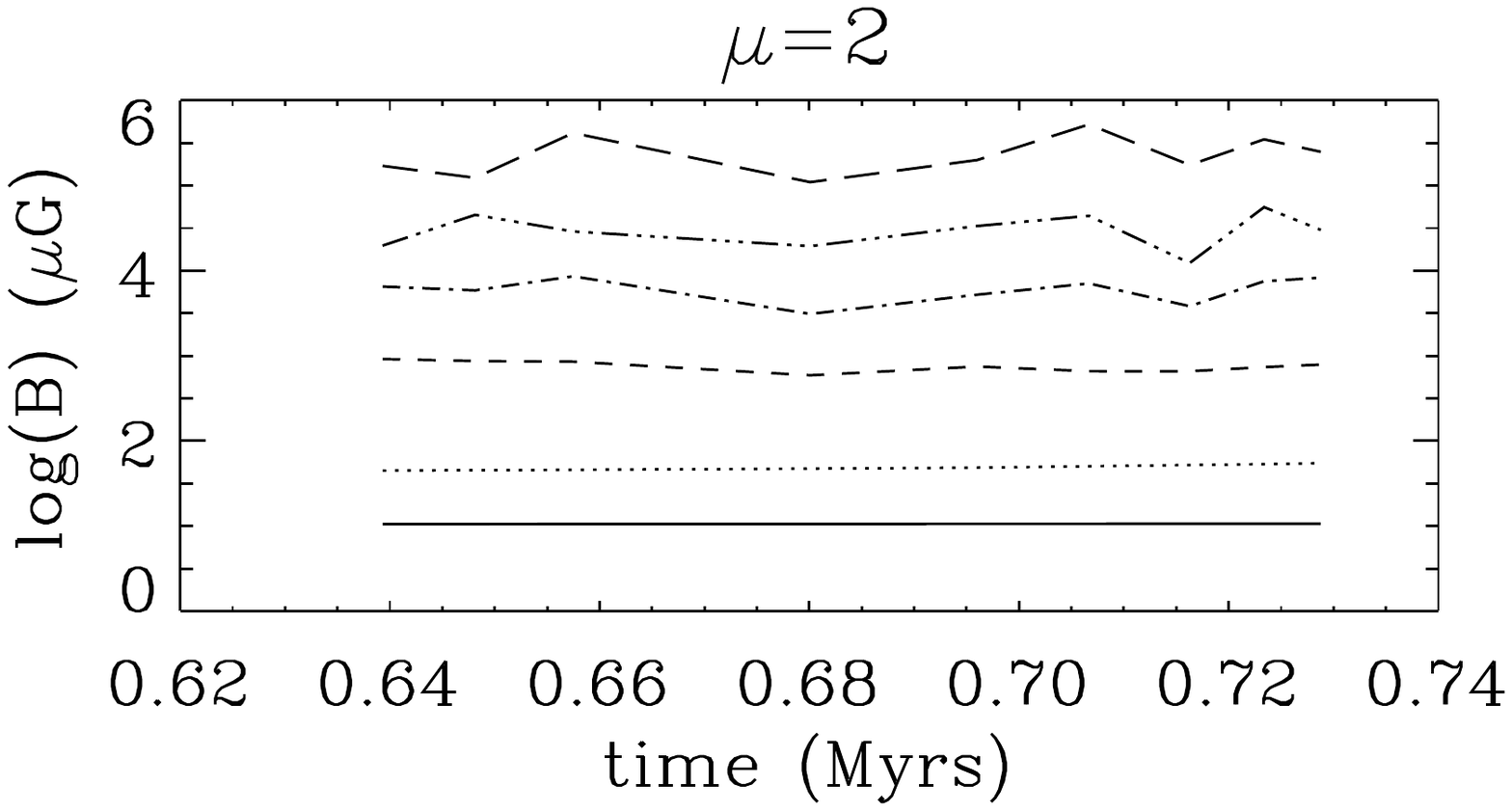}}
\end{picture}
\caption{Same as Fig.~\ref{mass}, except that the mean
 magnetic intensity is displayed.}
\label{mag}
\end{figure}


\setlength{\unitlength}{1cm}
\begin{figure} [t]
\begin{picture} (0,11)
\put(0,0){\includegraphics[width=7.5cm]{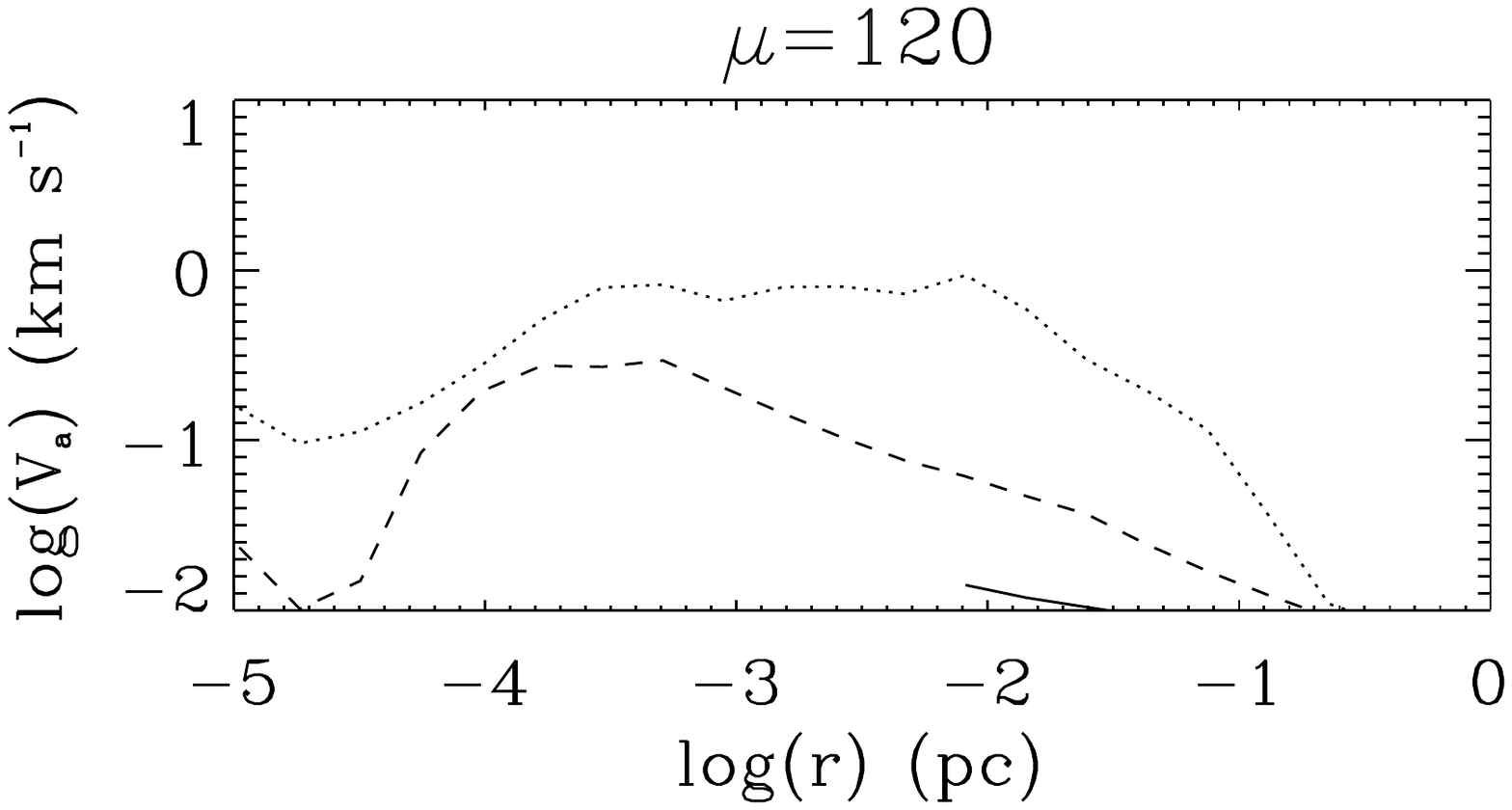}}
\put(0,3.5){\includegraphics[width=7.5cm]{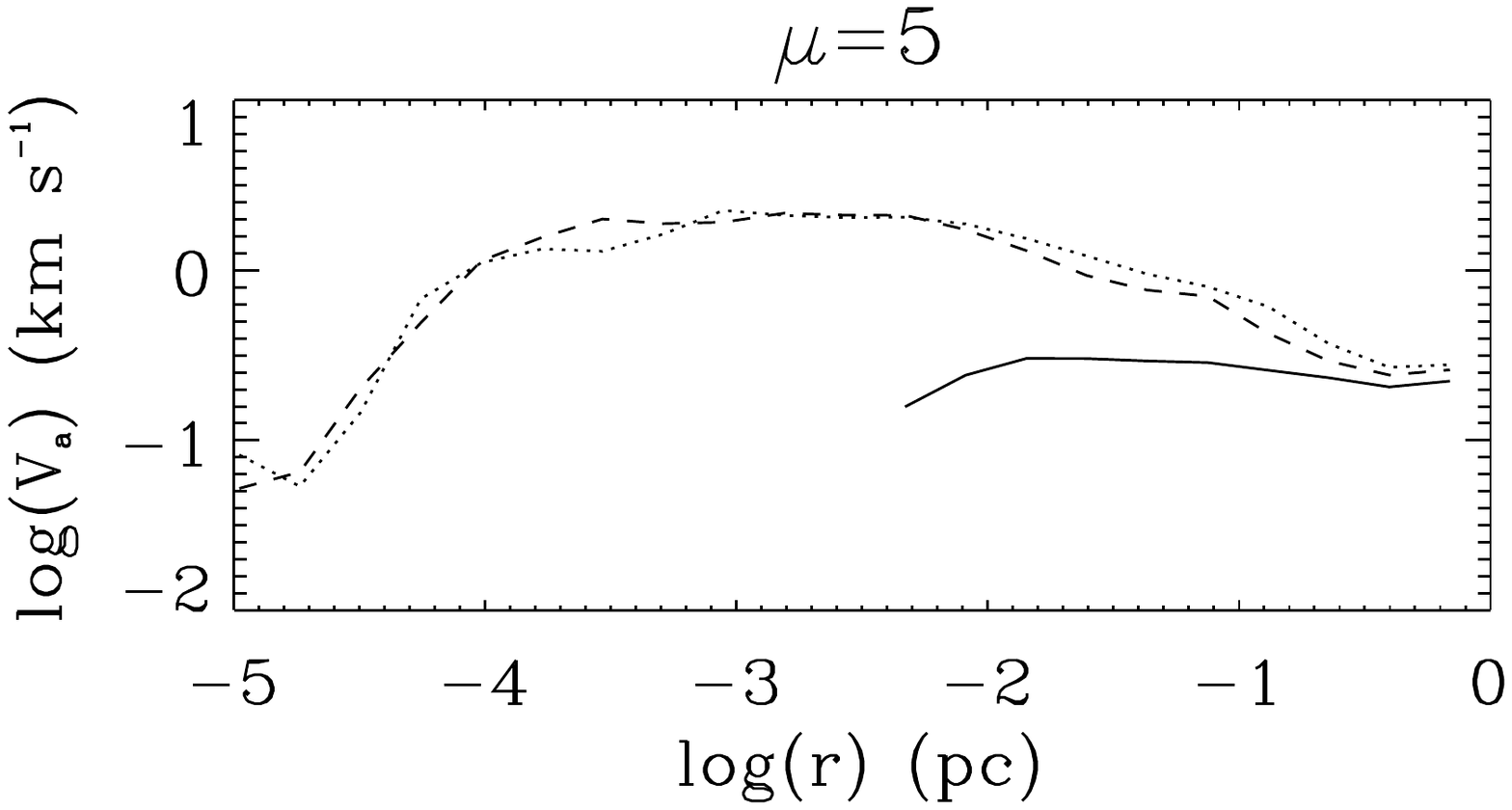}}
\put(0,7){\includegraphics[width=7.5cm]{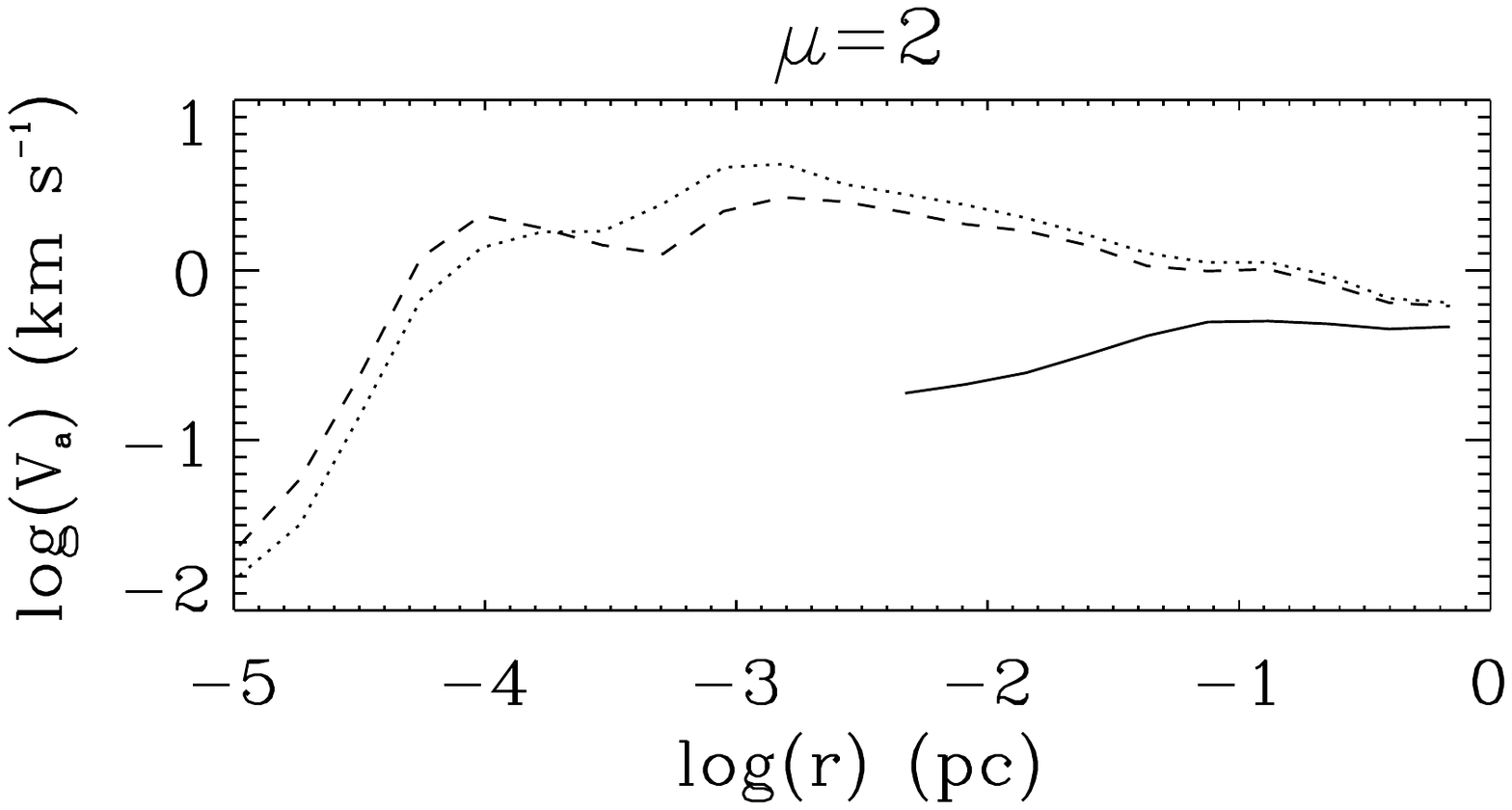}}
\end{picture}
\caption{Same as Fig.~\ref{density} except that the volume weighted 
mean Alfv\'en velocity 
is displayed.}
\label{alfven_vol}
\end{figure}

\setlength{\unitlength}{1cm}
\begin{figure} 
\begin{picture} (0,14)
\put(0,0){\includegraphics[width=7.5cm]{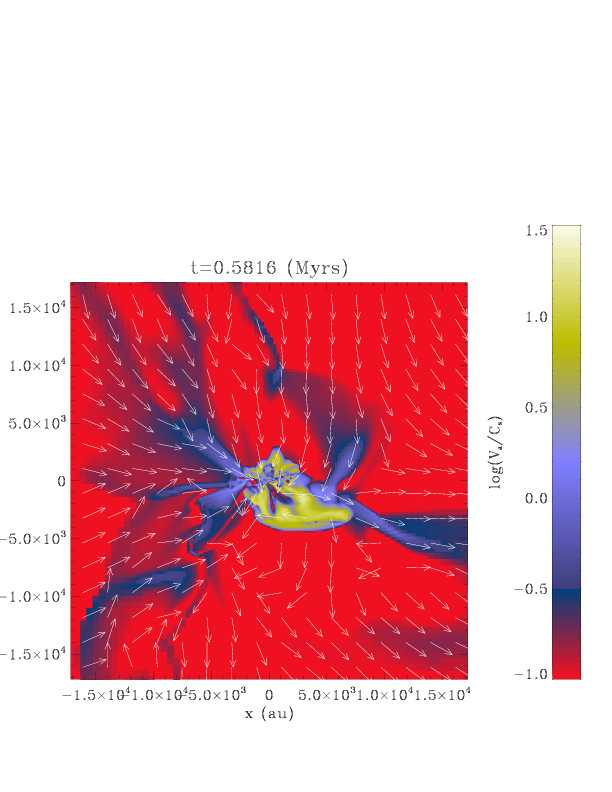}}
\put(0,7){\includegraphics[width=7.5cm]{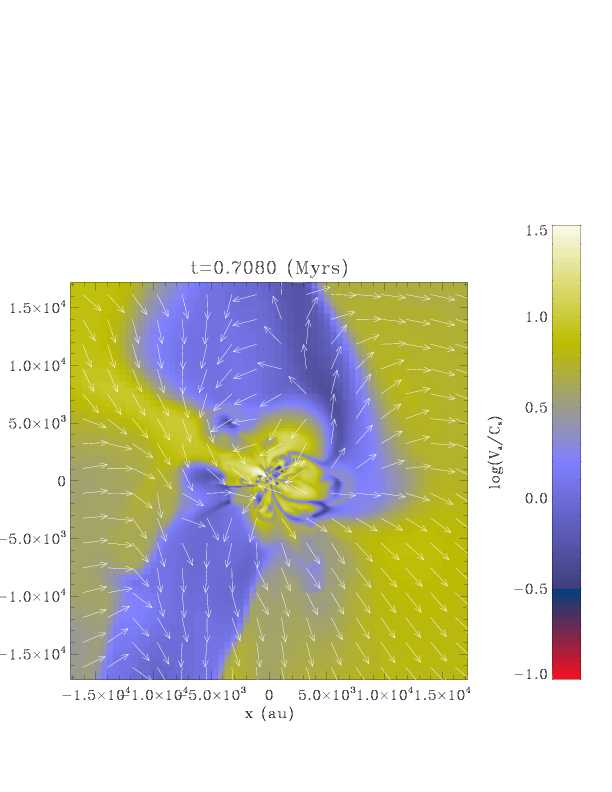}}
\end{picture}
\caption{Alfv\'en velocity in the xy plane. Top panel $\mu=2$.
Bottom panel $\mu=120$. The arrows represent the direction of the magnetic field. }
\label{alfven}
\end{figure}

\section{Initial conditions and numerical setup}

\subsection{Initial conditions}
We investigate the collapse of  hundred solar masses cores.
The initial conditions consist of a sphere whose profile resembles the 
observed cores and is given by $\rho(r) =  \rho_c / (1 + (r/r_0)^2)$.
We impose a density contrast of 10 between the central density 
and the edge density, $\rho_e$. Outside the cloud, a warm and diffuse medium 
of density $\rho_e / 10$ in pressure equilibrium with the cloud edge is set up.
The peak density is initially equal to $6.6 \times 10^3$ cm$^{-3}$ or 
 $1.4 \times 10^{-20}$ g cm$^{-3}$, 
corresponding to a freefall time of about 0.43 Myr.
The size of the core is initially equal to 1.35 pc while the central plateau has a radius of 
$r_0\simeq 0.22$ pc.
The temperature within the dense core is initially equal to $T_0=10$ K, leading to 
a thermal over gravitational energy ratio, $\alpha_{th}$, equal to about 0.12. 
At high density, a barotropic
equation of state is used to mimic the optically thick regime, and 
the temperature is then given by $T=T_0 (1 + (\rho/\rho_c)^\Gamma)$, where 
$\Gamma$ is  equal to $7/5$. 
The critical density is equal to $10^{-13}$ g cm$^{-3}$ or about 
$3 \times 10^{10}$ cm$^{-3}$.

The cloud is initially  threaded by a  
magnetic field along the x-axis, whose intensity is proportional to the column 
density through the cloud. The initial  degree
of magnetization is determined by the parameter $\mu$, the mass-to-flux over 
critical mass-to-flux ratio
equal to $\mu = (M/\phi) / (M_{crit}/\phi)$ where $M_{crit}/\phi = c_1/(3 \pi) (5/G)^{1/2}$
(Mouschovias \& Spitzer 1976). While Mouschovias \& Spitzer (1976)
infer  $c_1 \simeq 0.53$, we estimate in our case, which corresponds to a 
different magnetic configuration,  that $c_1 \simeq 1$.
Three degrees of magnetization are investigated, $\mu=120$, corresponding to a 
weak magnetic field, $\mu=5$ and $\mu=2$ close to the values of the order
of 1-4, which  have been observationally inferred (Crutcher 1999, Falgarone et al. 2008).
Finally, an internal {\it turbulent} velocity dispersion is initially given to the cores. 
The velocity field is obtained by imposing  a Kolmogorov power spectrum 
while the phases are randomly determined. 
Only one realization is explored at this stage. The turbulent energy is 
initially equal to about 20$\%$ of the gravitational one.

It is worth at this stage to express the amount of support that is initially 
provided to the clouds. Neglecting the surface terms, the 
virial theorem is
\begin{eqnarray}
\ddot{I} &=& 2 E_{{\rm therm}} + 2 E_{{\rm kin}} + E_{\rm grav} + E_{\rm mag} \\
\nonumber
 &=& 2 (E_{{\rm therm}} +  E_{{\rm kin}}) + E_{\rm grav} (1-\mu^{-2}), 
\label{virial}
\end{eqnarray}
since the magnetic energy, $E_{\rm mag}$ can be written as $-E_{{\rm grav}} \times \mu^{-2}$ (Lequeux 2005).
The conditions for virial equilibrium is  that $\ddot{I} \simeq 0$, thus:
\begin{eqnarray}
\alpha_{Vir} = { E_{{\rm therm}} +  E_{{\rm kin}} \over | E_{\rm grav} | (1-\mu^{-2}) } \simeq {1 \over 2}.
\end{eqnarray}
In the hydrodynamical case, $\alpha_{vir} \simeq 0.3$ and the cloud is therefore out of virial equilibrium 
by a factor of almost 2. In the $\mu=2$ case, $\alpha_{vir} \simeq 0.4$, which implies that the cloud is 
closer to equilibrium because of the magnetic field, which {\it dilutes} gravity.

Observationally, it is inferred that massive cores present motions, which are 
apparently not far from virial equilibrium (Bontemps et al. 2010, Wu et al. 2010). 
The values chosen here are close to but slightly below virial equilibrium.
We stress however that these values correspond to the initial conditions and evolve rapidly. 
In particular, gravity triggers large infall motions and tends to increase the 
ratio of kinetic over gravitational energy ratio (see e.g. Peretto et al. 2007).
Indeed  it is observationally difficult to separate the contributions of the systematic 
infall motions and the turbulent ones (Csengeri et al. 2010), in particular because massive cores are 
located at large distances.  It is therefore likely the case that the motions observed
in dense massive cores should not be entirely attributed to   turbulent support.
Finally, we note that higher values of the initial turbulent energy induce the 
formation of several collapsing regions within the clouds, which can be described 
as large scale fragmentation and could be seen as an ensemble of cores, rather than a single one. 
By contrast with the value adopted in this work, the 
cloud is undergoing a global contraction at large scale.

To characterize the initial state of the cloud, it is also worth  estimating the thermal and 
magnetic Jeans masses. To calculate the former, we rely on the expression
$M_J = \pi^{5/2}/6 C_s^3 G^{-3/2} \rho^{-1/2}$, obtained by defining the thermal Jeans mass
as the mass contained within a sphere of radius $\lambda_J/2$, $\lambda_J$ being the Jeans length.
This leads to $M / M_J = \pi^{-3}  (3 \sqrt{3}) (2 \alpha_{th} / 5)^{-3/2} \simeq 16$. Note that 
because the  contrast between the central and edge  densities is ten, the Jeans mass is about 
3 times smaller in the centre than near the cloud boundary.
 To estimate the initial magnetic Jeans mass, we follow  Li et al. (2010). 
The smallest pieces of gas, which are initially not supported by the magnetic field,
are typically critical. Let $l_{crit}$ be the characteristic size, we have 
$M / \phi  \simeq \rho_c l_{crit} / B_c \simeq (M / \phi)_{crit}$.
Because for the cloud 
$M_c / \phi_c \simeq  \rho_c l_c / B_c \simeq  \mu (M / \phi)_{crit}$ holds, we have 
$M / M_c = (l_{crit}/l_c)^3=\mu^{-3}=8$. Thus, when $\mu=2$ 
there are initially about twice as many thermal Jeans masses
than magnetic Jeans masses in the cloud.



\subsection{Numerical setup}

To carry out our numerical simulations, we employed RAMSES (Teyssier 2002, Fromang et al. 2006), 
an adaptive mesh refinement code that uses Godunov schemes to solve the MHD equations
and the constrained transport method to ensure that ${\rm div}B$ is maintained 
at zero within machine accuracy.
Initially the simulations start with an uniform grid of 256$^3$ cells corresponding 
to level 8 in RAMSES. 
Throughout the simulations 
the Jeans length is resolved with at least 10  cells up to the AMR level 16
for the low resolution calculations and 18 for the high resolution ones. 
This corresponds to a minimum resolution of about 8 AU in the first case and 
2 AU in the second case.
No other level is introduced because this leads to timesteps so small 
that advancing the simulations sufficiently becomes too prohibitive. 
For this reason the low resolution runs were performed for longer times
than the high resolution ones. Because the minimum Jeans mass in the simulation,
which is obtained for the density at which the gas becomes adiabatic, 
has a Jeans length which is about 20 AU, 
a reasonable numerical resolution is ensured regarding gravity. It is however worth 
stressing that turbulence and magnetic field may require the resolution 
of smaller spatial scales. Another difference between the two types of runs 
is that for the high resolution runs, the HLLD solver (Miyoshi \& Kuzano 2005) is employed, while 
for the lower resolution runs we use the HLL solver, which is more diffusive 
but permits  bigger timesteps. 

The combination of lower resolution runs and higher ones allows
us to test the numerical convergence in terms of the smaller scale
solved in the simulations and at the same time to obtain results for 
 longer physical times. Below,
we display the properties of the high resolution calculations 
and  where  necessary, we also display the  properties for the two types of runs. 
Note that  at this stage we did not explore  the influence of increasing the 
initial resolution or the number of cells per Jeans length.

We did not use sink particles at this stage meaning that the 
dense gas is prevented from collapsing  by the barotropic 
equation of state, which ensures that the thermal support stops the 
gravitational contraction.

The lower resolution simulations were performed on 32 CPU
while the higher resolution ones use 128 CPU.
Typically each low resolution simulation required about  25,000 CPU hours
while the high resolution ones took about 80,000-100,000 CPU hours.
The high resolution calculations 
have about 10$^7$ computing cells in total while the low resolution
one have about three times less.

\section{Core evolution  during collapse}
In this section we discuss various properties 
of the cores that are important to characterize their 
evolution and interpret the trends regarding the 
outflows and the core fragmentation that will be discussed in the next 
section.

\subsection{Mass evolution and density distribution}

Figure~\ref{mass} displays the total mass above various density 
thresholds as a function of time. 
As expected, the collapse time (estimated to be equal to about 0.52 Myr
in the $\mu=120$ case) increases with the magnetic 
intensity, which is a simple consequence of the magnetic support.  
The simulations 
are run up to about 0.1-0.2 freefall time after the formation 
of the first protostar for the high resolution simulations
and 0.4-0.5 freefall time for the low resolution calculations.
 By the end of the simulations, 
about 5-10 solar masses of gas, corresponding to 5-10 $\%$ of 
the gas within the massive core, is typically at densities 
higher than $10^{11}$ cm$^{-3}$ for the low 
resolution runs, while for the high resolution 
runs about 3 solar masses of gas have reached this density. 
This gas would have further collapsed up to 
stellar densities if the simulations could follow the
second collapse phase.
Note that as will be discussed later, the disk fragments 
and therefore the accretion rate is the total accretion rate
occurring on all fragments. 
Interestingly,  the accretion obviously  does not proceed
in the same way for the three simulations. In particular the 
fraction of dense gas is  smaller for the $\mu=2$
case than for the two less magnetized ones, which clearly is a 
consequence of the magnetic support. 

For $\mu=120$ and $\mu=5$ the accretion rate is initially of 
the order of $\simeq 10^{-4}$ M$_\odot$ yr$^{-1}$
and then drops to values of the order of $\simeq 10^{-5}$ M$_\odot$ yr$^{-1}$
while it stays close to this latter value when $\mu=2$. These values 
correspond to accretion rate 10 to 100 times higher than the 
canonical (Shu 1977) $C_s^3/G \simeq 2 \times 10^{-6}$ M$_\odot$ yr$^{-1}$ as 
already noted by Banerjee \& Pudritz (2007) in closer agreement with the 
accretion rate inferred by Larson (1969) and Penston (1969).
Our accretion  
rates are at least one order of magnitude lower than those considered  
in the fiducial case of McKee \& Tan (2003) of the collapse of a core with  
a mean mass surface density of $\simeq$1 g cm$^{-2}$. This is mostly because our  
core has an initial mass surface density, which is significantly  
smaller, $\simeq  10^{-2}$ g cm$^{-2}$ (the exact value depends on the time and the radius 
on which it is estimated) as shown in Fig.~\ref{col_dens}, which displays 
the column density through the core. We note that recent mid-IR extinction mapping studies  
have derived observed core mass surface densities in the range of    
several  $10^{-2}$ to several $10^{-1}$ g cm$^{-2}$ (Butler \& Tan 2009).

Figure~\ref{density} shows the mean gas density within a
 sphere of radius, $r$, centred at the cloud density maximum,
 as a function of $r$. The first timesteps,
which correspond to the thick solid line are before the formation 
of the first protostar. The thin solid lines show the density 
of the singular isothermal sphere $\rho_{sis} = c_s^2/ (2 \pi G r^2)$. 
Interestingly the density is about ten times higher than $\rho_{sis}$.
 As shown analytically by Shu (1977), densities significantly higher than 
 $\rho_{sis}$ are typical signatures of a very dynamical collapse in which 
the infall velocity is several times the sound speed. Indeed
the higher the infall velocity, the denser the envelope.  
A density  equal to about 10 times  $\rho_{sis}$ has also been 
found in  numerical simulations of a highly dynamical collapse
in which the infall is 2-3 times the sound speed 
(Hennebelle et al. 2003), which  agrees well with observations 
of fast collapsing cores (Belloche et al. 2006). 
In the 
inner part of the cloud, the density is as high as
$\simeq 10-20 \times \rho_{sis}$.  While the 
 density profile in the outer part is very close
to $r^{-2}$, it is slightly stiffer in the inner part, where it is 
about $\simeq r^{-2.3}$ below 1000 AU and even stiffer below 300 AU. This is because
 of the support provided by rotation and 
turbulence and can be qualitatively understood as follows. In the inner part, 
systematic infall motions are weak, meaning that the cloud is on average
not far from an equilibrium, which implies that:
\begin{eqnarray}
{G M(r) \over r^2} \simeq C_s^2 {\partial_r \rho \over \rho} + {V_\theta ^2 \over r},
\end{eqnarray}
where $V_\theta$ is the rotational support provided by systematic rotation, but also 
by the local rotation that can be provided by turbulence.
In the simplest case of rotation, it is generally found that because of angular momentum conservation, 
$V_{\theta} \propto r^{-\eta}$ with typically $\eta \simeq 0.2-0.5$. 
This is the case because as angular momentum is conserved (in the hydrodynamical axisymetrical
case), one gets $V_\theta \times r = r_0^2 \omega$, where $r_0$ is the initial position of the fluid 
particle, while $r$ is its position along time.  The mass
enclosed within the radius $r$, $M(r)$ is typically equal to a few times $ 4 \pi \rho _{sis} r^3$,
but mass conservation gives $M(r)=M(r_0) \propto r_0^3 $ (assuming spherical contraction).
Thus $r \propto r_0^{3}$ and consequently $V_\theta (r) \simeq r^{-1/3}$.

In the inner part, the thermal support can be neglected 
and one finds that $\rho \propto r^{-2(1+\eta)} \simeq r^{-2.66}$ close to the exponent 
obtained below 300 AU.

\subsection{Infall velocity}

Infall velocity is another important quantity to characterize the 
collapsing clouds. 
Figure~\ref{radial_velocity} shows the mean radial component of the 
velocity, $<v_r>  = (\sum \rho v_r dV) / ( \sum  \rho dV )$,
 as a function of radius. 
In the outer part of the cloud, it monotonically decreases to reach 
about 0.8 km s$^{-1}\simeq 4 C_s$, $C_s$ being the sound speed,
 in the $\mu=120$ case and about half this value for $\mu=2$. 
These high values are typical of very dynamical collapse as
described analytically by the Larson-Penston solution (Larson 1969, 
Penston 1969)
and have been observed in some prestellar condensations (e.g. di Francesco 
et al. 2001, Belloche et al. 2006). 

In the inner part, $r < 10^{-2}$ pc, however, the picture is very different. 
Instead of a coherent velocity field, strong fluctuations are dominating. 
This is a  consequence of the initial turbulence, which in particular
leads to a non vanishing angular momentum that is amplified as the collapse
proceeds. Their amplitude is  comparable to the infall velocity which 
clearly indicates that the collapse proceeds in a complex, non-axisymmetric 
manner.

\subsection{Angular momentum evolution}

Although no angular momentum is explicitly set up initially, the turbulent 
velocity field, which is initially given to the cores, possesses 
local and even global angular momentum. This angular momentum
 plays an important role in the cloud evolution and is therefore 
an important quantity to study. 
One difficulty however resides  in the choice  of the origin  
with respect to which the angular momentum is defined. A natural choice,
which we adopted in this study, 
is the cloud density peak. Another possible choice would be 
the cloud mass centre. However, this point is not necessarily corresponding
to the point where the first protostars or group of protostars collapse.
  
To compute the specific angular momentum, ${\bf J}$, 
we simply calculate its three 
components and then take its norm. For example the x-component of 
${\bf J}$ is given by 
$J_x = (\sum \rho (y v_z - z v_y) dV) / ( \sum  \rho dV )$
while ${\bf J}^2= J_x^2+J_y^2+J_z^2$.

Figure~\ref{mom} displays the evolution of the total specific angular 
momentum, $|{\bf J}|$, above the various density thresholds specified 
previously in the $\mu=120$ (bottom panels), $\mu=5$ (middle panels)
and $\mu=2$ (top panels) cases. 
The left column is for high resolution simulations while right column 
is for the low resolution calculations.

As expected $|{\bf J}|$ is almost always increasing with time 
and is larger for smaller density thresholds. This is simply because
 the angular momentum is larger in the outer part of the 
clouds, consequently as the collapse proceeds the material with larger angular momentum 
is continuously added to the dense material.
 
While the specific angular momentum does not vary significantly for
the density threshold $10^3$ cm$^{-3}$,
for all density thresholds higher than 10$^7$ cm$^{-3}$, the angular momentum decreases 
when the magnetic intensity increases. This is a  consequence
of the magnetic braking, which transports angular momentum from the 
inner dense part of the cloud towards the envelope. 
The dotted lines (corresponding to a density threshold of $10^5$ 
cm$^{-3}$) are particularly interesting. While the specific angular momentum  
increases with time in the $\mu=120$ case (bottom panel), it 
is almost flat in the $\mu=5$ case (middle panel), and 
even decreases in the $\mu=2$ case after $t \simeq 0.65$ Myr, which 
nicely illustrates the strong braking that occurs when 
the magnetic intensity is high.

\setlength{\unitlength}{1cm}
\begin{figure*}
\begin{picture}(0,13.5)
\put(0,6.5){\includegraphics[width=5.5cm]{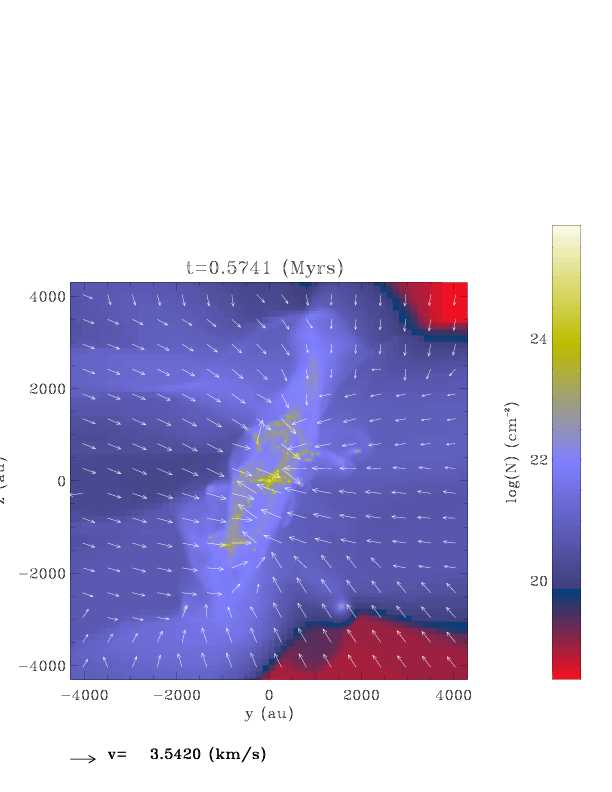}}
\put(0,0){\includegraphics[width=5.5cm]{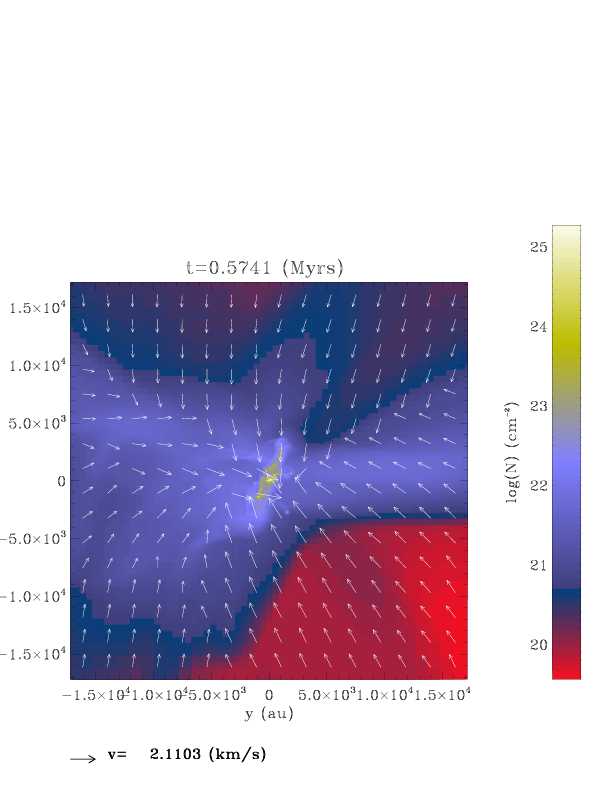}}
\put(6,6.5){\includegraphics[width=5.5cm]{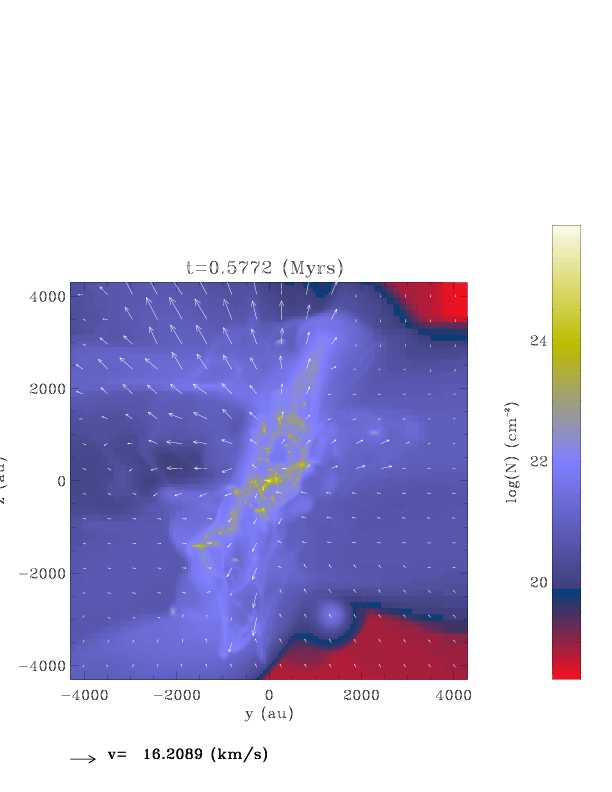}}
\put(6,0){\includegraphics[width=5.5cm]{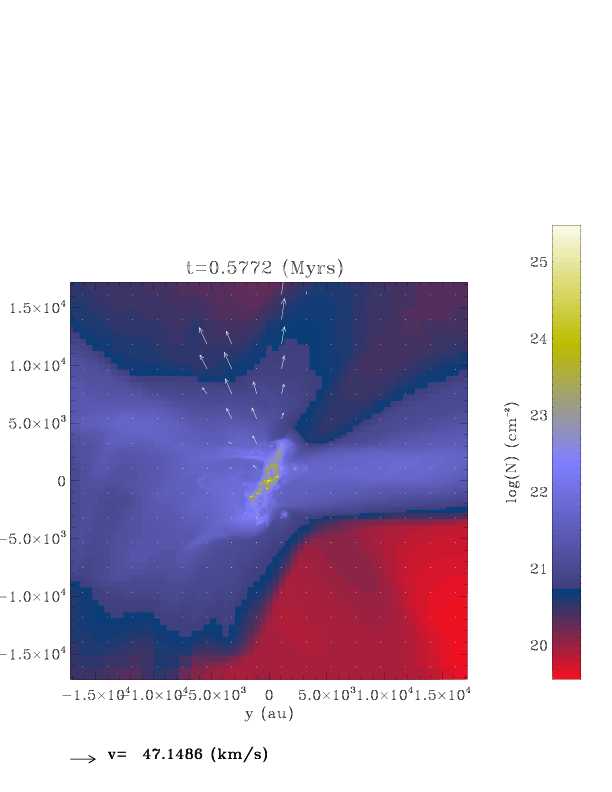}}
\put(12,6.5){\includegraphics[width=5.5cm]{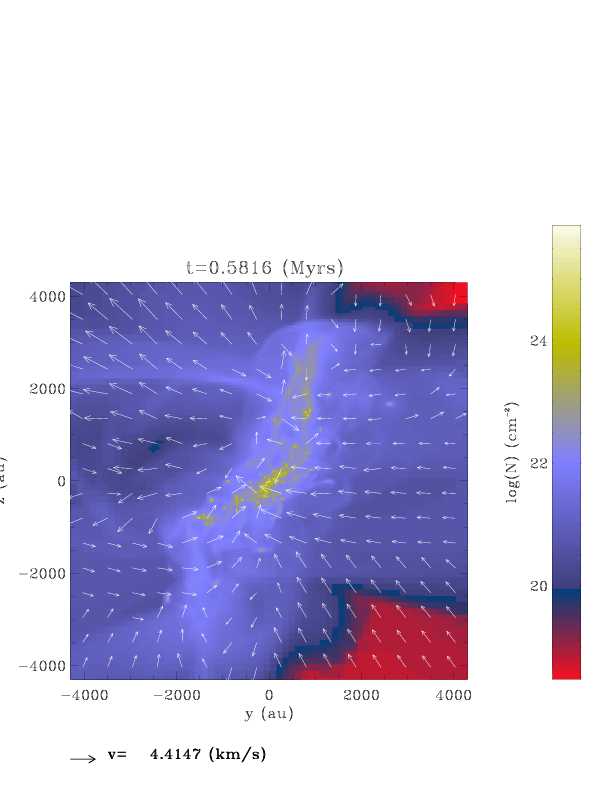}}
\put(12,0){\includegraphics[width=5.5cm]{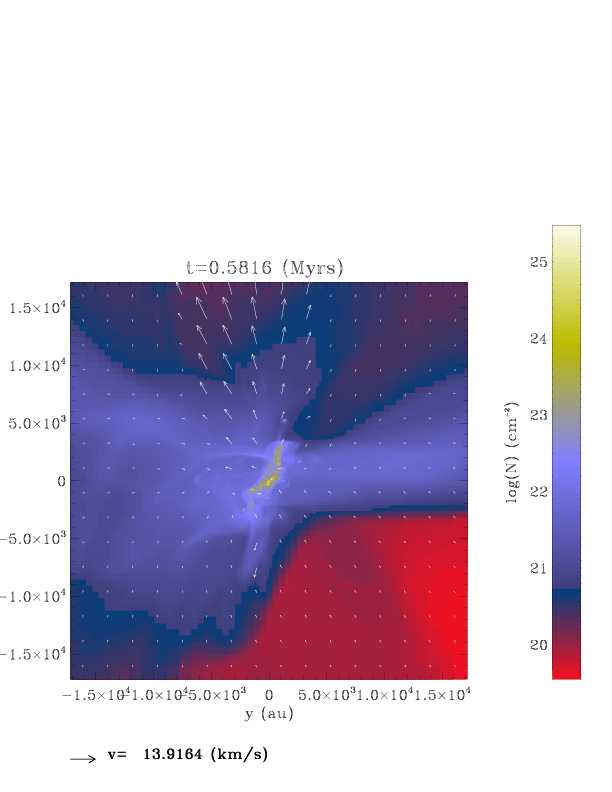}}
\end{picture}
\caption{ $\mu=120$. Column density and projected velocity field.}
\label{100_hy_out}
\end{figure*}

While we found for the $\mu=120$ simulation  a good 
agreement between the left and right panels, this is not the 
case for the more magnetized clouds ($\mu=5$ and $\mu=2$) 
for which the angular momentum 
of the high density gas ($\rho > 10^7$ cm$^{-3}$) is lower by a factor 
of about 3 (for example at $t=0.61$ Myr) for $\mu=5$
and by an even larger factor for $\mu=2$ (e.g. $t \simeq 0.70$ Myr).
Note that the sudden increase at 0.71 Myr is caused by the formation of
a new fragment far from the density peak (see top panel of Fig.~\ref{col_dens}), implying that our 
simple definition of angular momentum ceases to be valid.
The difference between the high and low resolution calculations,
 indicates that the lower resolution simulations underestimate 
the amount of magnetic braking as already discussed in Commer{\c c}on 
et al. (2010). Thus the results from the low resolution magnetized 
simulations must be considered with care.

Overall the specific 
angular momentum is about 1.5-2 times smaller in the $\mu=2$
case than in the $\mu=120$ case for the 
low resolution simulations and  larger 
than a factor 3 for the high resolution cases. Recalling that the 
centrifugal force is proportional to ${\bf J}^2$, this 
constitutes a very substantial difference.

Note however, that the angular momentum left appears nevertheless 
sufficient to lead to the formation of a centrifugally 
supported disk because the centrifugal radius is proportional to 
${\bf J}^2$. This is at variance with 
the conclusion that even low values of the 
magnetic field could entirely suppress the formation of a disk, 
as previously inferred by Allen et al. (2003), Galli et al. (2006), 
Price \& Bate (2007), Hennebelle \& Fromang (2008), and 
Mellon \& Li (2008, 2009).  Indeed, Hennebelle \& Ciardi (2009)
show that when the magnetic field is misaligned with 
the rotation axis, the magnetic braking is less 
efficient. This is because in the aligned case, 
the radial and azimuthal magnetic field components 
vanish in the equatorial plane, which  produces a strong magnetic compression, 
that in turn 
decreases the thickness of the pseudo-disk and produces
stiff gradients. When the magnetic field and the 
rotation axis are not aligned, the magnetic compression 
is less important because the radial and azimuthal magnetic 
components do not vanish any more in the equatorial plan.
This is obviously the case in this study because the initial 
velocity field is turbulent. Along the same line, the velocity dispersion 
likely contributes to make the pseudo-disk thicker, which 
may also decrease the efficiency of the magnetic braking.
Finally we note that it cannot be excluded  at this stage that 
because of numerical diffusivity the braking may be underestimated 
(see section 5.3) and the amount of specific angular momentum
could therefore be overestimated.

\setlength{\unitlength}{1cm}
\begin{figure*}
\begin{picture}(0,13.5)
\put(0,6.5){\includegraphics[width=5.5cm]{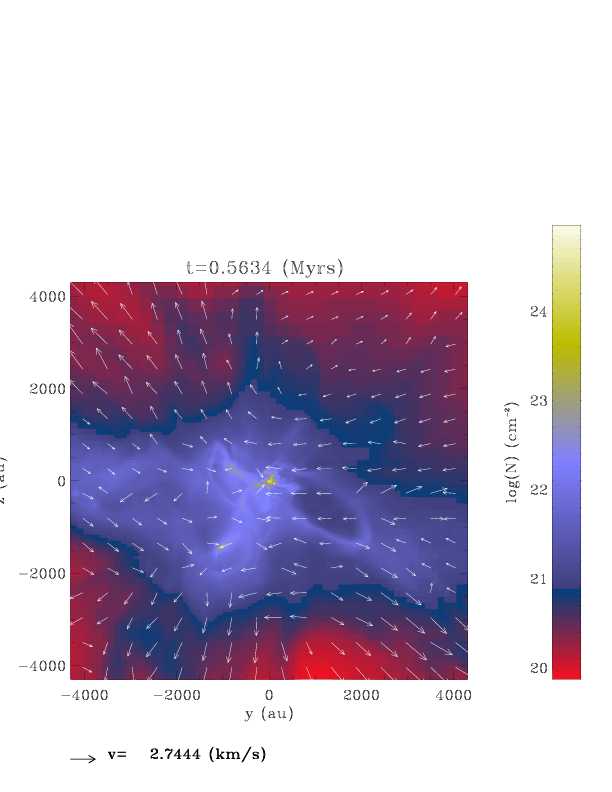}}
\put(0,0){\includegraphics[width=5.5cm]{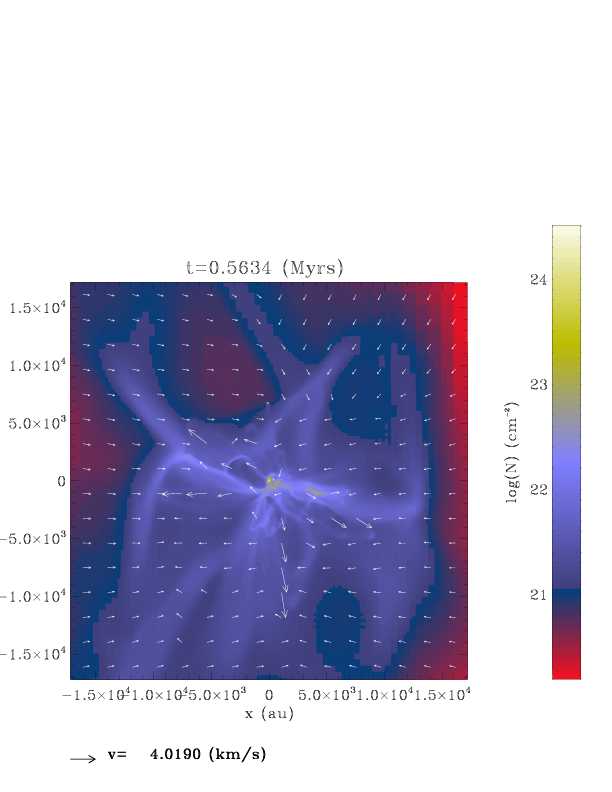}}
\put(6,6.5){\includegraphics[width=5.5cm]{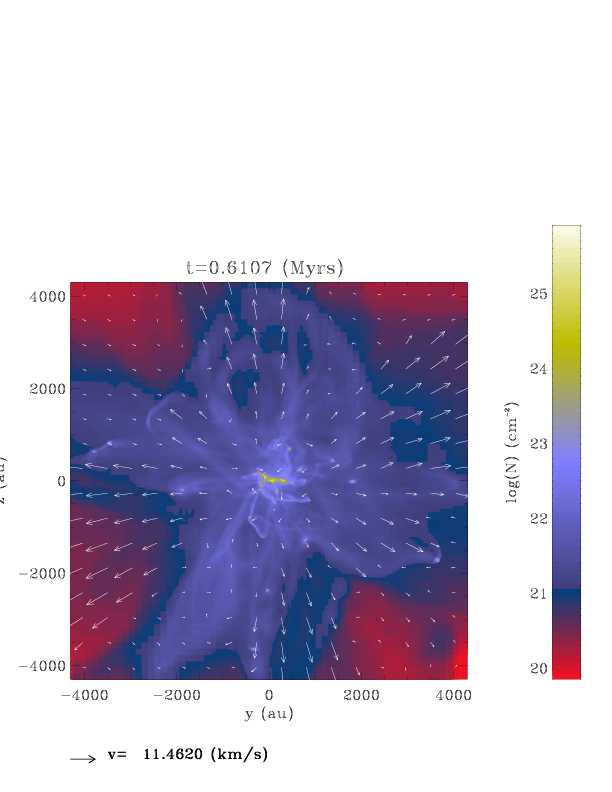}}
\put(6,0){\includegraphics[width=5.5cm]{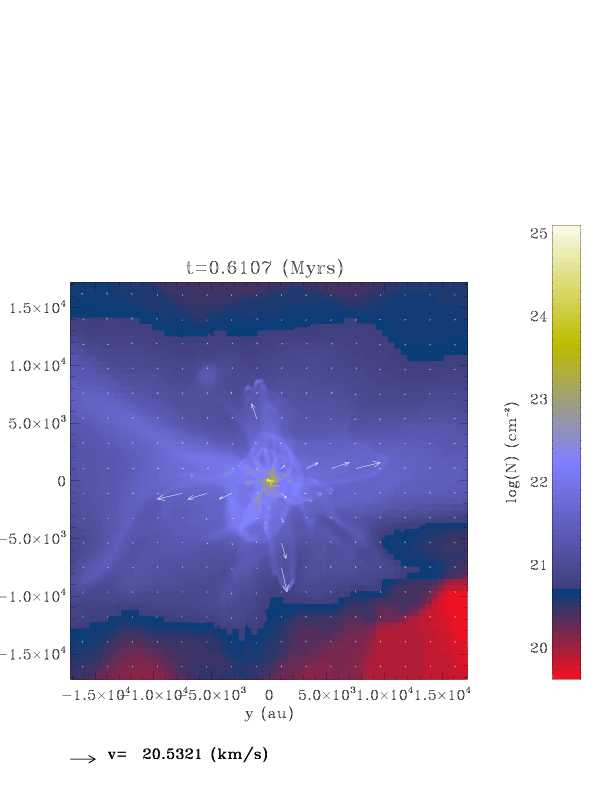}}
\put(12,6.5){\includegraphics[width=5.5cm]{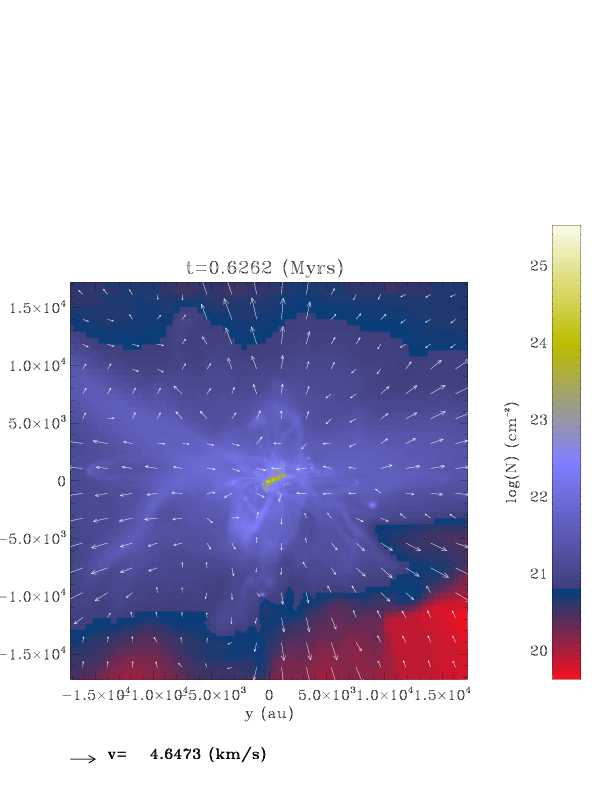}}
\put(12,0){\includegraphics[width=5.5cm]{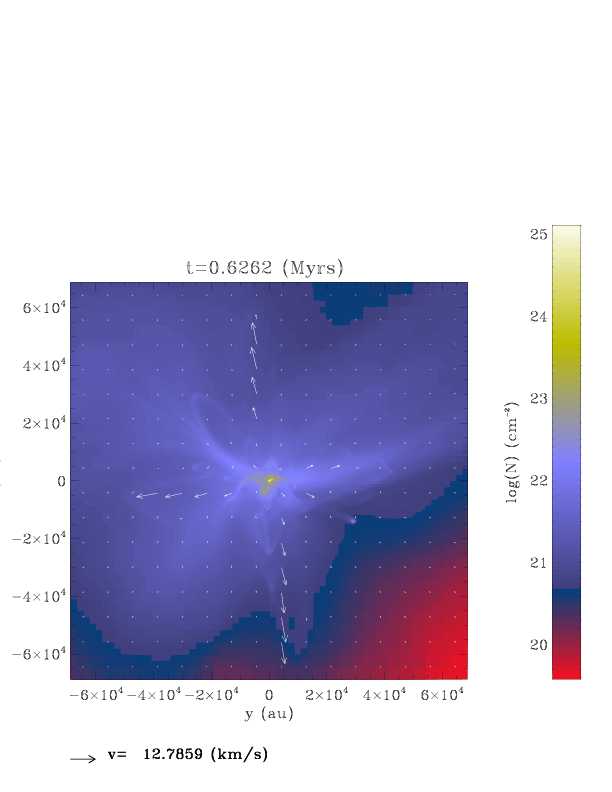}}
\end{picture}
\caption{ $\mu=5$. Column density and projected velocity field.}
\label{100_mu5_out}
\end{figure*}

\setlength{\unitlength}{1cm}
\begin{figure*} 
\begin{picture} (0,11)
\put(0,0){\includegraphics[width=7.5cm]{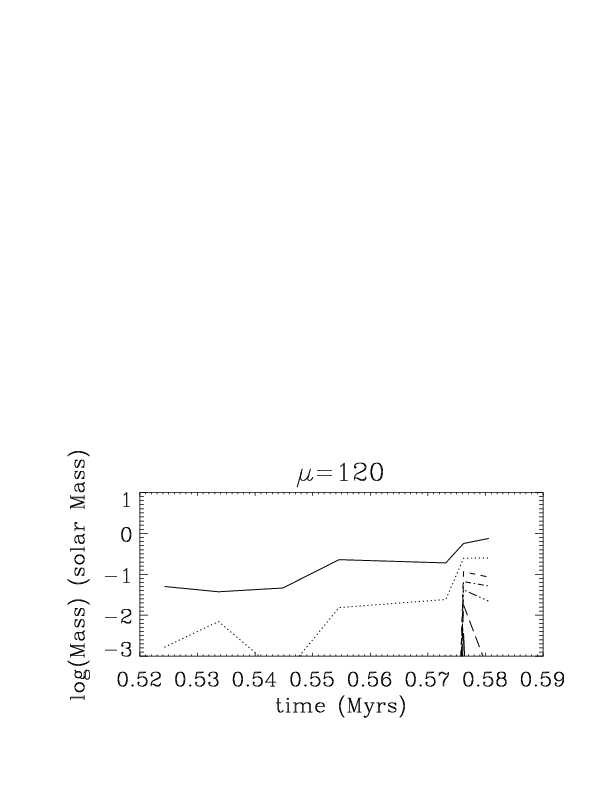}}
\put(8,0){\includegraphics[width=7.5cm]{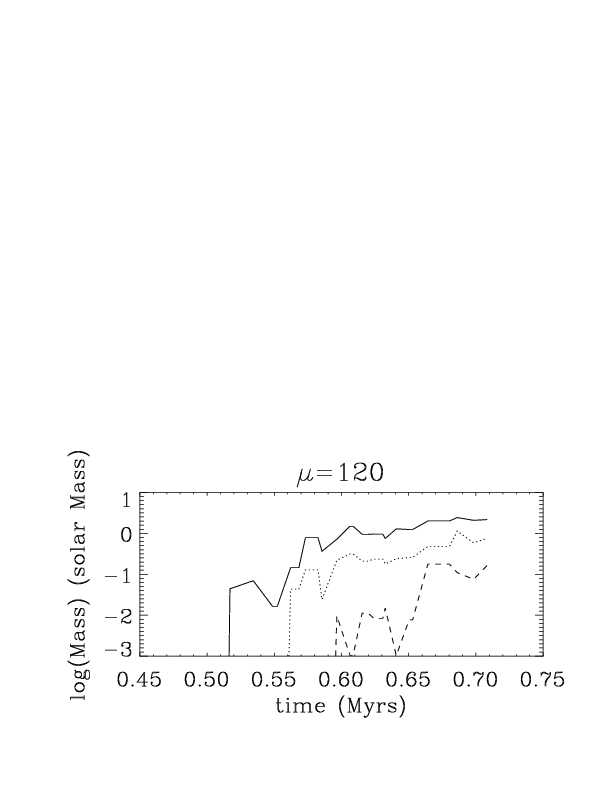}}
\put(0,3.5){\includegraphics[width=7.5cm]{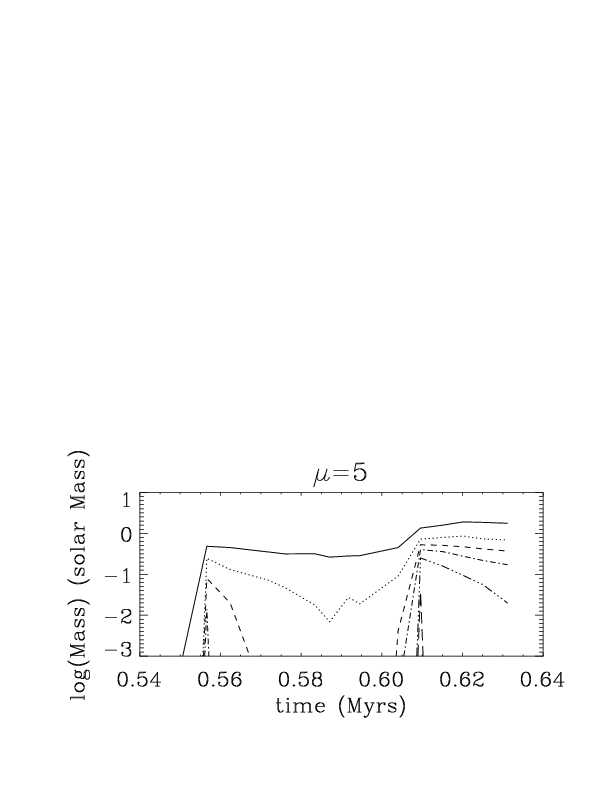}}
\put(8,3.5){\includegraphics[width=7.5cm]{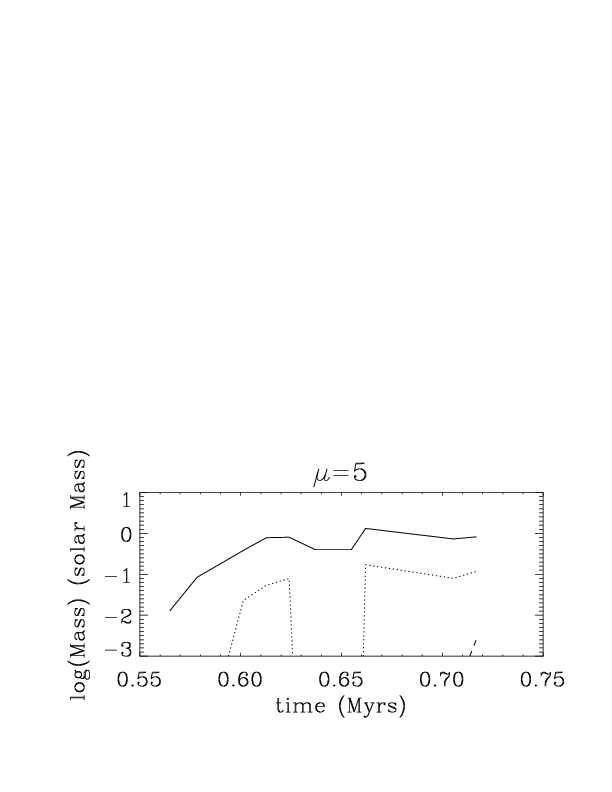}}
\put(0,7){\includegraphics[width=7.5cm]{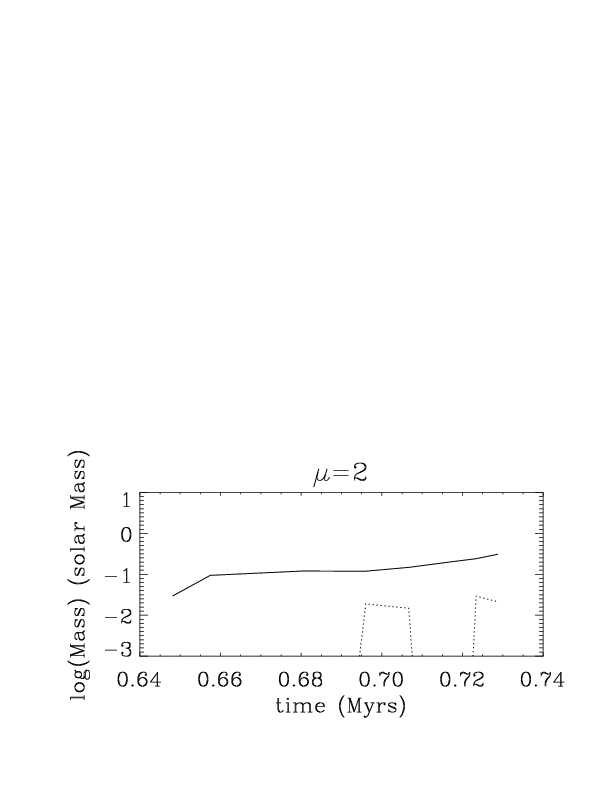}}
\put(8,7){\includegraphics[width=7.5cm]{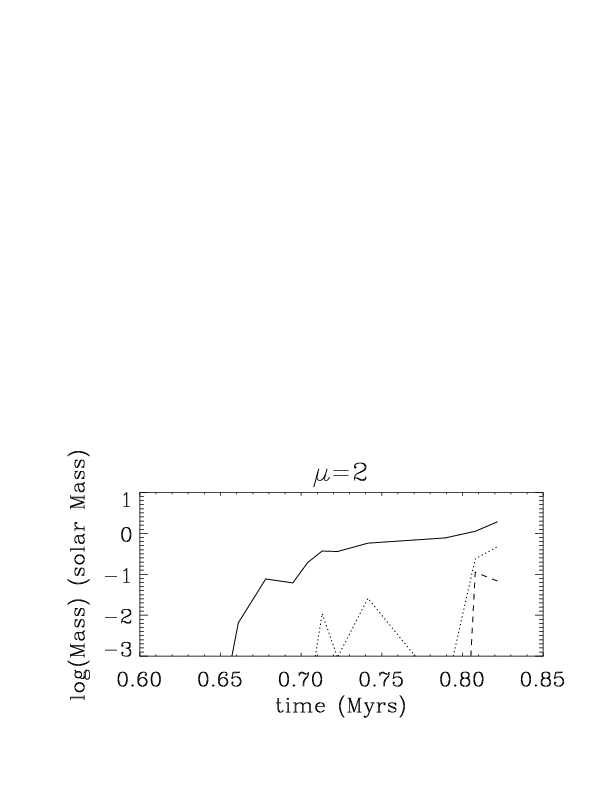}}
\end{picture}
\caption{Mass in the outflows as a function of time
for various velocity thresholds (see text). 
The left column shows the high resolution simulations, while 
the right column shows the lower resolution.
Top panels display the $\mu=2$ case, middle panels the $\mu=5$ ones, while
the bottom panel displays $\mu=120$.}
\label{mass_outflow}
\end{figure*}

\subsection{Magnetic field evolution}
The average magnetic intensity as a function of time is displayed in 
Fig.~\ref{mag} for various density thresholds. As expected,
the magnetic intensity increases with the density. Although 
for $\mu=2$  the magnetic intensity is about twice as high as
 for $\mu=5$,  at low density (solid line corresponding 
to a threshold of $10^3$ cm$^{-3}$), the magnetic intensity 
at higher density thresholds is nearly comparable for both cases. 
Similarly, while the magnetic intensity is extremely low 
for the density threshold $10^5$ cm$^{-3}$ in the case 
$\mu=120$, it is much closer to the values obtained for $\mu=5$ and 2
at higher densities, although it is still weaker by a factor of a few. 

This is because the magnetic field is less amplified 
when it is stronger because the gas tends to flow preferentially along
the field lines. Indeed, in the weak field case, one expects 
a nearly spherical contraction that leads to $B \propto \rho^{2/3}$, while 
when the field is stronger, $B \propto \rho^{1/2}$ (e.g. Basu 1997). Consequently, 
the magnetic intensity increases more rapidly when it is low than 
when it is high and tends to take a narrower range of values at higher 
density.

This is more clearly visible in Fig.~\ref{alfven_vol}, which  shows the 
mean Alfv\'en velocity, $<v_a> = < B / (\sqrt{4 \pi  \rho}) > = (\sum  v_a dV) / ( \sum   dV )$,
as a function of the radius. The mean Alfv\'en speed rapidly decreases toward the edge of the cloud 
in the outer part, in particular for $\mu=120$, and then reaches a maximum after 
which it tends to form a plateau, implying that $B \propto \rho^{1/2}$.  
For $\mu=2$  while the Alfv\'en speed is initially of the order of 
0.7 km s$^{-1}$, its value in the inner part is about   
 2-3 km s$^{-1}$ up to $r \simeq 10^{-4}$ pc, below which the magnetic intensity 
steeply drops. This latter behavior is caused by the numerical diffusion,
which becomes significant below ten computing cells and 
clearly shows the limit of these simulations. As the sound speed is about 
0.2 km s$^{-1}$, in the inner part of the collapsing cloud 
the magnetic support both in the $\mu=5$ and $\mu=2$ cases is largely dominating
over the thermal one.

The mean values displayed in Fig.~\ref{alfven_vol} do not reflect
the complexity of the magnetic field behaviour,  however. This is well 
illustrated by 
Fig.~\ref{alfven} which shows the Alfv\'en velocity in the xy plane for 
$\mu=120$ and $\mu=2$. In the first case, the Alfv\'en 
speed is, as expected very low while  at smaller scales, $r < 500$ AU, 
it dominates over the sound speed. Overall, it presents large 
fluctuations at all scales, which is a consequence 
of the weakness of the field. For $\mu=2$, the 
Alfv\'en velocity almost always dominates over the sound 
speed. Interestingly, there is a layer extending along the y-axis where the 
Alfv\'en velocity is lower by a factor of about 3 than in the surrounding 
medium. This  layer, which is the pseudo-disk, and extends nearly perpendicular to the initial direction 
of the magnetic field, is denser because of the magnetic 
compression in the x-direction induced by the pinching of the field lines (e.g. 
Li \& Shu 1996, Hennebelle \& Fromang 2008). This density 
enhancement is responsible for the slightly lower Alfv\'en velocity.
At smaller scales, $r < 500$ AU, the Alfv\'en velocity fluctuates significantly 
and the structure of the magnetic field is clearly much less ordered. 
As  is the case for uniformly rotating cloud, and even though the 
 angular momentum is not well conserved, the rotation motions become
dominant in the inner part of the cloud.

\section{Outflows}
The purpose of this section is to study the
 outflows that are launched in the numerical simulations.
Indeed, outflows spontaneously form in all simulations we run.

\subsection{Morphology and scales}
Figures~\ref{100_hy_out} and \ref{100_mu5_out} show three 
snapshots for $\mu=120$ and $\mu=5$. They display 
the column density along the x-axis integrated over a length
equal to the length of the map whose center is the 
density peak of the cloud. The arrows represent the 
velocity field obtained by taking along the line of sight the 
highest projected velocity (i.e. we select the velocity that 
 has the largest module in the yz-plane).  
In the top row the size of the snapshots
is about 8000 AU, while it is about 32000 AU for the bottom one
except  for the third column of Fig.~\ref{100_mu5_out}, for which 
the size of the snapshots is four times these values. 

While the first snapshot ($t=0.5741$ Myr) 
of Fig.~\ref{100_hy_out} shows no sign of 
outflows, in the second snapshot a relatively fast outflow is clearly 
evident in the upper part of the map ($z=0-4000$ AU). It has a broad 
angle of almost 90$^\circ$ (top panel) at 4000 AU and 45$^\circ$
at 1.5$\times 10^4$ AU. The outflow is not bipolar because it is almost entirely 
propagating towards the north with only a weak component propagating
 toward the south. The highest velocity which is as high 
as 47~km~s$^{-1}$ in the second snapshot, decreases with time and 
has dropped to about 13~km~s$^{-1}$ by the time of the third panel,
suggesting that the high speed is associated to a transient 
phase rather than a stationary stage.
 Although outflows and jets are a common feature of 
MHD collapse calculations (e.g. Machida et al. 2005, 
Banerjee \& Pudritz 2006, Mellon \& Li 2008, 
Hennebelle \& Fromang 2008, Ciardi \& Hennebelle 2010), 
which are thought to be caused by the magneto-centrifugal  mechanism 
(e.g. Blandford \& Payne 1982,   Pelletier \& Pudritz 1992,  Ferreira 1997
 and also Spruit 1996 for a discussion about the various interpretations
of the launching mechanism),  
it may sound surprising to see outflows being launched in 
a cloud that has such a small initial magnetic field. However,
as already discussed, the magnetic field is strongly amplified 
during the collapse (see Figs.~\ref{mag} and \ref{alfven_vol}). 
In a sense it is similar to the result of Machida et al. (2008), who 
treated the ohmic dissipation during the second collapse, and observed
in their simulation the launching of a strong jet induced 
by the rotation of the young protostar even though 
most of the magnetic flux has been lost by diffusion. 
 In this case, the weak magnetic field is rapidly twisted by the 
rotation and the toroidal magnetic pressure gradient efficiently
accelerates the flow (e.g. Spruit 1996).
The weak collimation and the strong asymmetry are 
 probably 
consequences of the weakness of the initial magnetic intensity though.

The situation is different for $\mu=5$. The first 
column reveals that the velocities are  lower ($\simeq$3-4 km s$^{-1}$)
and  that while the outflows tend to be more collimated, they tend to 
be more symmetrical with respect to the centre. Interestingly, we 
observe four thin flows instead of two. This is even  clearer
in the second snapshot, where the outflow is almost quadrupolar. 
These directions are significantly 
different from the outflow directions at time $0.563$ Myr. The velocities
are also about 4-5 times higher. The last snapshot shows that this outflow
propagates through the cloud (at time 0.62 Myr it reaches about 0.3 pc) 
and slows down.

For $\mu=2$,  outflows are also observed but their 
velocities are much lower and rarely exceed 3 km s$^{-1}$.
For this reason they are not displayed here although they can easily 
be seen in Fig.~\ref{100_mu2} (first column, bottom panel). 

It seems therefore that the outflows produced in the simulations are 
relatively fast for  low and intermediate magnetic intensities,
 slower for stronger fields, 
while in general intermittent and not bipolar. 
 The exact reason of this is not entirely clear but 
it may be that for stronger fields, because there is less angular momentum 
left in the cloud inner part because of the efficient braking, 
the twisting of the field lines is weaker and 
therefore the pressure gradient should be less steep.

Before turning to a 
quantitative description, we  find it useful to comment on the 
expected order of magnitude of the outflow velocity.
In the context of stationary, axisymmetric configurations, 
it has been established that (e.g. Pudritz et al. 2007)
\begin{eqnarray}
V \simeq \sqrt{2} \lambda_{mag} \sqrt{ {G M} \over{R} },
\label{vel_out}
\end{eqnarray}
where $M$ is the mass of the central object, $R$ is the 
radius from which the outflow is launched and $\lambda_{mag}$ is the 
magnetic lever arm, which typically is found to be of the order of 2-3.
This leads to
\begin{eqnarray}
V \simeq 3 \, {\rm km \, s^{-1}} \times \lambda_{mag}  \left( M \over  {1 M_\odot}\right)^{1/2}  
\left( R \over {100 AU} \right)^{-1/2}.
\label{vel_order}
\end{eqnarray}
The launching radius is not easily determined given the complexity of the 
flow. Visual inspection of Figs.~\ref{100_hy_out} and~\ref{100_mu5_out}
suggests that $R \simeq 1000$ AU is a reasonable order of magnitude. This 
is corroborated by Fig.~\ref{radial_velocity}, where it is shown that 
the radius of the inner region at which the velocity field is not 
dominated by systematic collapse is of the order of 1000 AU. 
The mass enclosed within this radius is of the order of 10 $M_\odot$, thus
a typical velocity for the outflows is of the order of 6-10 km s$^{-1}$.
It is worth stressing that the velocities fluctuate by orders of a few
around this simple estimate. This dispersion is easily 
accounted for given the uncertainties on the lever arm $\lambda_{\rm mag}$, 
the mass $M$, and the launching radius, $R$. One should also keep in mind 
that Eq.~(\ref{vel_order}) is inferred in the context of
 stationary and axisymmetric solutions, which is obviously not 
the case in our simulations.

\subsection{Masses and velocities}

To quantify the outflows more precisely, we computed the mass 
within the outflows as a function of time. As a criterion
to identify the mass they contain, we 
select the computational cells with a positive radial velocity  
that is higher than a given threshold. To avoid confusion 
with cells close to the density peak where high velocities
can also be achieved, we select cells whose distance from the density 
peak exceeds 1000 AU. We adopt six thresholds of 
1, 3, 5, 7, 10 and 20 km s$^{-1}$ (shown as  solid, dotted, 
dashed, dot-dashed, triple dot-dashed, long dashed respectively).
 Figure~\ref{mass_outflow}
shows the different masses as a function of time of the 
three cases $\mu=120, 5, 2$ and for the two resolutions (left column 
displays the high resolution runs).

At least two distinct episodes 
of ejection occur, for the three magnetic intensities, 
 the second leading to faster velocities. While in 
$\mu=2$ case, only a small mass is launched at velocities 
higher than 3 km s$^{-1}$, for $\mu=5$  almost 1 solar mass
of gas is ejected at a speed higher than 3 km s$^{-1}$ and at time 
0.61 Myr, more than 0.1 solar masses possess a velocity higher than 10 
km s$^{-1}$. Interestingly enough for $\mu=120$, the low resolution calculations reveal 
that at later times the mass in the outflows is typically larger by a factor of a few
while for  $\mu=2$, the mass at the end of the calculation
is about one orders of magnitude larger and seem to be still increasing with time.
For $\mu=120$ and $\mu=5$, the ejected mass does not seem to increase with time.

A comparison between the accreted mass (at density higher than 10$^{9-11}$ cm$^{-3}$)
indicates that the fraction of ejected mass over accreted mass, is of the order of 
one third in the high resolution models (except for $\mu=2$) and about one tenth in the 
low resolution one. These numbers are close to what Ciardi \& Hennebelle (2010) have 
been inferring for low mass cores. 

Altogether, the outflows are clearly not stationary and episodic. It is 
important to stress that while the general trends are similar for the low and 
the high resolution runs, the velocities, compared at the same physical time, 
are larger in the high resolution
case and that the outflows are more massive. This clearly means 
that numerical resolution plays an important role here. It is 
not excluded, and indeed even likely, that numerical convergence has not 
been reached yet and a better resolution may lead to even faster 
and more massive flows.

\setlength{\unitlength}{1cm}
\begin{figure*}
\begin{picture}(0,13.5)
\put(0,6.5){\includegraphics[width=5.5cm]{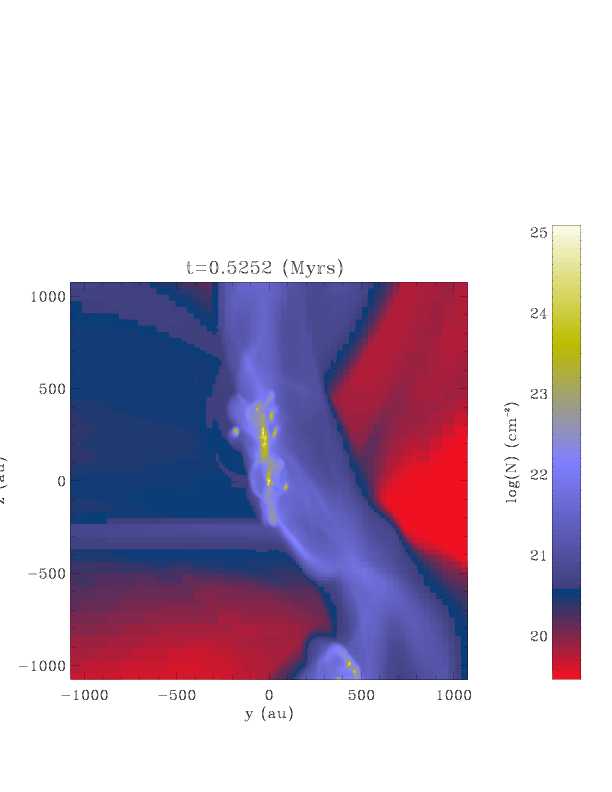}}
\put(0,0){\includegraphics[width=5.5cm]{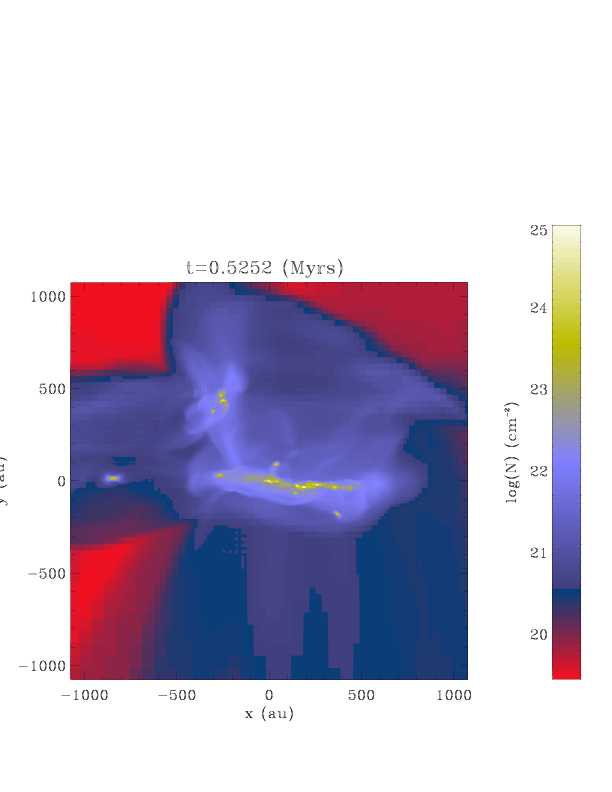}}
\put(6,6.5){\includegraphics[width=5.5cm]{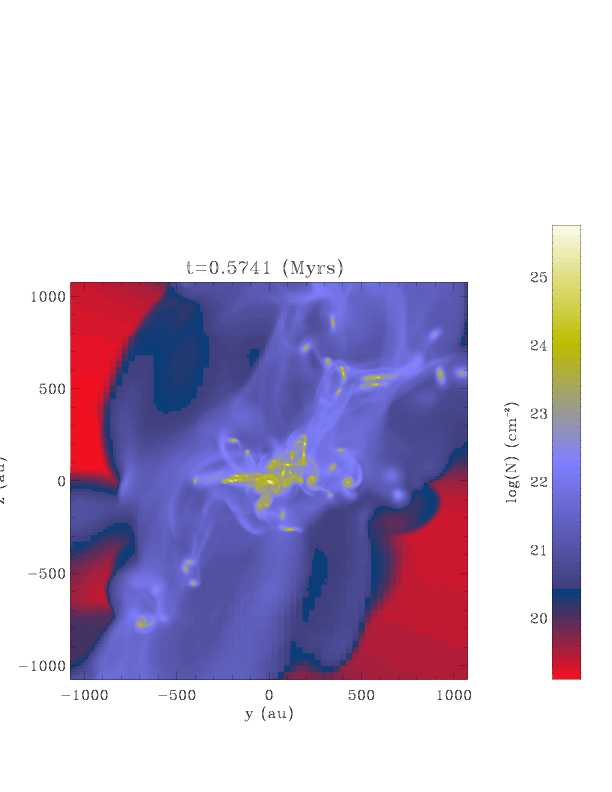}}
\put(6,0){\includegraphics[width=5.5cm]{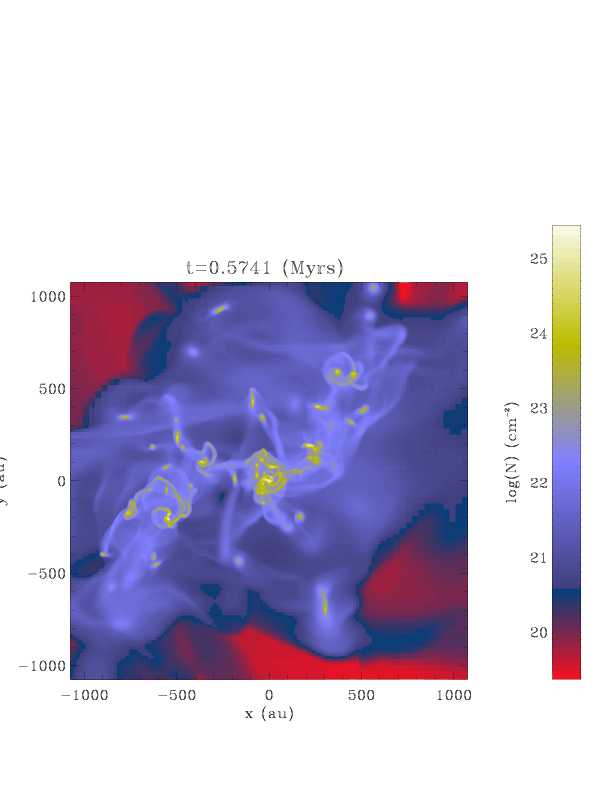}}
\put(12,6.5){\includegraphics[width=5.5cm]{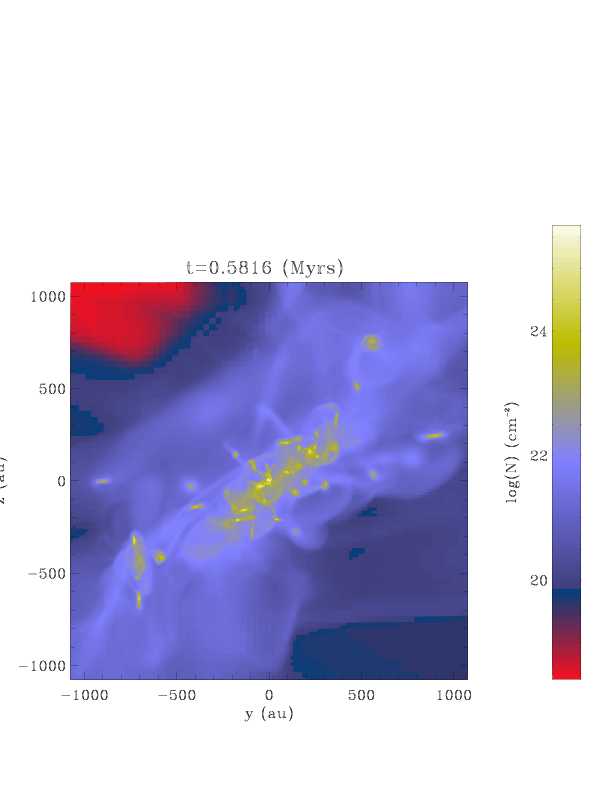}}
\put(12,0){\includegraphics[width=5.5cm]{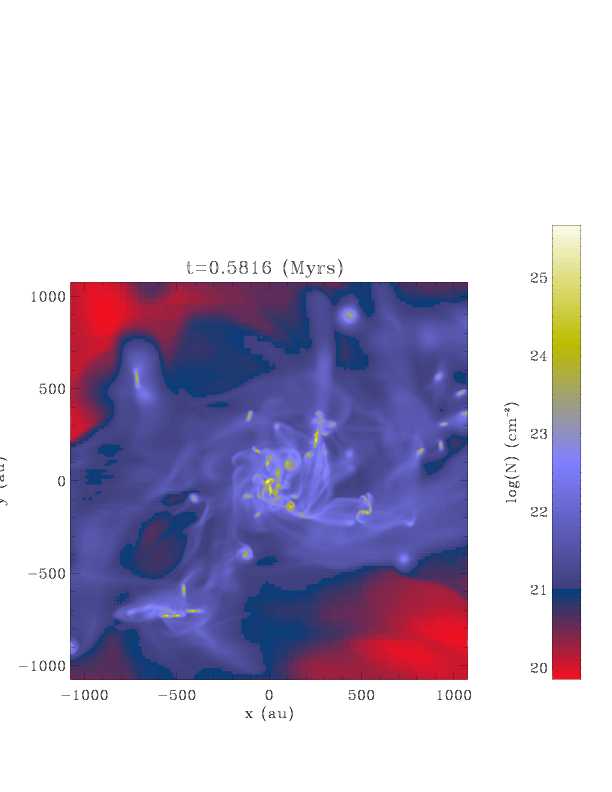}}
\end{picture}
\caption{ $\mu=120$. Column density.}
\label{100_hy}
\end{figure*}

\setlength{\unitlength}{1cm}
\begin{figure*}
\begin{picture}(0,13.5)
\put(0,6.5){\includegraphics[width=5.5cm]{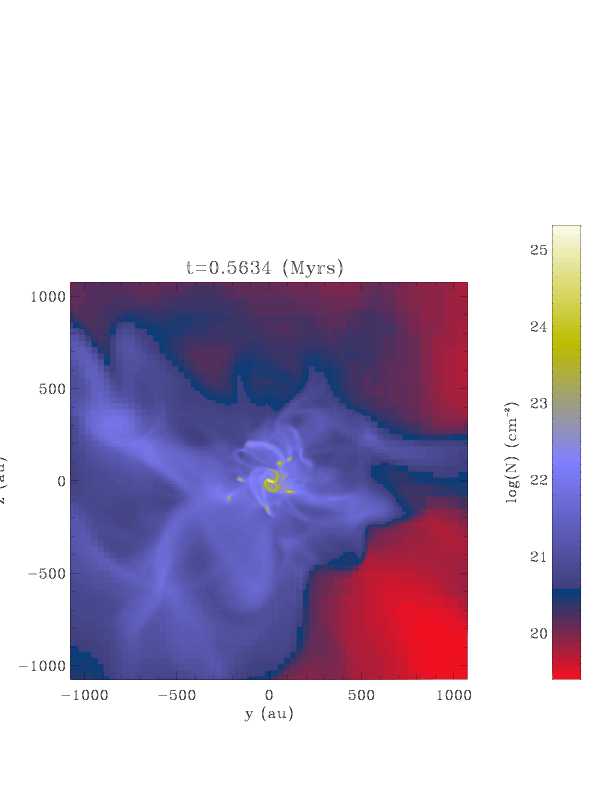}}
\put(0,0){\includegraphics[width=5.5cm]{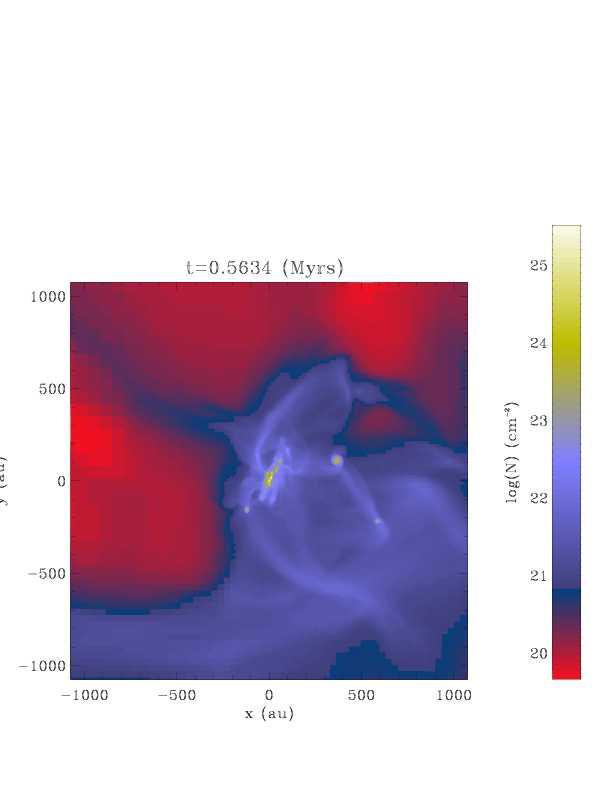}}
\put(6,6.5){\includegraphics[width=5.5cm]{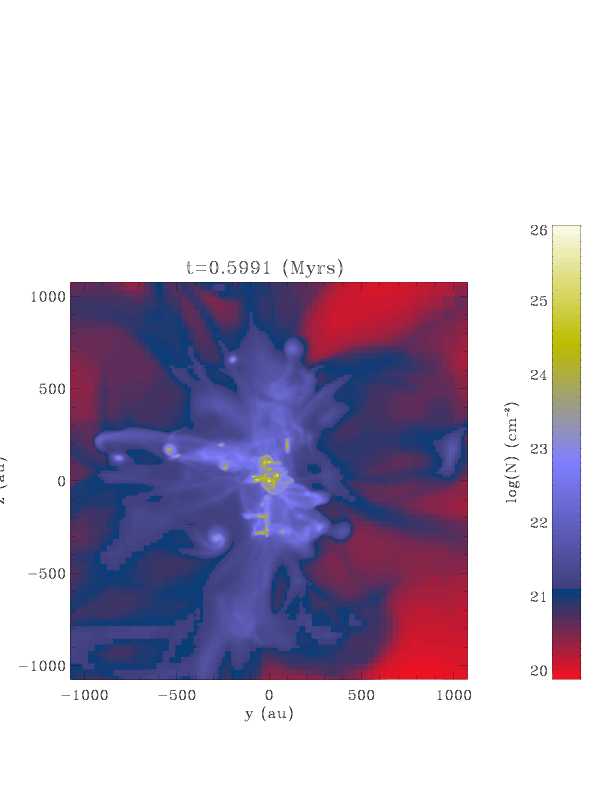}}
\put(6,0){\includegraphics[width=5.5cm]{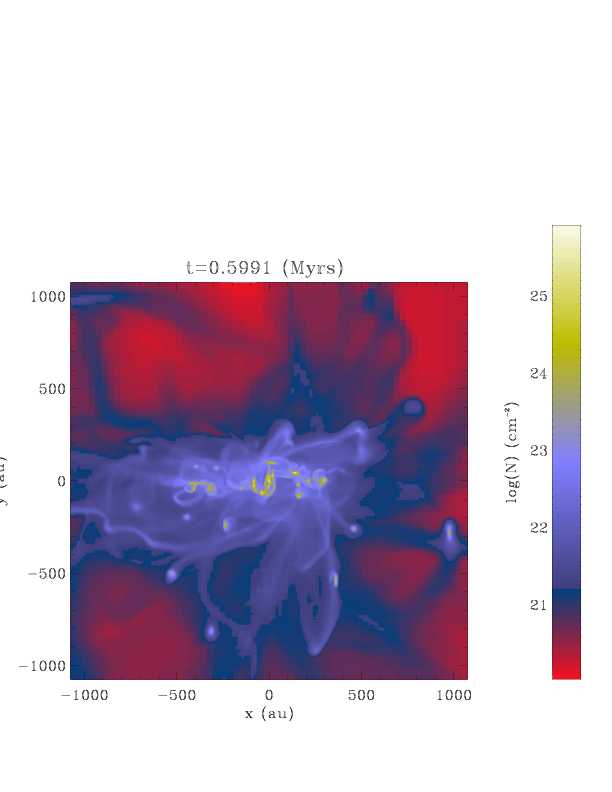}}
\put(12,6.5){\includegraphics[width=5.5cm]{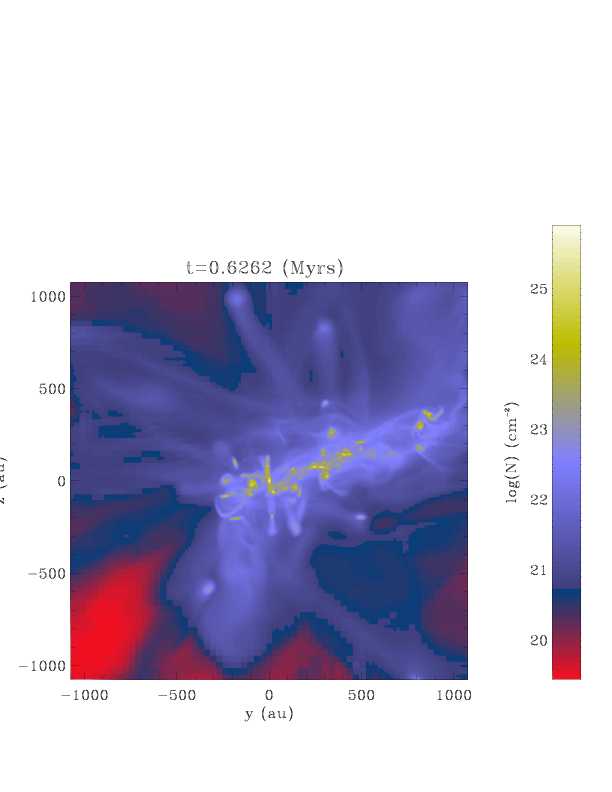}}
\put(12,0){\includegraphics[width=5.5cm]{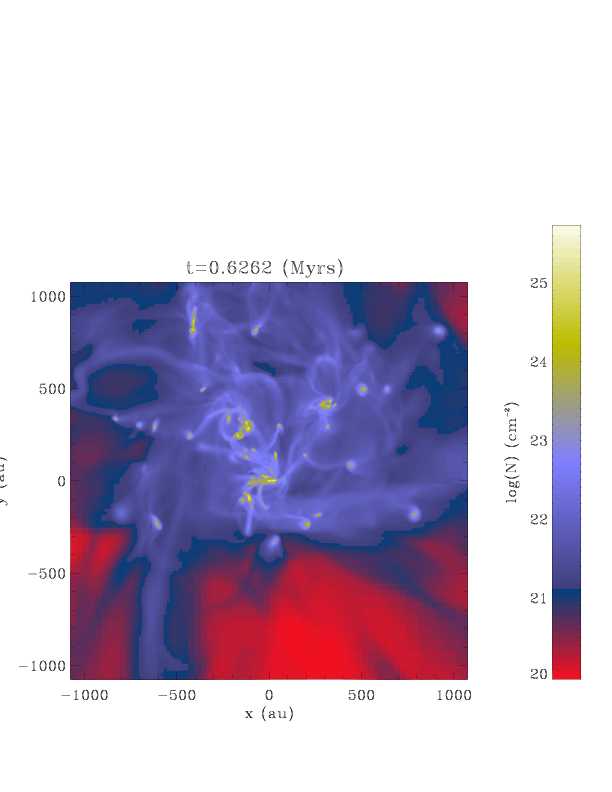}}
\end{picture}
\caption{$\mu=5$. Column density.}
\label{100_mu5}
\end{figure*}

\setlength{\unitlength}{1cm}
\begin{figure*}
\begin{picture}(0,13.5)
\put(0,6.5){\includegraphics[width=5.5cm]{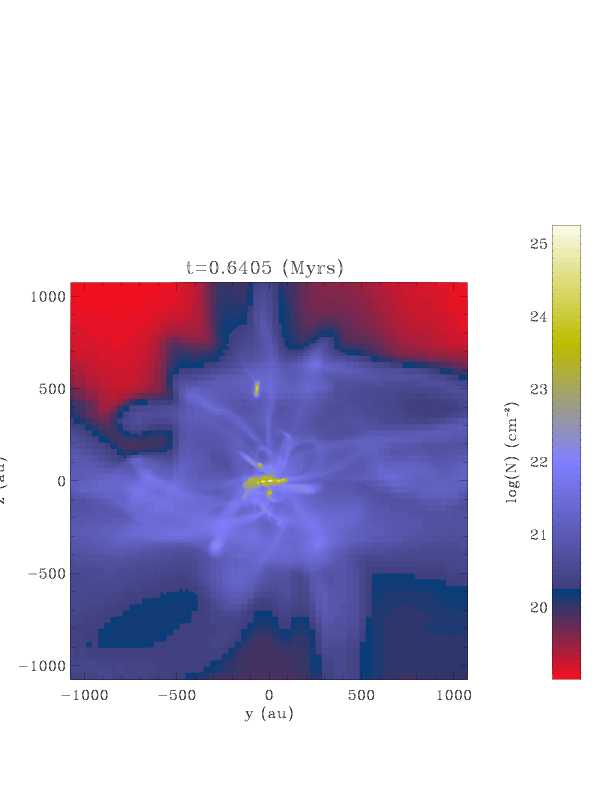}}
\put(0,0){\includegraphics[width=5.5cm]{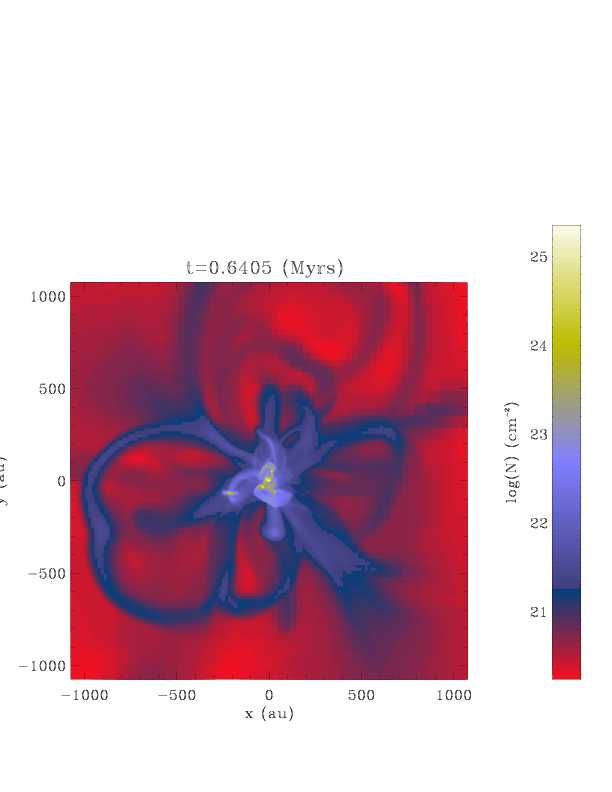}}
\put(6,6.5){\includegraphics[width=5.5cm]{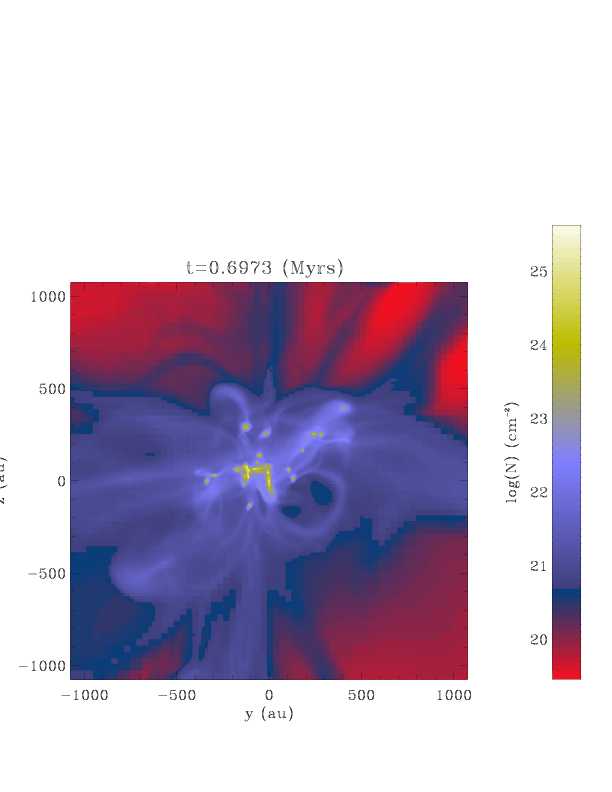}}
\put(6,0){\includegraphics[width=5.5cm]{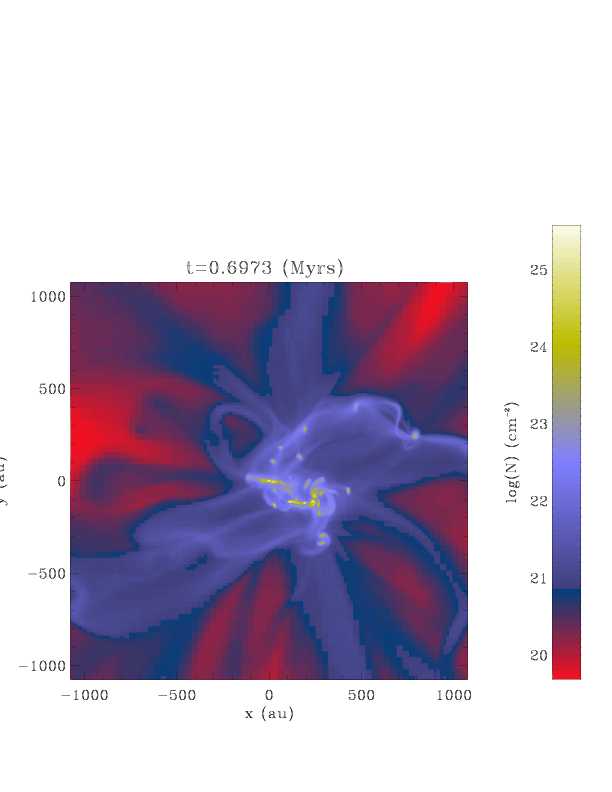}}
\put(12,6.5){\includegraphics[width=5.5cm]{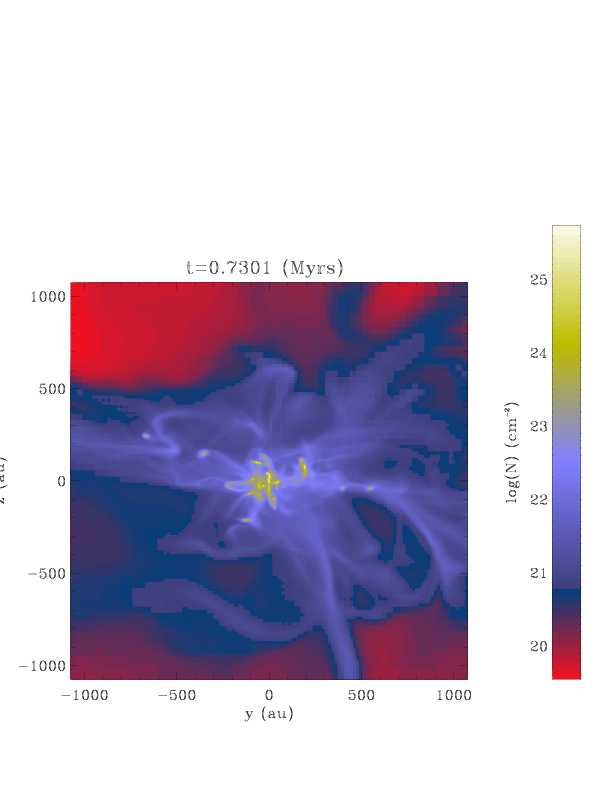}}
\put(12,0){\includegraphics[width=5.5cm]{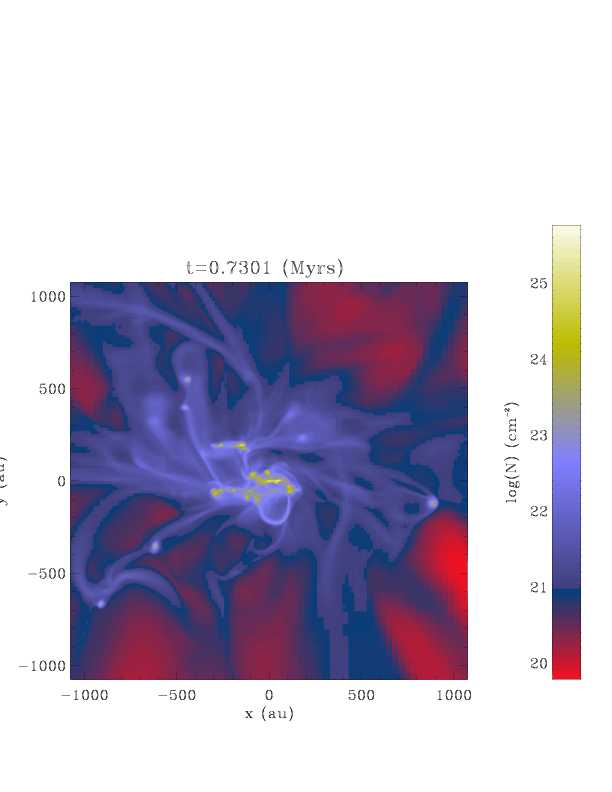}}
\end{picture}
\caption{$\mu=2$. Column density.}
\label{100_mu2}
\end{figure*}

\setlength{\unitlength}{1cm}
\begin{figure*} 
\begin{picture} (0,11)
\put(0,0){\includegraphics[width=7.5cm]{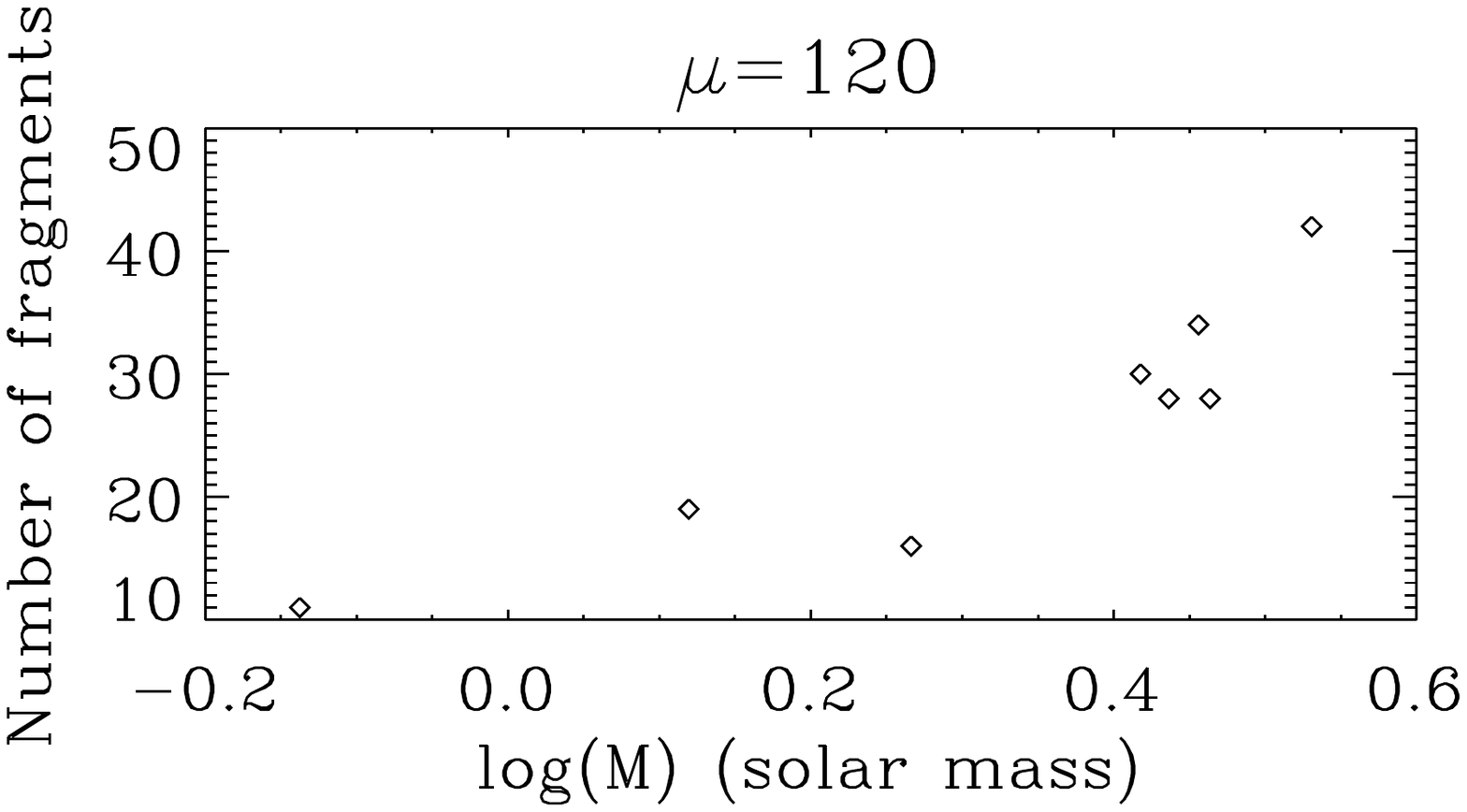}}
\put(8,0){\includegraphics[width=7.5cm]{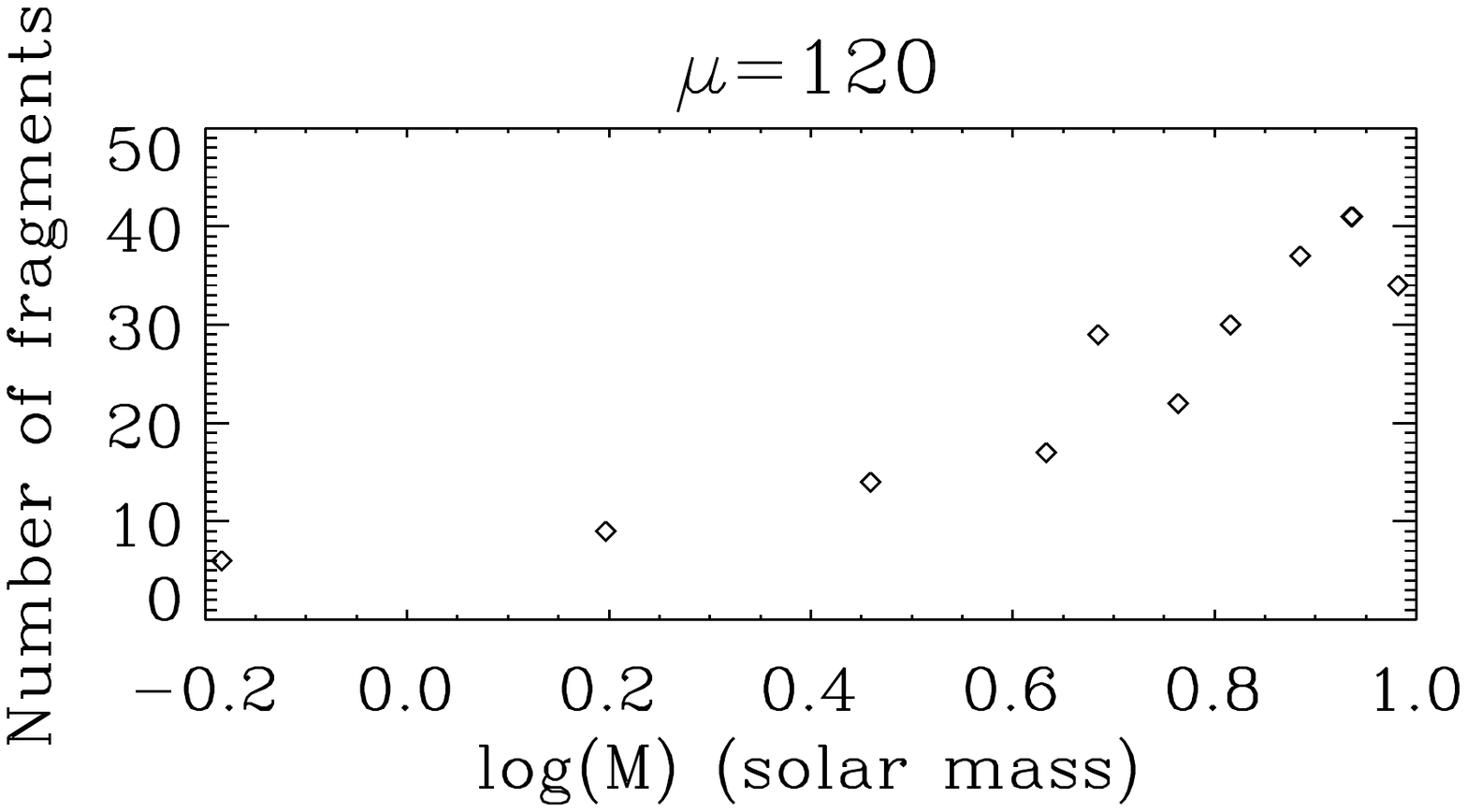}}
\put(0,3.5){\includegraphics[width=7.5cm]{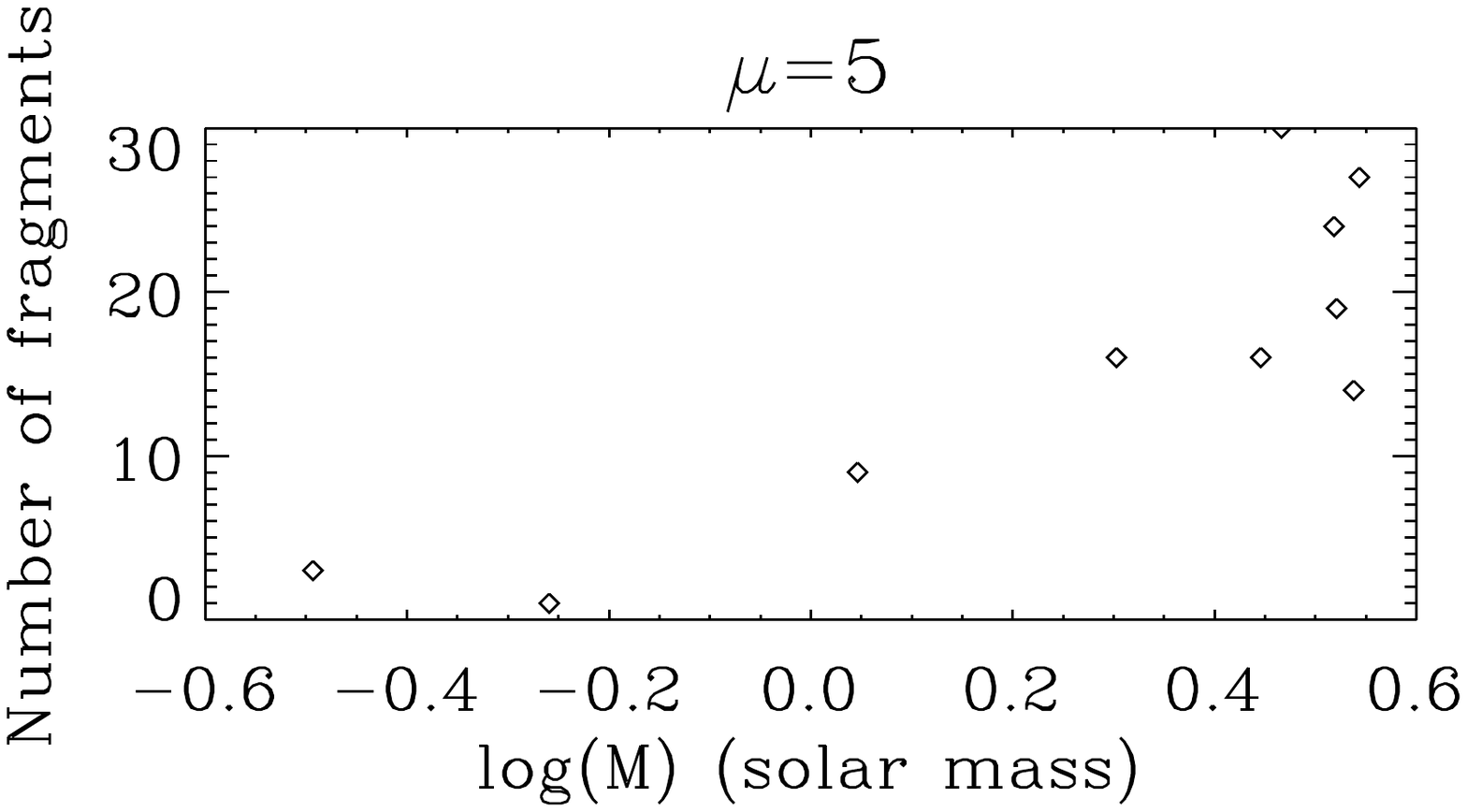}}
\put(8,3.5){\includegraphics[width=7.5cm]{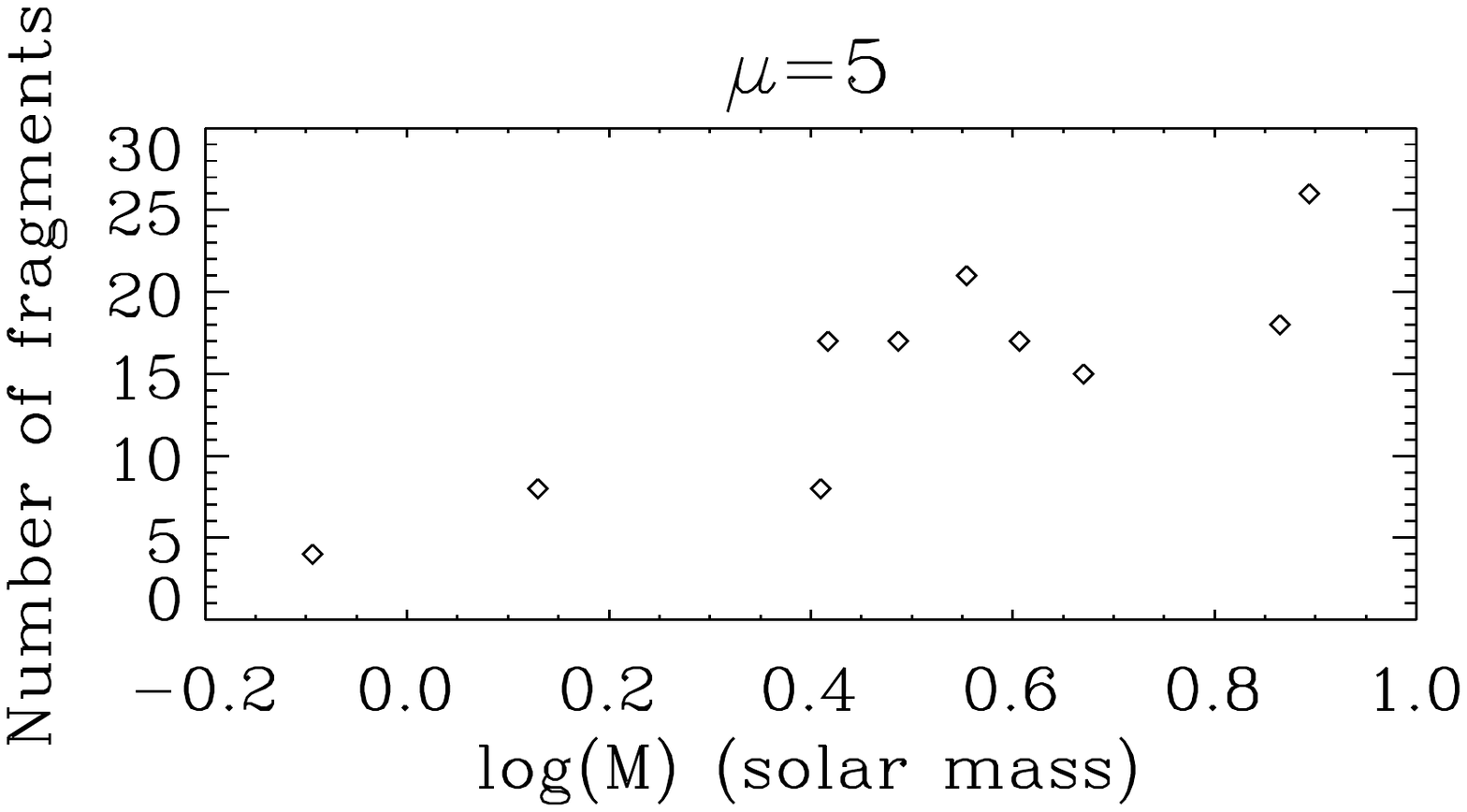}}
\put(0,7){\includegraphics[width=7.5cm]{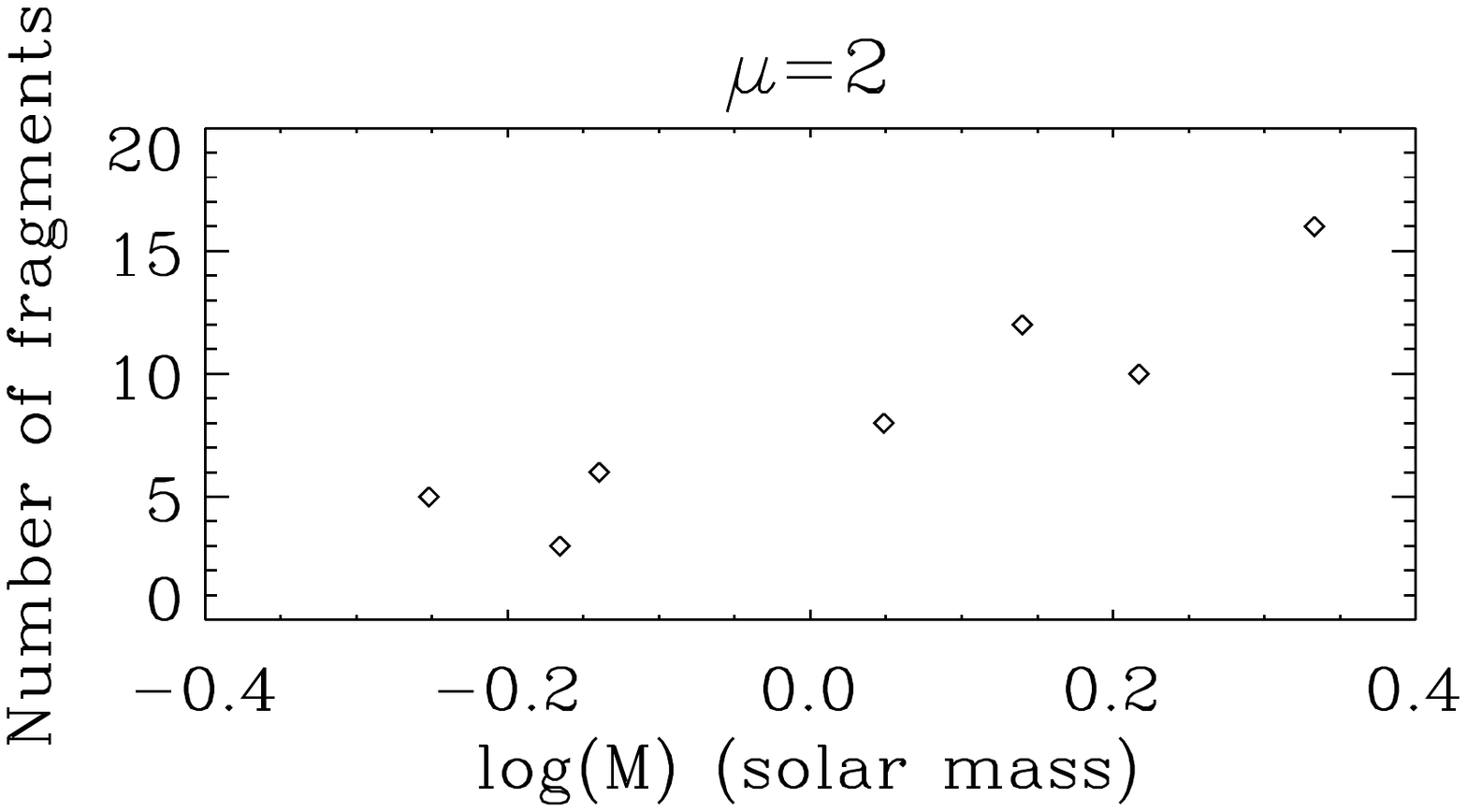}}
\put(8,7){\includegraphics[width=7.5cm]{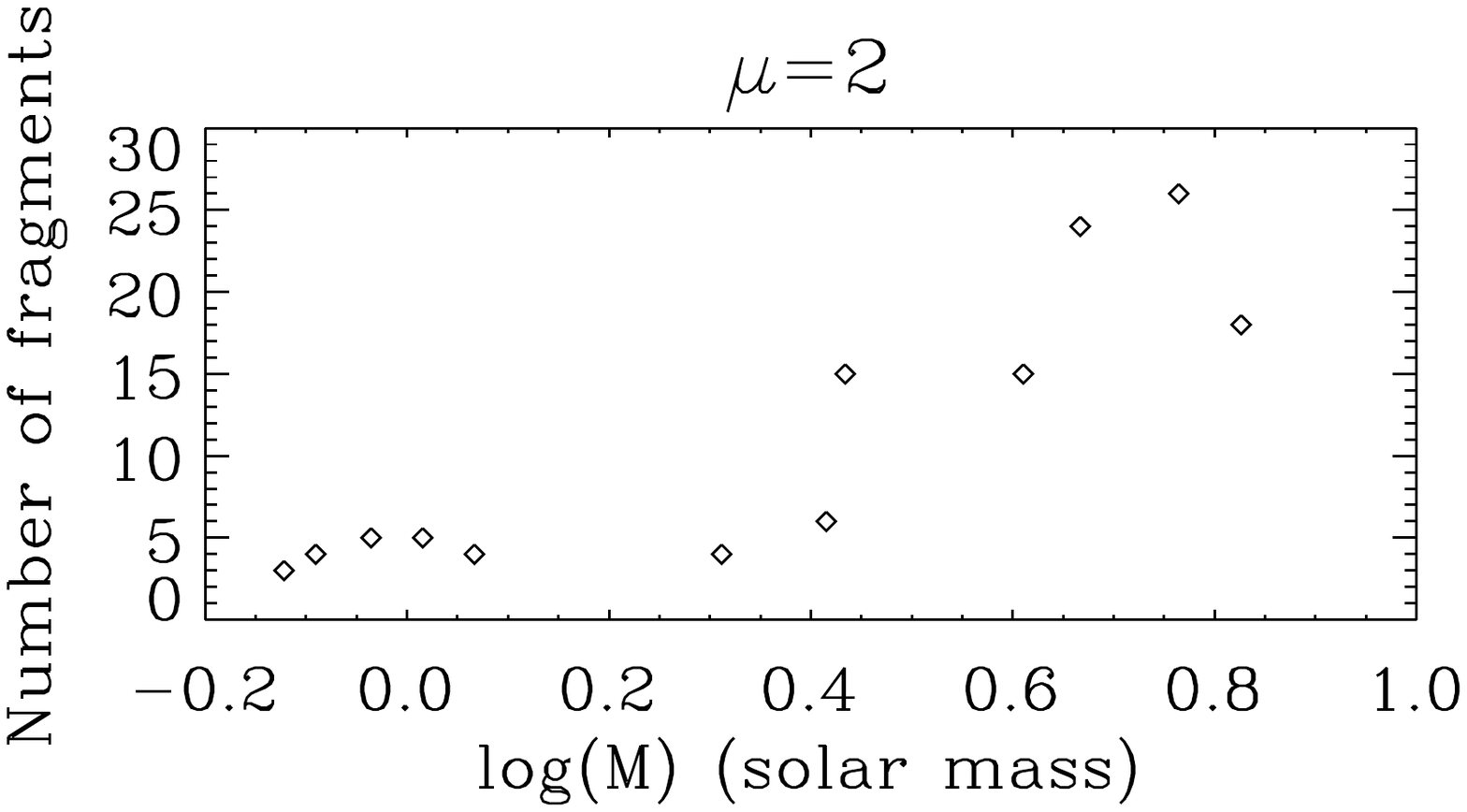}}
\end{picture}
\caption{Number of fragments more massive than $10^{-2} \, M_\odot$ 
versus total mass within fragments. The left column shows the high resolution
 simulations, while 
the right column shows the lower resolution.
Top panels display the $\mu=2$ case, middle panels the $\mu=5$ ones, while
bottom panels display the $\mu=120$ case. }
\label{number_fragment}
\end{figure*}

\setlength{\unitlength}{1cm}
\begin{figure*} 
\begin{picture} (0,11)
\put(0,0){\includegraphics[width=7.5cm]{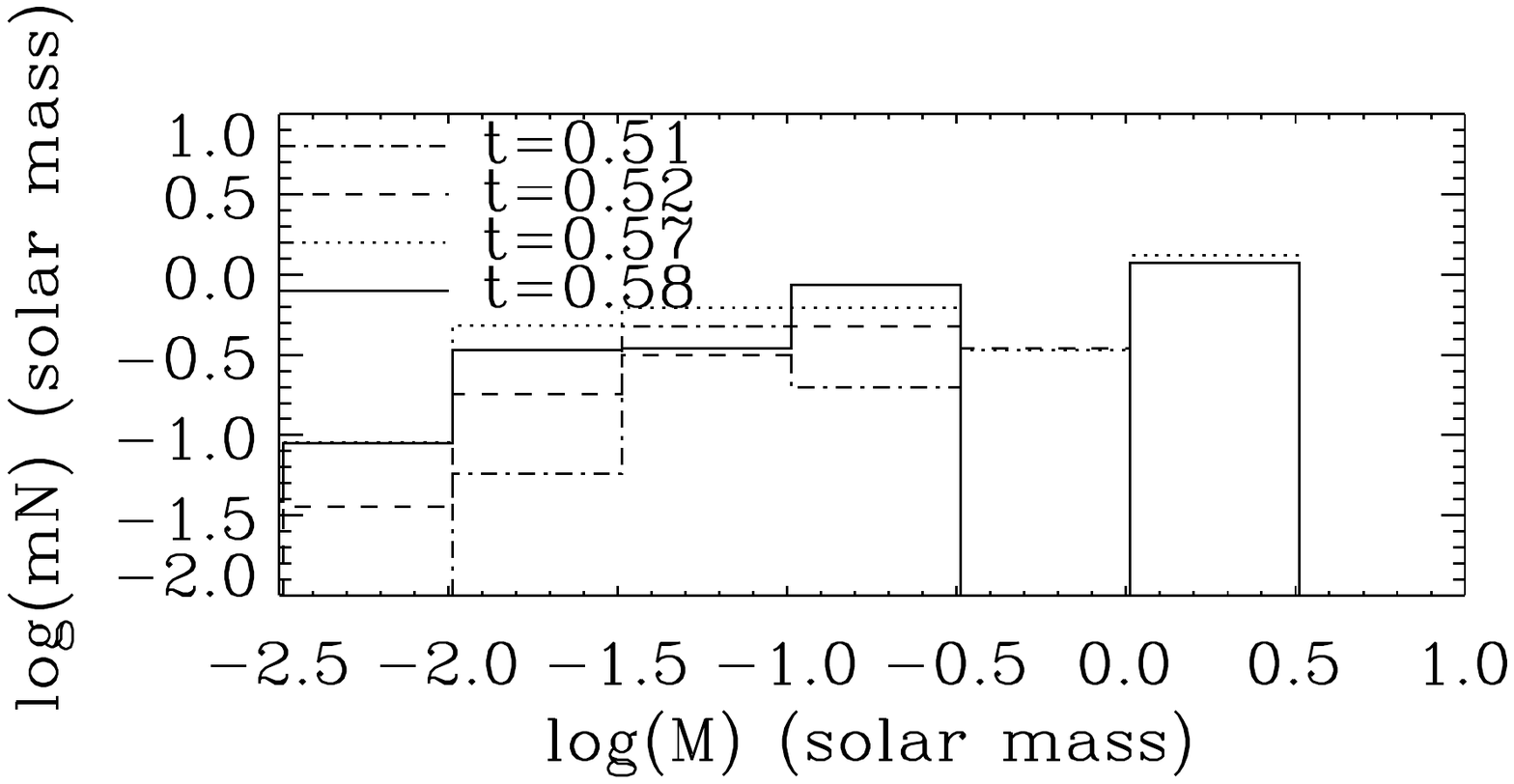}}
\put(8,0){\includegraphics[width=7.5cm]{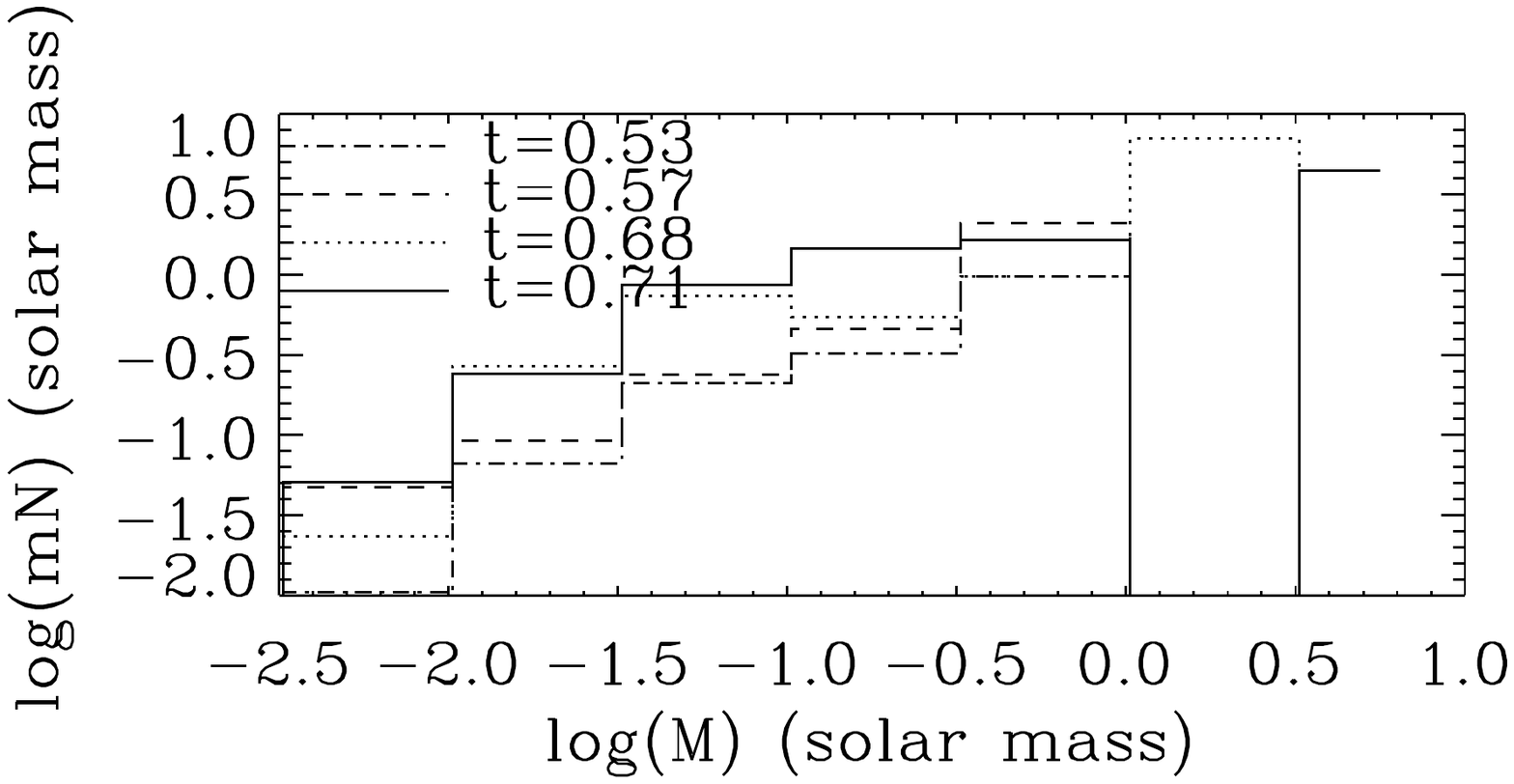}}
\put(0,3.5){\includegraphics[width=7.5cm]{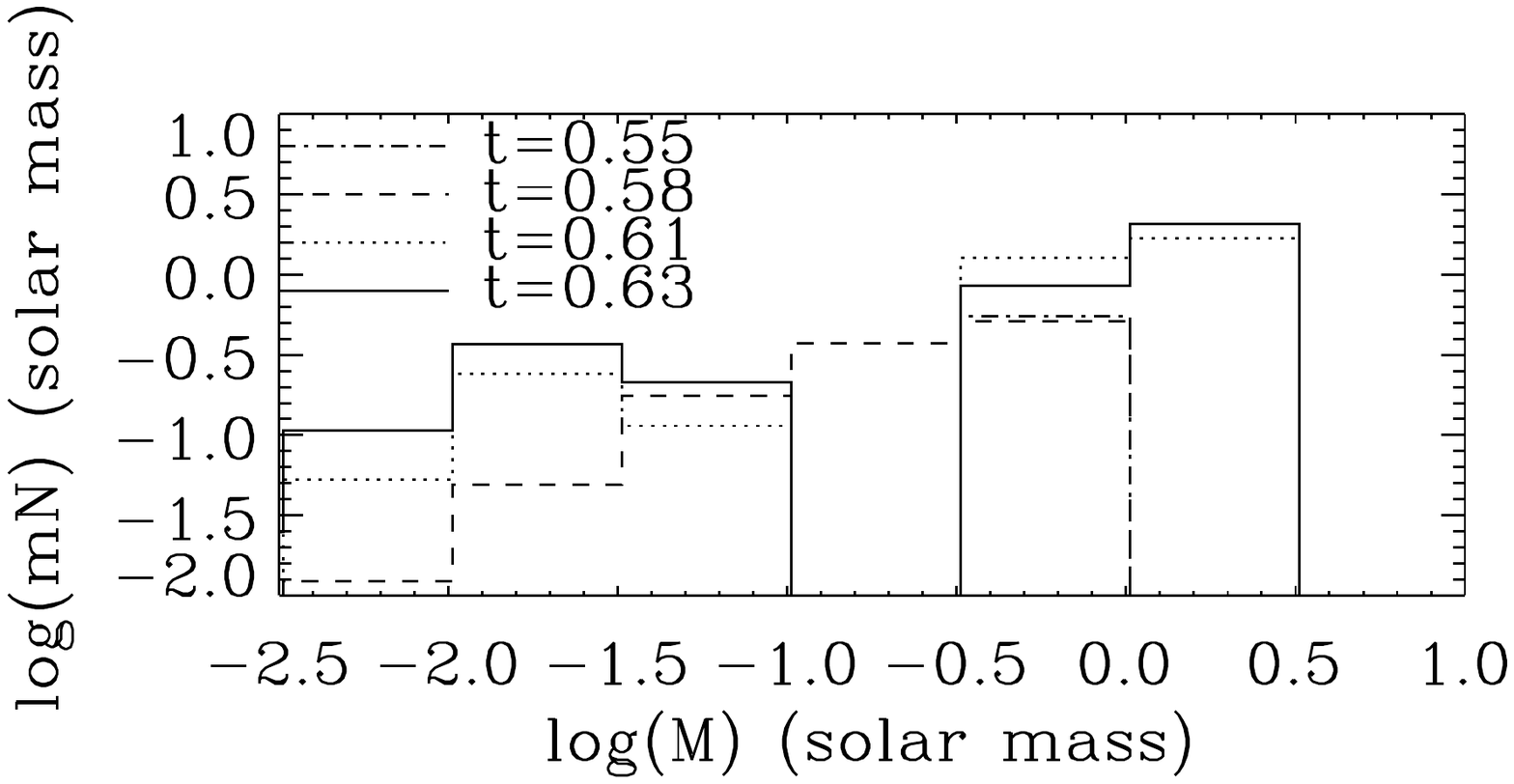}}
\put(8,3.5){\includegraphics[width=7.5cm]{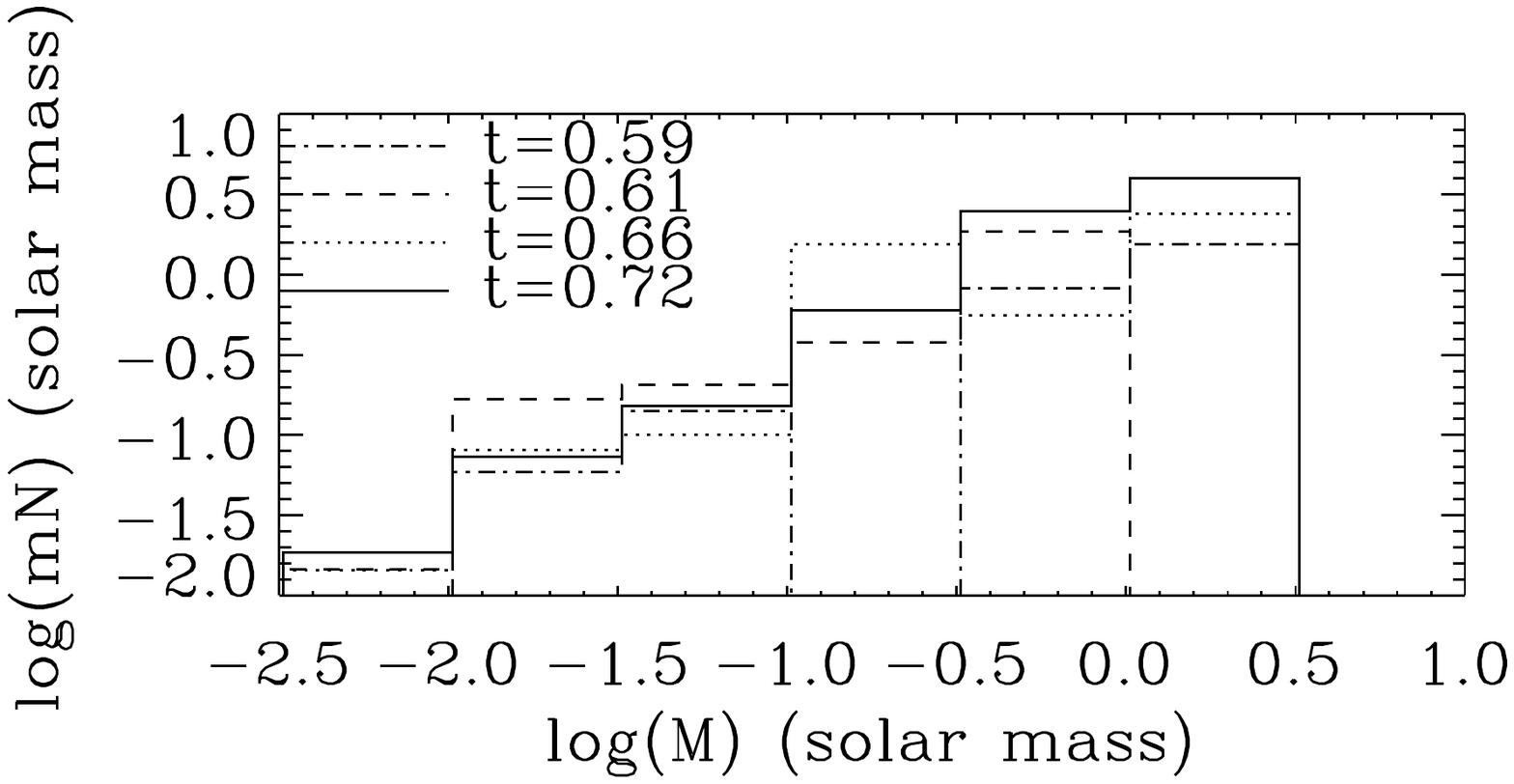}}
\put(0,7){\includegraphics[width=7.5cm]{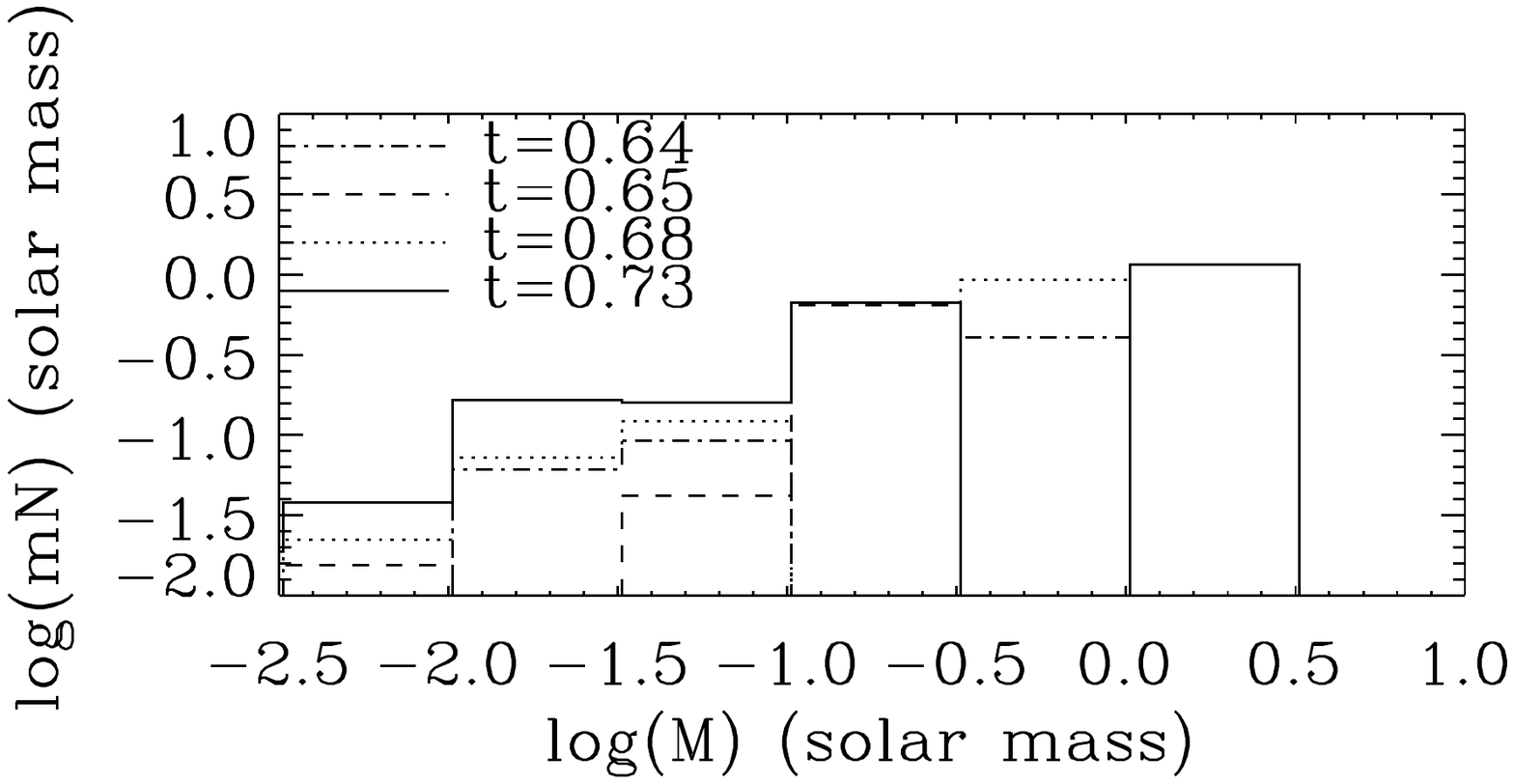}}
\put(8,7){\includegraphics[width=7.5cm]{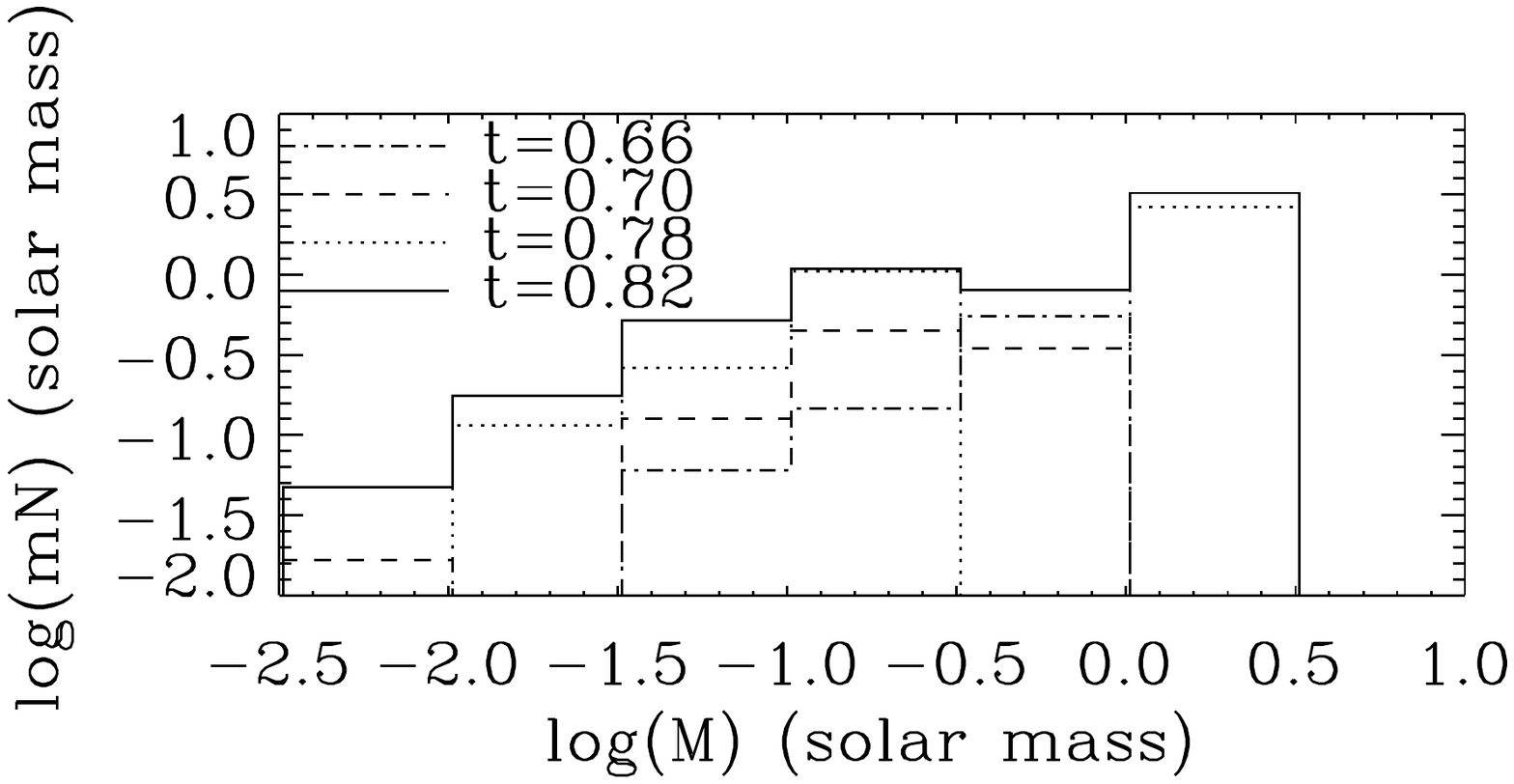}}
\end{picture}
\caption{Total mass of fragments per interval of mass for various time
steps. The left column shows the high resolution simulations, while 
the right column shows the lower resolution.
Top panels display the $\mu=2$ case, middle panels the $\mu=5$ ones, while
the bottom panels display the $\mu=120$ case. Times are given in Myr.}
\label{fragment}
\end{figure*}

Although a detailed comparison does not seem to be possible at this stage, it is
worth mentioning that observationally  outflows in massive
cores have been studied in details (e.g. Arce et al. 2007).
In general, a broad variety of flows have been observed and we restrict our attention 
to the study of Beuther et al. (2002b), which present similarities
with our results.
They find  multiples outflows, well collimated, containing about 10 $M_\odot$
and with velocities of the order of a few 10 km s$^{-1}$. Our outflows contained
about ten times less mass and are less rapid on average, though velocities of that 
orders are reached. It should be the case that a better agreement could be obtained at later 
stage (as suggested by the low resolution calculations) since both the mass and the velocities
 increase as the collapse proceed. Another related issue is the fact that 
the regions they observe has stars more massive ($\simeq 10 M_\odot$) than the 
{\it stars} present in our simulations. That may indicate again that we have to wait for 
longer time or that the core in the simulations are less massive than the regions 
observed by Beuther et al. (2002b). 
  It is particularly 
interesting to note that in the $\mu=5$ case, the outflows
are well collimated and nearly quadrupolar, a feature also mentioned by Beuther et al. (2002b). 

Finally, it should be made clear  that, as we do not treat the second collapse and the 
formation of the protostar itself, we do not form the jets as is  the 
case for example in the study of Banerjee \& Pudritz (2006) and Machida et al. (2008).
The jets have much faster velocities than the outflows and would therefore 
trigger further outwards motions in the cloud. The question as to whether the 
jets are driving the observed outflows and constitute the primary source for the outflows
is not settled yet. Would this be the case, 
then the outflows produced in this way should dominate over the outflows 
obtained in this work which are directly launched at large scales
through the magneto-centrifugal mechanism.

\section{Fragmentation}
In this section, we discuss the fragmentation which occurs in the simulations. 
As we do not have sink particles at this stage, we identify the dense clumps 
using a simple density threshold of $10^{11}$ cm$^{-3}$. To construct the clumps
we use a friend of friend algorithm, that is all cells above the density 
threshold, which are spatially connected are assigned to belong to the same entity. 
The mass of the clumps can then be obtained by summing over the 
constituting cells. One of the drawback of this method is the fact that the clumps
can merge while the stars, that would have formed if the collapse would have been properly 
followed up to the formation of the protostars, may have not. This problem could  
partially be solved if sink particles were used (Bate \& Burkert 1997, Krumholz et al. 2004, Federrath et al. 2010). 
However, sinks may alter significantly the 
evolution of the calculations in particular in the presence of magnetic field and we do
not  use them at this stage.

\subsection{Qualitative description}
Figures~\ref{100_hy} ($\mu=120$ case), \ref{100_mu5} ($\mu=5$) and \ref{100_mu2}
($\mu=2$)  show 3 snapshots of
the cloud column density, around the density peak, integrated along the z-axis (top row) and 
along the x-axis. Each image represents a length of 2000 AU. The first time displayed  
is close to the formation of the first protostars (typically $10^4$ years) and the third corresponds to 
the last timesteps of the simulations (these panels are zoom of Fig.~\ref{col_dens}), which is about 6-7$\times 10^4$ years  
after the formation of the first protostar while the second 
is  intermediate.

In the three cases, it is seen that many objects form and that their numbers increase 
with time as  accretion proceeds. This  is relatively unsurprising giving that 
the thermal over gravitational energy ratio is initially equal to about 0.12 implying that 
the cloud contains about 16 Jeans masses at the beginning.

In the $\mu=120$ case, the objects are relatively distant from each 
other and the distance of the more distant objects is of the order of 1000 AU. This length
roughly corresponds to the radius below which the infall velocity is  smaller
or comparable to
 the velocity fluctuations as shown by Fig.~\ref{radial_velocity}. This implies 
that fragmentation occurs when some sort of dynamical equilibrium, or at least non-uniformly 
collapsing region, is 
reached as it is the case in the inner region of radius $1000-2000$ AU. 
By comparison it is seen that the fragments are significantly closer in the magnetized cases 
as suggested by Fig.~\ref{radial_velocity}, which shows that the size of the central region where the 
velocity is essentially random, is typically 3 to nearly 10 times smaller. As  discussed previously,
this is a consequence of the magnetic braking which extracts angular momentum from the inner region as shown by 
Fig.~\ref{mom}. 

The comparison
between the three cases also reveals that the $\mu=120$ case fragments more and earlier
than  the magnetized cases, in particular the $\mu=2$ case fragments significantly less 
than the $\mu=120$ case. Although a quantitative analytical estimate is mandatory here, it is 
qualitatively not surprising. First as revealed by Fig.~\ref{alfven_vol}, the magnetic field is 
 higher in the central part for the $\mu=5$ and $\mu=2$ cases than for the $\mu=120$ case
by a factor of about 3. As the magnetic support dominates the thermal one, this implies 
that the total support is indeed few times larger. Second, as mentioned above the angular momentum is 
smaller in the magnetized cases while large angular momentum tends to favor fragmentation (e.g. Miyama 1992, 
Machida et al. 2005). 

We stress nevertheless that even if reduced, the magnetic field is, for the case explored here, 
not suppressing fragmentation. This is expected since, as mentioned earlier, the cloud 
contains about 20 Jeans masses initially and the large turbulence present in the cloud 
triggers large density perturbations (remembering that the rms Mach number is of the order of 2 initially).
Indeed, previous authors have concluded that while a magnetic field can easily quench 
rotationally driven fragmentation (Price \& Bate 2007, Hennebelle \& Teyssier 2008), 
that is the fragmentation of massive self-gravitating disks, large perturbations can lead
to fragmentation even when the magnetic field is relatively strong. This is because while 
the magnetic field is strongly amplified by the differential rotation, this is not the case 
when  isolated Jeans masses collapse individually. 

Finally, we note that for $\mu=2$, fragments at much larger distances form (see top panel of Fig.~\ref{col_dens})
than in the less magnetized cases. This is because the $\mu=2$ case is closer to an equilibrium and 
collapses less rapidly, leaving time for the fluctuations induced by turbulence to develop. We note that 
this behavior is reminiscent of the recent observations by Bontemps et al. (2010),
 who find that massive 
core are fragmented at scales of a few thousands of AU in a few ($\simeq 1-3$) objects.

\subsection{Quantitative estimate}
We now present a more quantitative analysis of the fragment distribution.

Figure~\ref{number_fragment} displays the number of fragments more massive than 
$10^{-2}$ solar masses as a function of the total mass, $M_f$, within the fragments,
that is $M_f$ is equal to the sum of all the fragment masses. 
Several interesting trends can be inferred. 
First, as expected, the number of fragments clearly tends to increase with $M_f$. 
There are fluctuations that are caused by  clump merging and also 
by our algorithm that is based on a simple density threshold, 
which is used to identify the clumps.
Second, at the end of the high resolution 
runs, the number of fragments is about 2.5 times less for $\mu=2$  than 
for $\mu=120$. As both simulations have been run for about the same physical  
time after the formation of the first protostar, this implies that the more magnetized
case fragments later. Third, in terms of mass, obviously for the same
value of $M_f$, the number of fragments is lower for $\mu=2$ and $\mu=5$ 
than for $\mu=120$ by a factor of about $\simeq 1.5-2$. This indicates that for the 
same amount of mass, there are less fragments when the magnetic field is significant.
The same trends are also seen in the low resolution calculations although slightly 
less clear. Because the number of fragments is slightly 
larger in the high resolution case (although only fragments more massive than $10^{-2}$ solar 
masses are shown) and as already shown by Fig.~\ref{mom}, numerical resolution 
 clearly appears to be an issue here. It is therefore possible that higher resolution 
calculations could show a stronger difference between low magnetized and highly
magnetized calculations.

Figure~\ref{fragment} shows the mass spectrum, more precisely the mass per interval of mass, 
 for the six simulations and for four different times. The later time corresponds to the 
last timesteps calculated, the first is close to the moment when 
protostar formation begins and the two others are intermediate. 
The fragment masses are distributed between 
$3 \times 10^{-3} \; M_\odot$ and 3 $M_\odot$. 
For both set of simulations, high and low resolution, 
the trends are the same. The number of fragments is larger for 
$\mu=120$  and decreases as magnetic intensity increases (see also Fig.~\ref{number_fragment}
where the number of fragments can be seen).
The mass is slightly more concentrated on the more massive fragments in the MHD cases.

Although performing a  quantitative analysis of the fragment distribution 
(e.g. Hennebelle \& Chabrier 2008, 2009)
seems difficult at this stage, the decrease of the fragmentation with magnetic 
intensity seems to be attributable to two different processes, as already discussed.
First the angular momentum is larger for $\mu=120$  and second 
the magnetic support is obviously stronger for  $\mu=5$ and $\mu=2$.
This latter effect can be understood in the following way. As discussed 
in Mouschovias \& Spitzer (1976), the highest external pressure for which  
a cloud of mass $M$ and temperature $T$ can be supported is 
given by
\begin{eqnarray}
P_{{\rm ext}} \propto {{ (kT / m_p)^4} \over {G^3 M^2 (1-\mu^{-2})^3}},
\label{press_mous}
\end{eqnarray}
$m_p$ being the mean mass per gas particle. Above this pressure, a cloud of mass $M$
collapses. Thus, the pressure needed to produce 
fragments of mass $M$ increases when $\mu$ decreases. For
$\mu=2$, one finds that the pressure must typically be about 2.4 times higher. 
This value is, however a lower limit because as $\mu$  is the mass-to-flux ratio
of the fragment of mass $M$ and not the mass-to-flux of the whole cloud, 
$\mu$ must be lower than 2
 \footnote{It is strictly equal to the  mass-to-flux ratio of the whole cloud 
if all the matter originally contained in the flux tube that threads the fragment 
is contained within the fragment}. This means that the external pressure required to form 
a fragment of mass $M$ must  be even  higher.
If, for example 3/4 of the mass within the flux tube has contracted and is 
available to form a new fragment, the mass-to-flux ratio of the 
material is about 1.5 and the external pressure would be of the order of 6
times its value in the hydrodynamical case. 
Note that Eq.~(\ref{press_mous}) stems from the virial theorem as shown
on Eq.~(\ref{virial}), which in particular does not distinguish
between the magnetic tension and the magnetic pressure. More detailed 
analyse by Li \& Shu (1997) and more recently by Lizano et al. (2010) 
while finding  the rescaling of the gravitational term by a factor $(1-\mu^2)$ 
as well, infer that the sound speed should also be modified, 
$C_s^2 \rightarrow \Theta C_s^2$.
Li \& Shu (1997)  infer that $\Theta$ depends on $\mu$ as well as 
on the ratio of the horizontal over vertical components  of the gravitational field, 
but always remains below 2 (and would be smaller in our case as $\mu =2$), while Lizano et al. (2010) infer 
that $\Theta=1+V_a^2/C_s^2$.  The differences seem to arise from different assumptions 
linked to the thin disk approximation. In the present calculation, the Alfv\'en speed
in the inner fragmenting region is typically a few times (up to 10) higher
than the sound speed. As the Jeans mass is proportional to $C_s^{3}$, this would 
have a very significant impact on the effective Jeans mass that should be multiplied 
by a factor $\simeq 10^3$, which does not seem to be the case in the simulation, 
 while following the Li \& Shu (1997) approach, we would get 
an increase of the Jeans mass, due to the effective sound speed being larger,  by less 
than a factor $2^{3/2} \simeq 3$. Because the geometry adopted in these works (thin disk 
geometry) is different from the complex turbulent case that is  
characteristic of our simulations, 
and given that the coefficient renormalising the sound speed is uncertain,
 we adopt the simplest prescription stated by Eq.~(\ref{press_mous}),
keeping in mind that this is presumably a lower value.

As shown in Fig.~\ref{density}, the mean density roughly scales 
as $r^{-2}$ and is comparable in the three cases explored here. This implies
that fragments of mass $M$ should form at  radii a few times smaller 
 for $\mu=2$  than for $\mu=120$, a reasonable estimate being   
$(1-\mu^{-2})^{3/2}$ with $\mu<2$, which suggests a factor of the order $\simeq$1.5-2.5.
Since, as already mentioned, the density is roughly proportional to $r^{-2}$, the mass
contained within a given radius, $r$, is roughly scaling as $r$. This implies 
that the mass available for forming fragments of mass $M$ is also 
1.5-2.5 times smaller for $\mu=2$  than for $\mu=120$. Thus a factor 
of the order of 2 on the number of fragments. Note that taking into account 
a factor, $\Theta$, of the order of the one proposed by Li \& Shu (1997), we may
expect that the pressure in Eq.~(\ref{press_mous}) is  multiplied 
by a factor $\Theta^4$ and consequently that the radius at which fragments of mass $M$ could form
is divided by a factor $\Theta^2$ implying that the  available mass is also 
decreased by the same factor.
In this case, the number of fragments should be additionally reduced by a factor of less than 4 (which corresponds
to the highest value of $\Theta=2$ inferred by Li \& Shu 1997).
Altogether, this suggests that the number of fragments that we observe is higher than what 
is theoretically expected.
As discussed below, significant flux leakage is probably 
reducing the effect of the magnetic field.

It is worth mentioning that the number of initial magnetic Jeans masses 
is about half the number of thermal one, which could offer an alternative explanation. 
However, the masses of the objects formed during the collapse are much smaller than 
the initial value of the Jeans mass, and it is unclear to which extent they
can be related.

\setlength{\unitlength}{1cm}
\begin{figure} 
\begin{picture} (0,12)
\put(0,0){\includegraphics[width=7.5cm]{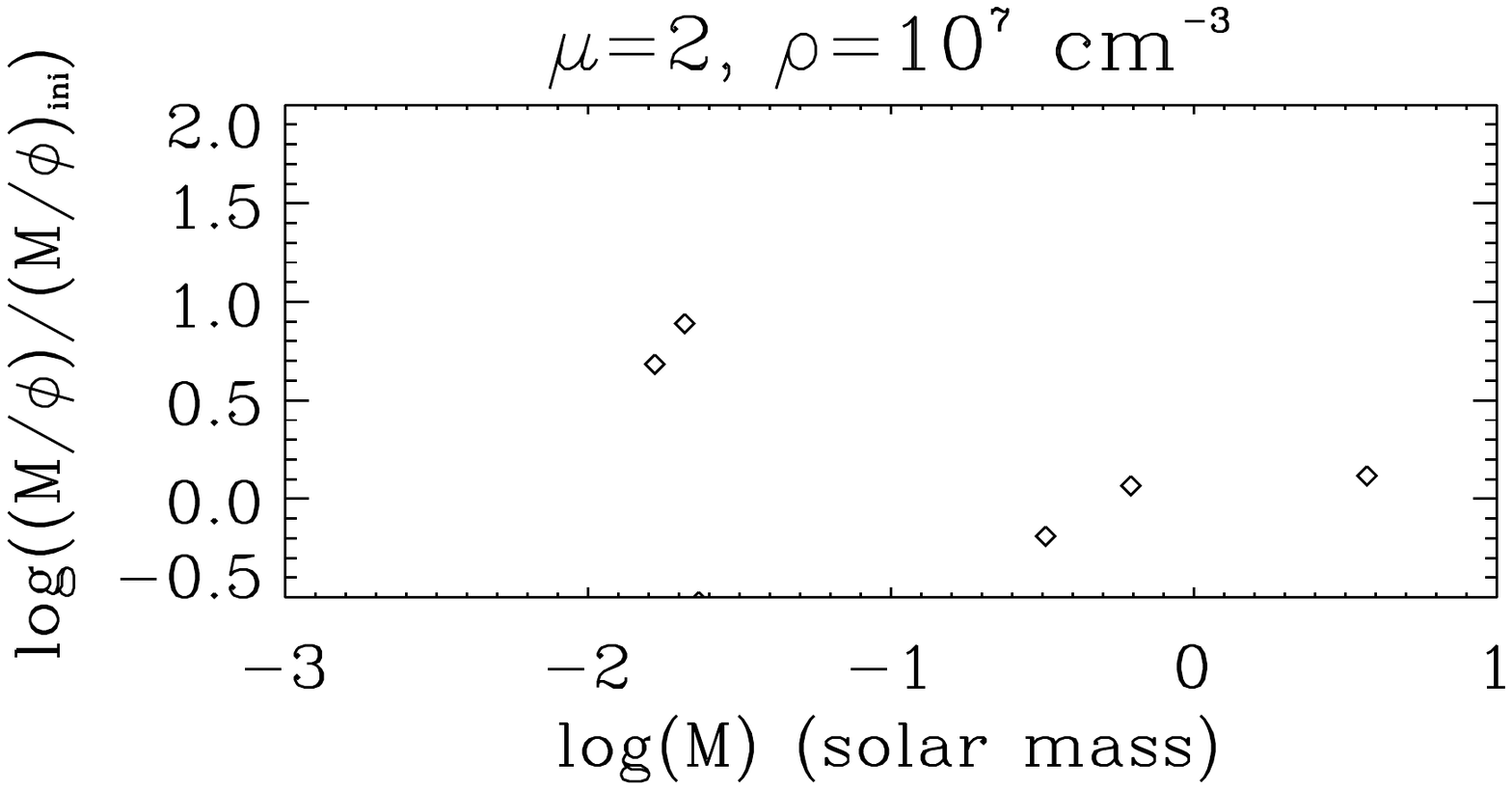}}
\put(0,4){\includegraphics[width=7.5cm]{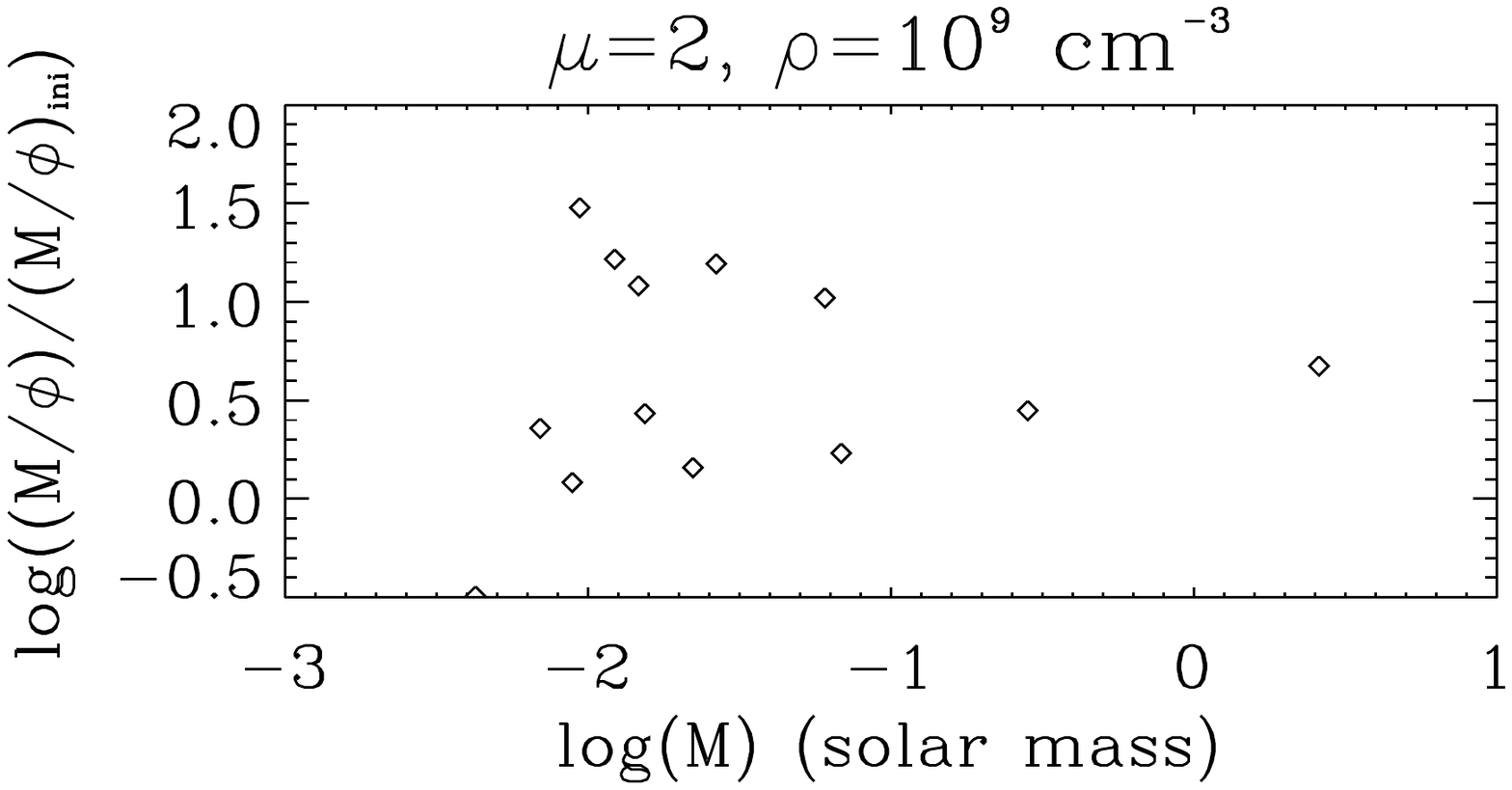}}
\put(0,8){\includegraphics[width=7.5cm]{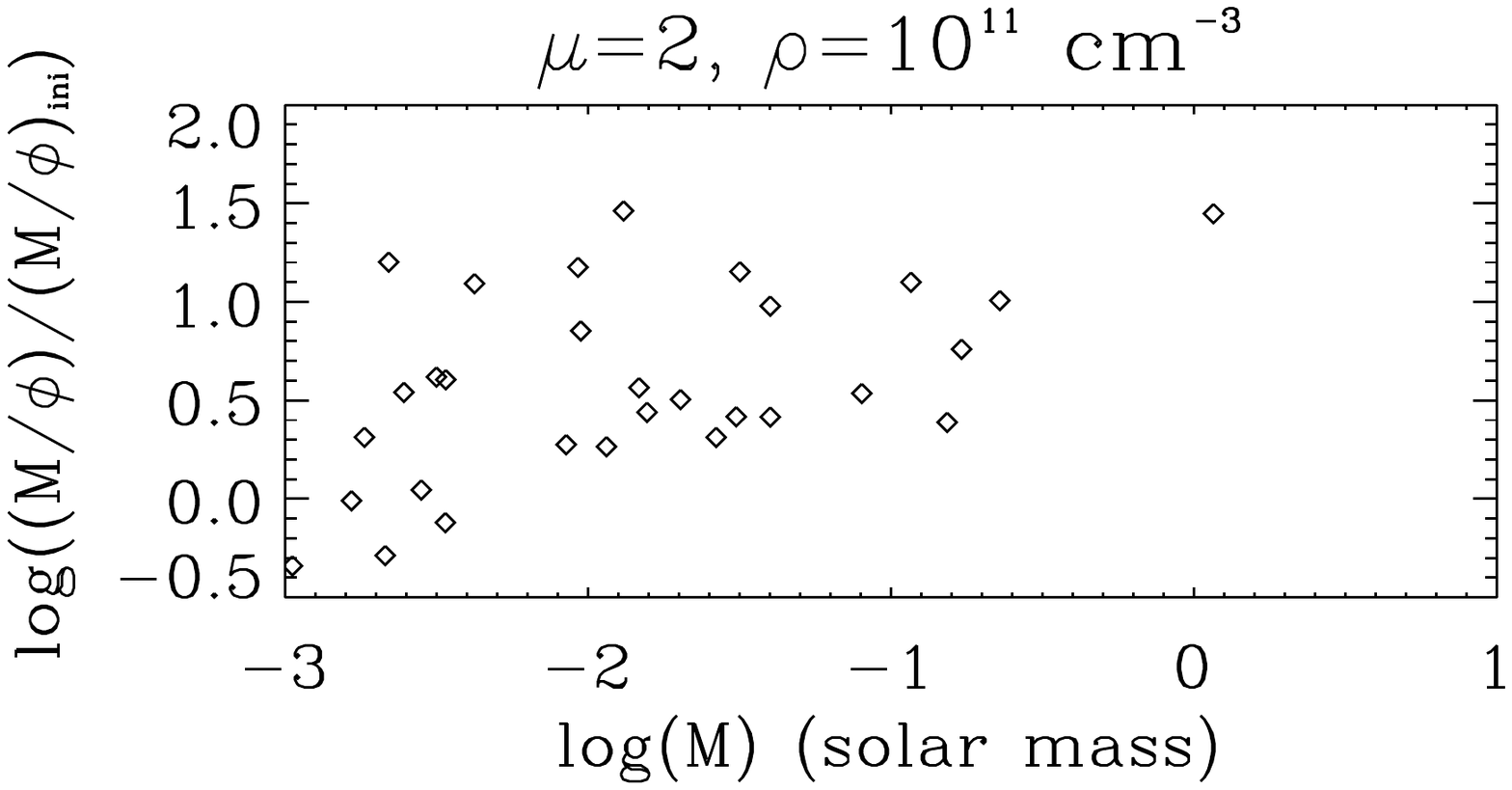}}
\end{picture}
\caption{Mass-to-flux over initial mass-to-flux ratio of the fragments 
as a function of mass for   $\mu=2$ at time 
$t=0.7301$ Myr in the high resolution run calculations. 
Three density thresholds are considered. For a threshold of $10^7$ cm$^{-3}$, 
the mass-to-flux of the fragments is very close to the initial value, while
it is typically 10 to 30 times higher for a threshold equal to $10^{11}$ cm$^{-3}$,
indicating that most of the magnetic flux was lost by numerical diffusion.}
\label{mass-to-flux}
\end{figure}

\subsection{Magnetic flux within fragments and numerical diffusion}
To  further characterize the fragments, we need to 
estimate  the magnetic flux, which  threads them. 
To calculate the magnetic flux, we proceed as follows. 
For each fragment, we calculate the magnetic flux of all the surface
parallel to the $x=0$, $y=0$, $z=0$, $x=y$, $x=z$ and $y=z$ planes 
intersecting  the volume of the fragment. The 
magnetic flux is then defined as the maximum of all these 
fluxes. 

Figure~\ref{mass-to-flux} shows the values of the 
mass-to-flux ratio over the initial mass-to-flux ratio in the fragments
for three different density thresholds, $10^7$, $10^9$, and $10^{11}$ cm$^{-3}$.
Obviously, it is typically close to 1 for $10^7$, which indicates good 
field freezing but about 10-30 for $10^{11}$ cm$^{-3}$, which shows that most of the magnetic flux
has been lost. Indeed, as already mentioned, 
the mass-to-flux ratio of any fluid particle could not be higher than 2 if 
magnetic flux was conserved. This implies that as already shown in 
Fig.~\ref{alfven_vol}, the numerical diffusion leads to strong magnetic flux losses
at scales smaller than $\simeq 20$ AU. Clearly, this raises the question 
as to whether the formation of most of the fragments would not have been 
prevented if the magnetic flux had been conserved.  In particular, the second panel
indicates that at densities of about $10^9$ cm$^{-3}$, that is to say more 
than an order of magnitude in density before the gas becomes adiabatic, 
about two-thirds of the initial magnetic flux have been lost. As it is 
typically at these densities that fragmentation occurs (see for example the difference
between the second and third panels of Fig.~\ref{mass-to-flux}), 
this may indicate that the fragmentation is indeed  overestimated in the $\mu=2$
case.
On the other hand, the reasonable similarity between 
the high and low resolution calculations suggests that this may  not be too severe a problem
but this question should be solved when it will be possible to 
performed higher resolution calculations.  Note that the lower resolution calculations
show very similar trends regarding the mass-to-flux values. However, it is 
worth recalling that in this work we have not varied the number of cells per 
Jeans length from the first amr levels, but simply allowed for two more amr levels 
for the high resolution runs. Thus the possible dependence of magnetic 
diffusivity with numerical resolution during the first stage of collapse 
remains to be investigated. 

In reality, it is expected that significant flux leakages 
 either induced by 
ambipolar diffusion  and/or ohmic dissipation 
(Nakano et al. 2002, Tassis \& Mouschovias 2005, Kunz \& Mouschovias 2010)
possibly enhanced by turbulence (Santos-Lima et al. 2010),
 occur. However, the question as to whether the numerical diffusion captures these effects 
accurately enough is open. In particular, it may be the case that significant flux leakage 
is indeed occurring, but at higher densities.
 The amount and the efficiency of these processes 
should be thoroughly quantified in future studies.

\section{Discussion}
Here we provide some discussion about the aspects of the work that should be improved
in future studies. We also speculate on the consequences this may have. 

The choice of the initial conditions is obviously crucial in this problem. 
Given the large amount of CPU necessary to run each case, it was impossible
to explore the influence of initial thermal energy, turbulent energy, and 
rotation. Varying the mass would also be necessary in the future. 
Different realizations of the turbulent velocity field should ideally 
be tested as well as different realizations of the initial magnetic field.
We also stress that in our initial conditions,  magnetic, velocity, and
density fields are set up independently, i.e. in reality fluctuations 
of magnetic and density fields should be correlated with the velocity fluctuations.
We did not consider any well organized rotation field, which could lead to a systematic
growth of a magnetic toroidal component and could possibly modify our conclusion.
Finally, fragmentation strongly depends on the initial density profile
(see e.g. Girichidis et al. 2010).

An important aspect that we did not attempt to address here is the 
statistics of the binary systems, that will form. As is evident from 
Figs.~\ref{100_hy}, \ref{100_mu5} and \ref{100_mu2}, the size 
of the regions where fragments form as well as the available angular momentum 
are quite different for the three values of the magnetic intensity.
What consequences this may have on the binary properties is 
an open question. This could however constitute an interesting test to know 
whether there is a preferred magnetization for high mass cores. This, as already discussed,
requires the use of sink particles. Before introducing them, it should 
however be investigated  how sink particles behave  in  the presence of 
a magnetic field.

Perhaps the most important restriction of the present study is 
the lack of radiative transfer, which has been demonstrated
to play an important role during the collapse of massive cores
(Yorke \& Sonnhalter 2002, Krumholz et al. 2007, Bate 2009) and even low mass cores
(Offner et al. 2009, Commer{\c c}on et al. 2010, Tomida et al. 2010). 
As discussed in the introduction, the role of the radiative transfer 
has been investigated by Krumholz (2007, 2010), Bate (2009), 
Kuiper et al. (2010), Peters et al. (abcd),
and they conclude that it has an impact on the gas temperature because of the heating
induced by the accretion luminosity.
The exact importance of this effect  remains
a matter of debate and could indeed vary with initial conditions. 
  When magnetic field and radiation are taken 
into account, Commer{\c con} et al. (2010) have suggested that 
the impact of the radiative feedback may be even larger, because 
as the magnetic braking removes angular momentum, the accretion rate is higher. 
That is, the trend inferred in this work regarding the reduced fragmentation
may possibly be amplified because the stronger accretion onto fewer objects will 
trigger a stronger radiation field that may consequently tend to reduce 
 the fragmentation even more. 
Another interesting effect may be caused by the magnetic outflows. As investigated 
by Krumholz et al. (2005), the outflows may channel the radiation and 
possibly modify its effect on the surrounding gas. The thermal and the radiative 
pressure should in principle add up to the Lorenz force and produce 
faster outflows. 

 We would like to reiterate that as only the first collapse is treated in this work,
 the outflows produced in our simulations are directly 
launched at large scales and are not the consequence of the entrainment from a fast wind
generated at small scales. It is possible that the flows produced that way are 
too slow, but one should also keep in mind, as pointed out in this study, 
that numerical resolution 
may be an important factor in getting higher velocities.

Finally, we stress that performing  integration over a longer timescale
is an important issue as already discussed. Because high numerical 
resolution is really needed here, this constitutes a severe 
problem.

\section{Conclusion}

 We  performed high resolution numerical simulations of collapsing 
magnetized and turbulent hundred solar masses cores assuming  a barotropic equation of state using the RAMSES code.
 Three different magnetic intensities
corresponding to mass-to-flux ratio, $\mu$ equal to 120, 5, and 2 were explored. 
The simulations 
were repeated with two different resolutions 
to investigate the impact of the numerical method and the issue of numerical
convergence.
These simulations confirm the strong impact that the magnetic field has, in particular 
regarding the byproduct of the collapse. 

The main effects of  the magnetic field are 
i) to  significantly reduce the angular momentum in the inner part of the cloud, 
ii) to launch episodic and relatively fast outflows, even when the value 
of the magnetic intensity is initially weak, iii) to reduce the fragmentation of the cloud 
in several objects (by about a factor 2 when $\mu$, the mass-to-flux ratio is equal to 2).

While the collapse is relatively well organized in the outer part of the cloud
and exhibites a classical $r^{-2}$ density profile,
the inner part is very turbulent and the infall is dominated by high velocity
fluctuations.  In this region, the density profile is  steeper 
and typically goes as $r^{-\simeq 2.5}$.
The magnetic field is amplified by gravitational 
contraction that leads to roughly $B \propto \sqrt{\rho}$, which in turn implies 
that the Alfv\'en velocity is nearly constant on average although it significantly
fluctuates at all scales. When the magnetic field is very weak ($\mu=120$), 
the amplification is stronger which makes in the cloud inner part
 the Alfv\'en speed of the order of the sound speed. 

The outflows appear to be episodic and are usually non-bipolar.
Not only their velocities evolve with time, but there are events of 
intense ejections followed by periods without significant outflow motions. 
 The typical velocity of these flows is of the order of 1-3 km s$^{-1}$ but 
much higher velocities (5 to 10 times higher) 
can be reached for a small fraction of the mass, in particular 
when the field is weak. The total mass they carry is 
 of the order of a tenth to a few solar masses, depending on the time
and the resolution. There is a clear 
influence of the numerical resolution, implying that 
these numbers are {\it probably}  underestimated.

The strongly magnetized clouds tend to fragment  less (factor 1.5-2)
than the weakly magnetized ones, implying that the mass is more 
concentrated in the more massive stars. The region in which fragmentation 
occurs is also  more compact when the magnetic intensity is stronger.
We stress however that numerical diffusion is clearly reducing the magnetic 
flux in the dense
part of the clouds, which makes it possible that the fragmentation is indeed 
overestimated for $\mu=5$ and 2.

\section{Acknowledgments}
We thank the anonymous referee for comments, which helped to improve the original
version of this work.
PH thanks Daniele Galli for enlighting discussions about the role of the 
magnetic field in the  collapse of  prestellar cores.
RSK thanks Robi Banerjee, Rainer Beck, Florian
B\"{u}rzle, Paul Clark, Christoph Federrath, Simon Glover, Dominik Schleicher,
Kandaswamy Subramanian, and Sharanya Sur for stimulating discussions on magnetized
collapse and the small-scale dynamo.
This work was granted access to HPC resources of CINES under the 
allocation x2009042036 made by GENCI (Grand Equipement National de Calcul Intensif).
PH, MJ, and RSK acknowledge financial support from the Institut des sciences de l'univers (CNRS) and
the German Bundesministerium f\"ur Bildung und Forschung via the ASTRONET project STAR FORMAT.
RSK acknowledges financial support from the Baden-W{\"u}rttemberg
Stiftung via their programme International Collaboration II
(grant P-LS-SPII/18). R.S.K. furthermore gives thanks for subsidies
from the Deutsche Forschungsgemeinschaft (DFG) under grants no. KL 1358/1,
KL 1358/4, KL 1359/5, KL 1358/10, and KL 1358/11, as well as from a Frontier
grant of Heidelberg University sponsored by the German Excellence Initiative.
MRK acknowledges support from an  
Alfred P. Sloan Fellowship; the US National Science Foundation through  
grants AST-0807739 and CAREER-0955300; and NASA through Astrophysics  
Theory and Fundamental Physics grant NNX09AK31G and through a Spitzer  
Space Telescope Theoretical Research Program grant.
The research of BC is supported by the postdoctoral fellowships from
Max-Planck-Institut f\"{u}r Astronomie.
JCT acknowledges support from NSF CAREER grant AST-0645412 and NASA  
Astrophysics Theory and Fundamental Physics grant ATP09-0094.

\appendix

\end{document}